\documentclass[fleqn,usenatbib,usedcolumn]{mnras}

\pdfoutput=1

\usepackage{graphicx} 

\usepackage{newtxtext,newtxmath}

\usepackage[T1]{fontenc}

\usepackage{advdate}

\DeclareRobustCommand{\VAN}[3]{#2}
\let\VANthebibliography\thebibliography
\def\thebibliography{\DeclareRobustCommand{\VAN}[3]{##3}\VANthebibliography}

\usepackage{graphicx}	

\newcommand{\HD}{HD~163296}
\newcommand{\lmla}{\texttt{Lm}$\alpha$6}	
\newcommand{\hmha}{\texttt{Hm}$\alpha$5}	
\newcommand{\hmls}{\texttt{HmL}$\Sigma$}	
\newcommand{\mjup}{\mbox{$M_{\mathrm{J}}$}} 
\newcommand{\ms}{\mbox{$M_{\star}$}}        
\newcommand{\AU}{\mbox{au}}                 
\newcommand{\Rhill}{\mbox{$R_\mathrm{H}$}}  
\newcommand{\sigu}{\mbox{$\mathrm{g\,cm}^{-2}$}} 
\newcommand{\rhou}{\mbox{$\mathrm{g\,cm}^{-3}$}} 

\title[Second-generation dust in \HD]{%
Second-generation dust in planetary systems: The case of \HD\thanks{Accepted for publication in Monthly Notices of the Royal Astronomical Society Main Journal. This version of the manuscript was prepared and typeset by the authors.}}

\author[D'Angelo and Marzari]{
Gennaro D'Angelo,$^{1}$\thanks{E-mail: gennaro@lanl.gov}
and Francesco Marzari$^{2}$\thanks{E-mail: francesco.marzari@pd.infn.it}
\\
$^{1}$Theoretical Division, Los Alamos National Laboratory, Los Alamos, NM 87545, USA\\
$^{2}$Department of Physics and Astronomy, University of Padova, via Marzolo 8, I-35131, Padova, Italy\\
}

\date{Accepted 2 November 2021. Received 17 October 2021; in original form 23 August 2021}

\pubyear{\the\year}

\begin{document}
\label{firstpage}
\pagerange{\pageref{firstpage}--\pageref{lastpage}}
\maketitle

\begin{abstract}
Observations indicate that large, dust-laden protoplanetary discs are common.
Some features, like gaps, rings and spirals, suggest they may host young 
planets, which can excite the orbits of nearby leftover planetesimals.
Energetic collisions among these bodies can lead to the production
of second-generation dust.
Grains produced by collisions may have a dynamical behaviour different 
from that of first-generation, primordial dust out of which planetesimals 
and planets formed.
We aim to study these differences for the \HD\ system and determine 
whether dynamical signatures in the mixture of the two dust populations
can help separate their contributions.
We use three-dimensional (3-D) hydrodynamic models to describe 
the gaseous disc with three, Saturn- to Jupiter-mass, embedded planets. 
Dust grains, of sizes $1\,\mu\mathrm{m}$--$1\,\mathrm{mm}$, are 
treated as Lagrangean particles with resolved thermodynamics and mass loss.
Initial disc and planet configurations are derived from  observation-based 
work, which indicates low gas viscosity.
The 3-D approach also allows us to detect the formation of vortices induced
by Rossby waves, where dust becomes concentrated and may contribute to
planetesimal formation.
We find that the main differences in the dynamical behaviour of first-
and second-generation dust occur in the vertical distribution.
The two populations have similar distributions around the disc
mid-plane, although second-generation dust shows longer residence times
close to the radial locations of the planets' gas gaps. 
Sedimentation rates of $\mu\mathrm{m}$-size grains are comparable to
or lower than the production rates by planetesimals' collisions, making
this population potentially observable. These outcomes can be extended 
to similar systems harbouring giant planets.
\end{abstract}

\begin{keywords}
accretion, accretion discs --
methods: numerical ---
planets and satellites: gaseous planets ---
planet–disc interactions ---
protoplanetary discs ---
stars: individual: HD~163296
\end{keywords}



\section{Introduction}
\label{sec:Intro}

\HD\ is a young, Herbig Ae star surrounded by a protoplanetary disc
that extends out to at least $450\,\AU$
\citep[e.g.,][]{wisniewski2008,rich2019}.
ALMA imaging at millimeter wavelengths detected three concentric, bright 
ring-like features in the disc \citep{isella2016,muro-arena2018},
located beyond $50$--$60$~\AU\ from the star 
\citep[see also][and references therein]{rab2020}.
Concentric dark regions were also detected interior of these bright rings. 
Observations of a fourth, wider ring (at around $330\,\AU$) were reported 
by \citet{rich2020}.
The concentric dark features reflect some amount of depletion, or gaps, in 
the distribution of dust and gas around the star, whereas bright rings are
likely regions of enhanced dust concentration.

One possible mechanism that can produce the observed disc structures 
relies on the presence of planets at or around those orbital locations.
Evolutionary models by \citet{baraffe2003} suggest luminosity-based 
detection limits in the range from $1.5$ to $4.5$ times Jupiter's mass, 
\mjup, with a mass limit of $\approx 2\,\mjup$, for the outer planet
\citep[associated to the third gap,][]{rich2019}.
Upper limits of a few \mjup\ for direct imaging were also reported by 
\citet[][]{mesa2019}.
Modelling of the disc dust distribution suggests the presence of smaller, 
Saturn-mass planets in the second and third gap \citep[][]{isella2016}. 
At those same locations, modelling of the gas rotation 
curves, tracked by CO emission, suggests the presence of $\approx 1\,\mjup$ 
planets \citep{teague2018}.
The presence of three planets, in the mass range 
$\approx 0.5$--$0.6\,\mjup$, was proposed by \citet{liu2018} to reproduce
observations of the three inner gaps.

In all the cases mentioned above, modelling involved hydrodynamic simulations 
of disc-planet tidal interactions.
It should be pointed out, however, that other physical processes, such as
grain growth at condensation fronts of particular gas species, may as well 
be responsible for the observed ring features \citep[see][]{vandermarel2018}.
Moreover, any physical process capable of producing radial variations of 
the gas turbulence viscosity \citep[see, e.g.,][]{ohashi2019} or
perturbations of the gravity field 
\citep[e.g., aided by dust back-reaction,][]{takahashi2014}
may also be able to produce radial features in the disc.

Current observations of thermal emission suggest that \HD\ contains dust
grains out to about $200\,\AU$ \citep[e.g.,][]{dullemond2020}. Part of this
dust is likely primordial, i.e., it was in the medium out of which
the system formed. Yet, if giant planets as those predicted by some models
exist, second-generation dust may also be found. 
In fact, the formation of giant planets requires the presence of km-size
(and larger) planetesimals 
\citep[e.g.,][and references therein]{alibert2018,voelkel2020,gennaro2021},
which must have emerged from the primordial dust.
Leftover planetesimals that were not incorporated into the planets continue
orbiting the star, and those moving in the proximity of the planets have 
their orbits excited by the planets' gravity. 
This process may lead to collisional comminution of planetesimals 
\citep[e.g,][]{turrini2019} and, hence, to the production of
second-generation dust. 

At an age of $7.6$ million years \citep{vioque2018}, the primordial
dust may have aged to an extent such to render it compositionally different 
from  younger, collisionally-formed dust. Yet, distinguishing the two
populations based on composition may prove difficult, if not impossible,
since collisions would continuously add new dust to the system and, thus,
second-generation grains would have a wide range of ages. 
The size distribution of the two populations may also be different,
but the difference would depend on poorly known details on the outcome
of collisional comminution.
Coagulation and weathering would also contribute to homogenise the species.

The focus of this study does not rest on the possible compositional or size
differences between the two dust populations, but rather on their possible 
dynamical differences.
In fact, gravitational interactions between planetesimals and the planets
would raise orbital eccentricity and inclination of the former
\citep[][]{turrini2019}. 
Therefore, whereas primordial dust would be expected to orbit mostly in 
(or close to) the disc mid-plane, and with an eccentricity distribution
determined by the gas motion, collisions among planetesimals may generate 
dust grains with much broader distributions of both eccentricities and
inclinations.

One purpose of this study is to analyse, through hydrodynamic modelling,
what differences can arise in the distributions of dust species that
may represent a primordial population and a second-generation, collisionally
produced population of particles, orbiting in the disc region that harbours
the predicted planets.
Another purpose is to determine the extent to which an admixture of the 
two populations can bear dynamical signatures that may allow us 
to separate their contributions.

In the following, we describe the physical properties adopted
for the system in Section~\ref{sec:PDS}. The computational methods are
outlined in Section~\ref{sec:CM} and the results of the models are
described and discussed in Section~\ref{sec:R}. Our conclusions are
presented in Section~\ref{sec:C}.

\section{Physical Description of the System}
\label{sec:PDS}

The physical description of the system is mainly based on the work presented
by \citet{isella2016}, \citet{liu2018}, and \citet{turrini2019}.
We assume a stellar distance $d\approx 122\,\mathrm{pc}$ and a stellar mass
$\ms=2.3\,M_{\sun}$ 
(\href{http://exoplanet.eu/catalog/hd_163296_b/}{http://exoplanet.eu}).
Although these parameters were revised following Gaia observations,
$d=101.5\pm 2\,\mathrm{pc}$ and $\ms=1.83\pm 0.09\,M_{\sun}$
\citep{vioque2018}, they are nonetheless representative of the system 
and do not impact the general conclusions of this study, which simply 
uses \HD\ as a proxy for a category of protoplanetary discs.

The initial surface density of the disc's gas is approximated to 
\citep{isella2016}
\begin{equation}
 \Sigma(t=0)=\Sigma_{0}\left(\frac{r_{0}}{r}\right)^{4/5}%
                  \exp{\left[-\left(\frac{r}{3.3r_{0}}\right)^{6/5}\right]},
 \label{eq:S}
\end{equation}
in which $r_{0}$ is a reference radius that depends on the stellar
parameters.
Because of the assumed distance, we set $r_{0}=50\,\AU$ (the revised distance
would result in a smaller value, $r_{0}=41.6\,\AU$).
Therefore, one orbital period at $r=r_{0}$ takes $233.13$ years.
The reference value of the initial density is $\Sigma_{0}=14.23\,\sigu$,
and the disc comprises $\approx 0.037\,M_{\star}$ of gas out to $500\,\AU$.
The power-law used for the surface density profile is somewhat shallower
than that adopted by, e.g., \citet[][]{dullemond2020}, although
the exponential cut-off is in the range examined by the latter authors.
Nonetheless, in light of \citeauthor{dullemond2020} results, a value
$\Sigma_{0}=6.0\,\sigu$ is also applied in order to mimic their profile
in the region $50\lesssim r\lesssim250\,\AU$ (note that their total
gas mass is $\approx 0.039\,M_{\star}$, close to the value stated above).

The temperature structure in the disc, taken as height-independent, is
\begin{equation}
 T=T_{0}\sqrt{\frac{r_{0}}{r}},
 \label{eq:T}
\end{equation}
in which $T_{0}=33.94\,\mathrm{K}$, as the mid-plane temperature chosen
by \citet{isella2016}.
We note that shallower temperature profiles have been proposed for this
system. In particular, \citet{flaherty2015} proposed a temperature power
index $\approx 0.3$ whereas \citet[][]{dullemond2020} proposed a much
flatter profile, $T\propto (1/r)^{0.14}$.
Nonetheless, the latter authors estimated a mid-plane temperature at
$r\approx 100\,\AU$ of about $25\,\mathrm{K}$, only a few degrees lower
than the temperature applied here.
Equation~(\ref{eq:T}) places the ice sublimation line at around $2.5\,\AU$
($T\approx 150\,\mathrm{K}$), at which radial location $1\,\mathrm{mm}$ ice 
particles would take less than an orbital period to evaporate.

According to Equation~(\ref{eq:T}), the disc is flared and its aspect ratio 
is given by 
\begin{equation}
 h=h_{0}\left(\frac{r}{r_{0}}\right)^{1/4},
 \label{eq:h}
\end{equation}
where $h_{0}=0.0537$.
The initial density of the gas scales as $\rho\propto \Sigma/H$ 
in the radial direction and as $\rho\propto \exp{[-(\cot{\theta}/h)^2]}$ 
in the meridional direction ($\theta$ is the co-latitude angle,
$\theta=\pi/2$ at the mid-plane).

We apply a simple prescription for the kinematic viscosity of the disc's 
gas, $\nu=\alpha H^{2} \Omega$, where $H=hr$ is the pressure scale-height,
$\Omega$ is the Keplerian orbital frequency of the gas, and $\alpha$ is 
the Shakura \& Sunyaev turbulence parameter \citep{S&S1973}.
Disc models have been proposed in which $\alpha$ varies with radius 
\citep[e.g.,][]{isella2016,liu2018}. 
Here we opt for a constant $\alpha$ since this study does not test 
different viscosity prescriptions or temperature distributions.

\begin{table}
	\centering
	\caption{Parameters applied to the cases investigated in this
	study. The table lists the turbulence parameter $\alpha$ of the
	disc's gas, the mass of the planets, and their semi-major axes.}
	\label{tab:cases}
	\begin{tabular}{llll} 
		\hline
		Model & \lmla & \hmha & \hmls\\
		\hline
		$\alpha$      & $10^{-6}$            & $10^{-5}$            & $10^{-5}$\\
		$\Sigma_{0}$ [\sigu]%
		              & $14.23$              & $14.23$              & $6.0$\\
		$m_{1}$/\ms   & $1.91\times 10^{-4}$ & $1.91\times 10^{-4}$ & $1.91\times 10^{-4}$\\
		$m_{2}$/\ms   & $1.91\times 10^{-4}$ & $4.15\times 10^{-4}$ & $4.15\times 10^{-4}$\\
		$m_{3}$/\ms   & $2.41\times 10^{-4}$ & $4.15\times 10^{-4}$ & $4.15\times 10^{-4}$\\
		$a_{1}/r_{0}$ & $1.2$                & $1.2$                & $1.2$\\
		$a_{2}/r_{0}$ & $2.1$                & $2.1$                & $2.1$\\
		$a_{3}/r_{0}$ & $3.2$                & $3.2$                & $3.2$\\
		\hline
	\end{tabular}
\end{table}
It is assumed that three planets orbit in the disc, with orbital radii
as indicated in Table~\ref{tab:cases}. We consider three basic scenarios,
two of which (\lmla\ and \hmha) by varying the mass of the planets 
\citep[``low'' and ``high'' masses, see][]{turrini2019} and
the $\alpha$ parameter, in the range of values considered by
\citet[][]{liu2018}.
In particular, the favoured model of the latter authors predicts depletion
factors in the gas density gaps around the inner two planets of somewhat 
less than $10$ and $3$, respectively (and smaller still around the third
planet). Note that these authors used a colder gas to model the disc,
which is more prone to open gaps by tidal torques.
As mentioned, a third model also considers a lower gas surface density 
(\hmls) which may affect dust dynamics. This model may also be viewed as
representing a later stage in the evolution of the system (i.e., an older,
more depleted disc). 
The planets interact gravitationally among themselves and with the star.
They act on the disc (gaseous and solid component) but do not feel 
the disc's gravity.

\section{Computational Methods}
\label{sec:CM}

We use the numerical methods described in 
\citet[][and references therein]{gennaro2012} 
to model the gas evolution and those presented in \citet{gennaro2015}
to model the evolution of solids. These methods were also applied in
\citet{marzari2019} and are briefly outlined in the next two sections.

\subsection{Gas Evolution}
\label{sec:GE}

The evolution of the gaseous disc is simulated by means of three-dimensional
(3D) hydrodynamic calculations. We use a spherical polar reference frame 
$\{\mathcal{O}; R, \theta, \phi\}$ whose origin $\mathcal{O}$ is attached
to the star. Quantity $R$ indicates the polar radius whereas
$r=R\sin{\theta}$ is the cylindrical radius. The disc domain extends
radially from $r=0.1\,r_{0}$ to $10\,r_{0}$ and $2\pi$ in azimuth ($\phi$)
around the star. The co-latitude angle $\theta$ is $\le \pi/2$ and the disc
is symmetric relative to its mid-plane, $\theta=\pi/2$.
The disc extends $12^{\circ}$ above the mid-plane, thus including
about $4$ pressure scale-heights $H$ at the reference radius $r_{0}$.
The coordinate system rotates around the axis $\theta=0$, perpendicular
to the mid-plane, at a constant angular velocity
$\Omega_{f}=\sqrt{GM_{\star}/r^{3}_{0}}$.

The gravitational potential in the disc is given by
\begin{equation}
 \Phi=\Phi_{\star}+\sum_{n}\Phi_{n}+\Phi_{\mathrm{I}},
 \label{eq:PHI}
\end{equation}
where $\Phi_{\star}=\sqrt{GM_{\star}/R^{3}}$ is the star's potential,
$\Phi_{n}$ is the potential generated by planet ``$n$'',
and $\Phi_{\mathrm{I}}$ is the indirect potential due to non-inertial
forces caused by the motion of the system's origin, $\mathcal{O}$.
To avoid singularities, the potential $\Phi_{n}$ includes a softening length
equal to $1/5$ of the planet's Hill radius.

The equations of motion of a viscous fluid are solved with an Eulerian, 
finite-difference code that is second-order accurate in both space and time. 
Orbital advection \citep{masset2000} is applied to increase the time-step 
of the calculations \citep[see][]{gennaro2012}.
The disc is represented by a grid comprising $662\times 22\times 422$ 
grid cells, in the $R$, $\theta$, and $\phi$ direction, respectively. 
Reflective boundary conditions are applied at both radial boundaries and
at the disc surface, whereas symmetry boundary conditions are applied at
the mid-plane.
The equations of motion of the planets in the disc are solved in 
the rotating reference frame as discussed in \citet[][]{gennaro2012}.
Although some amount of gas accretion on the planets is expected,
accretion of gas is neglected (and largely irrelevant for the purposes 
of this study). 

To gauge resolution sensitivity, some portions of the evolution were 
also performed at a higher resolution, 
$1332\times 42\times 842$ grid elements, for each of the three models 
in Table~\ref{tab:cases}. Comparisons of the results at the two
resolutions yield good agreement (see discussion in
Appendix~\ref{sec:DSP}). 

\subsection{Solids' Evolution}
\label{sec:SE}

Solids are treated as Lagrangian particles. 
They interact with the gas, the star and the planets, but not with each
other. Therefore, results can be re-scaled to an arbitrary number of
particles, as long as the solids' mass is small compared to the gas mass
in each grid cell. The calculations model particle dynamics, its thermal
state, and mass evolution 
\citep[through evaporation, for details see][]{gennaro2015}.
Dust is assumed to be ice (mass density $\rho_{s}=1\,\rhou$), although
an admixture containing a small percentage of silicates is expected
to result in a similar dynamical and thermal behaviour.

The initial distribution of dust comprises $22400$ particles, equally
distributed in four size bins of radius $R_{s}=1\,\mu\mathrm{m}$,
$10\,\mu\mathrm{m}$, $100\,\mu\mathrm{m}$, and $1\,\mathrm{mm}$.
This range of sizes is supported by radiative transfer modelling of \HD\ 
\citep[e.g.,][]{ohashi2019}.
A simulation with a higher number of particles is presented in
Appendix~\ref{sec:DSP}. The comparison with the case modelling 
fewer particles yields consistent outcomes and indicates that 
the particle number does not introduce any significant 
bias. Details and further discussion on the initial distributions 
of the particles' eccentricity, $e_{s}$, and inclination, $i_{s}$, 
are also provided in the Appendix, including a test on the orbital
excitation of planetesimals via interactions with the planets.
Although collisions are not modelled, we tested that the orbits of 
planetesimals in the disc's planet region become readily eccentric
and inclined.

Solids that move close to the planets, within about $1.6$ times 
the current radius of Jupiter, are regarded as accreted and removed 
from the calculation.
Giant planets can achieve these radii during their contraction phase, 
when gas accretion is limited by the disc supply 
\citep[e.g.,][]{lissauer2009,gennaro2021}, as expected for the age 
of this system. 

The production of dust is not modelled in this study. It is assumed 
that collisional comminution of planetesimals leads to fragmentation
down to small scales, so that $\mu\mathrm{m}$- to $\mathrm{mm}$-size
grains are generated. No assumption is made on mass partition among
various grain sizes although, as discussed in the next sections,
dust at the lower end of the range is more likely to trace its 
origin (i.e., first- versus second-generation).
It is worth noticing that calculations modelling planetesimal
impacts do not necessarily capture the full size distribution
of the fragments arising from collisions, from hundreds of kilometres 
(or more) to microns (or less). 
For example, \citet{weidenschilling2010} computed the collisional
evolution of large planetesimals down to metre-size bodies and then
extrapolated the size distribution to $\mu\mathrm{m}$-size grains.
\citet{turrini2019} applied a similar approach and our working
hypothesis relies on these results.

\section{Results}
\label{sec:R}

\subsection{Planet-Induced Perturbations}
\label{sec:PP}

\begin{figure*}
\centering%
\resizebox{2.08\columnwidth}{!}{\includegraphics[clip]{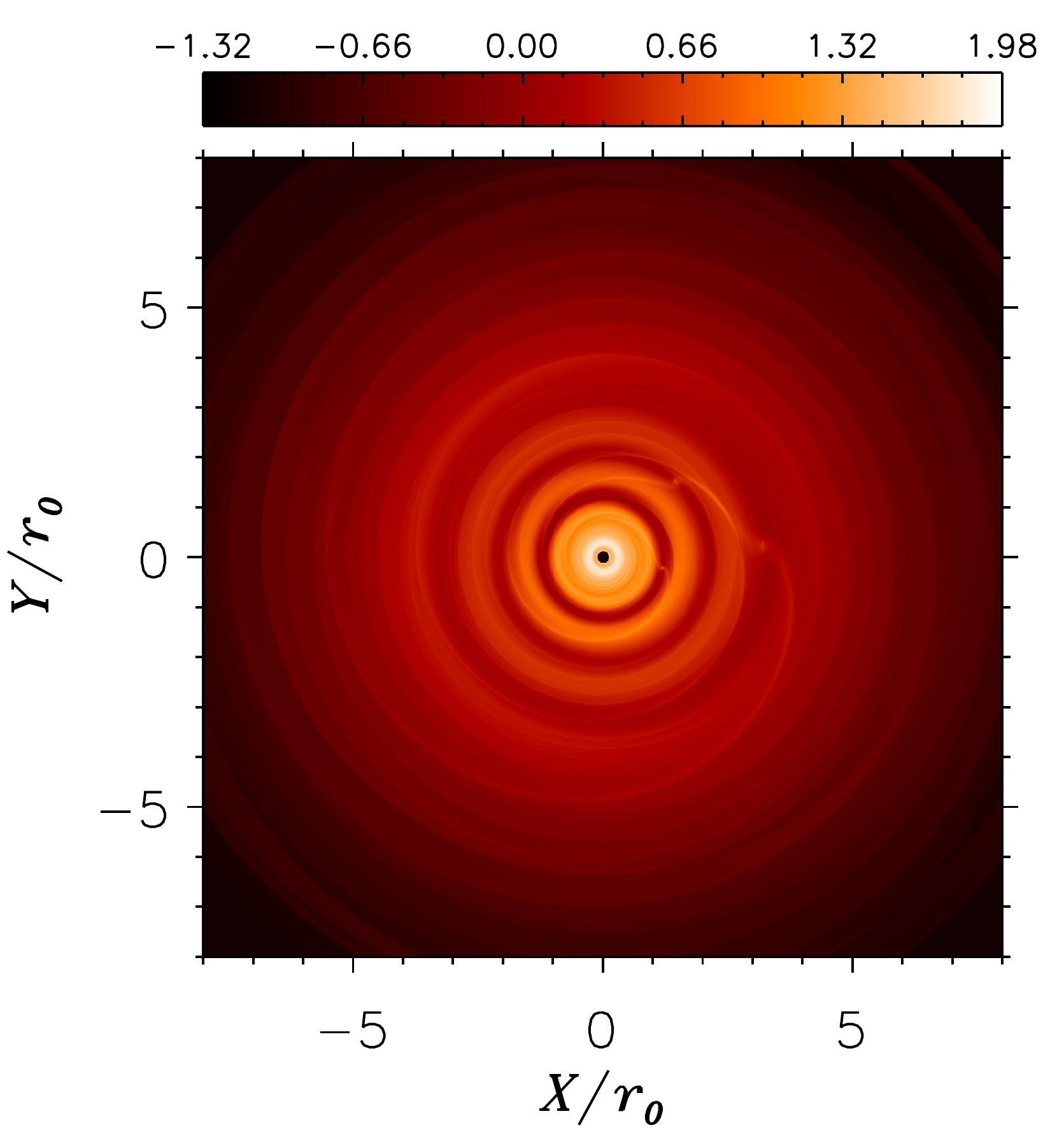}%
                             \includegraphics[clip]{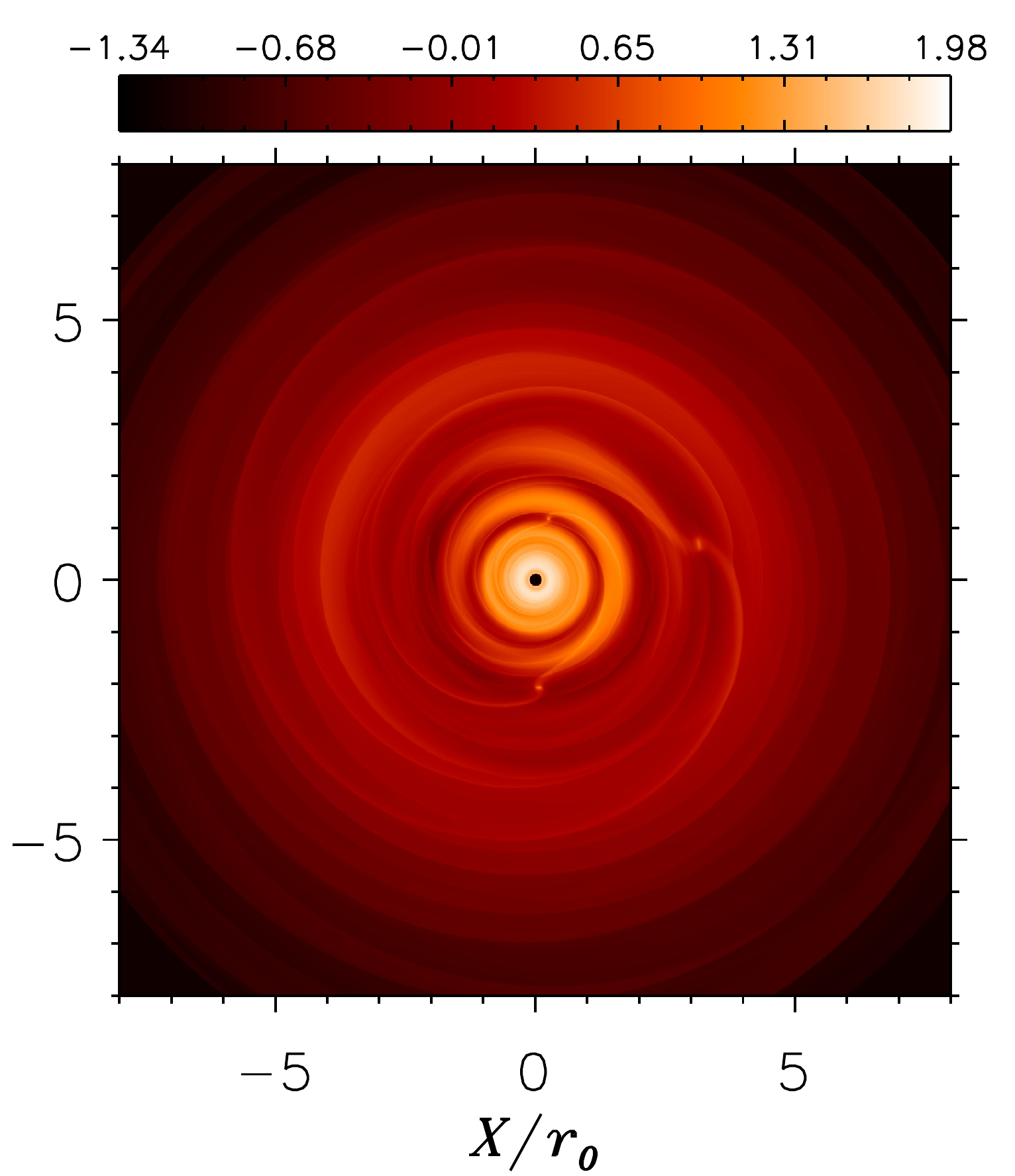}%
                             \includegraphics[clip]{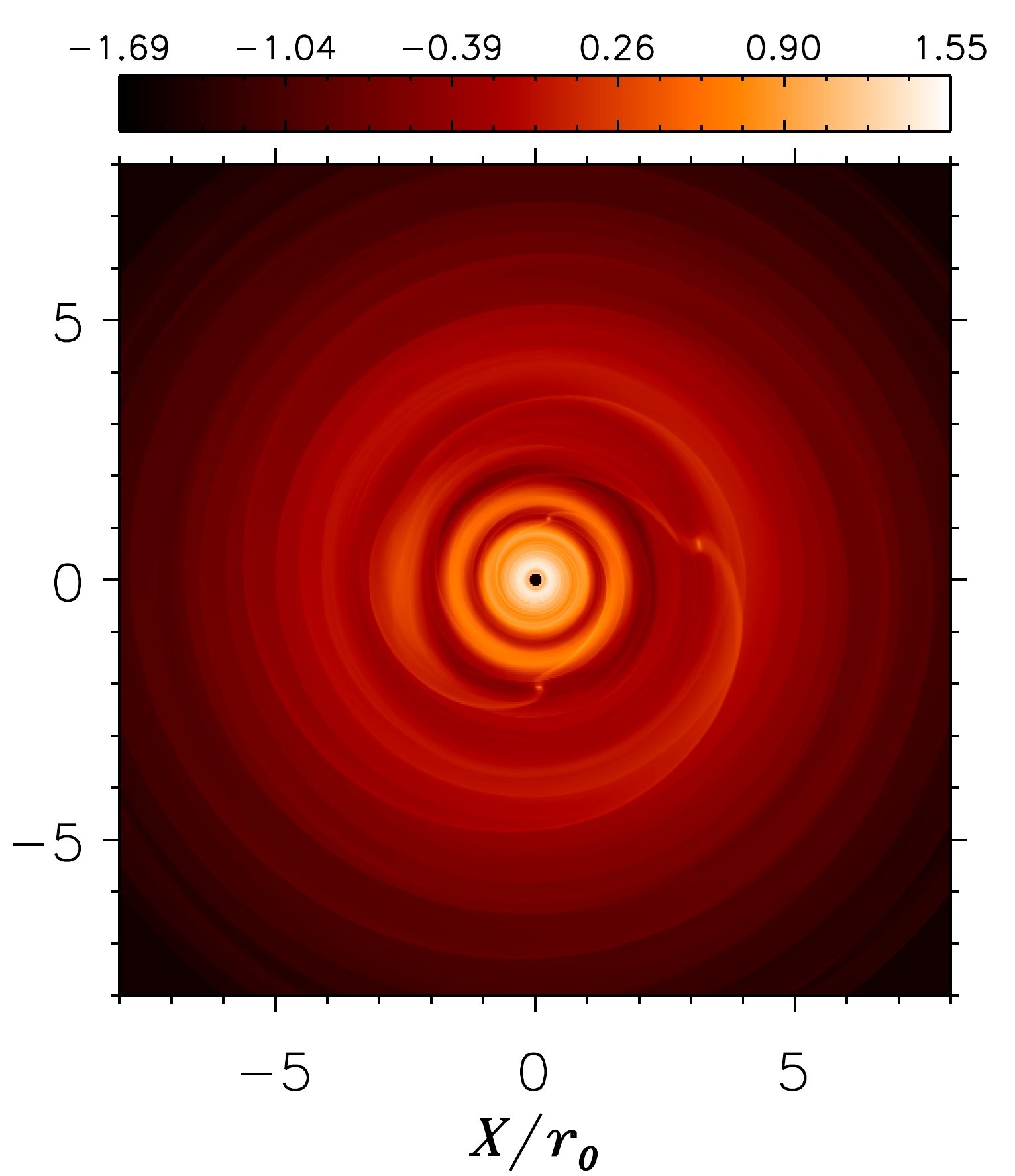}}
\caption{%
        Gas surface density ($\Sigma$) after $2600$ orbital periods 
        of evolution at $r=r_{0}$ ($t\approx 6\times 10^{5}$ years) 
        in model \lmla\ (left), \hmha\ (centre), and \hmls\ (right). 
        The colour scale represents $\log{\Sigma}$, where $\Sigma$ is
        in units of \sigu.
        }
\label{fig:img}
\end{figure*}

\begin{figure*}
\centering%
\resizebox{2.08\columnwidth}{!}{\includegraphics[clip]{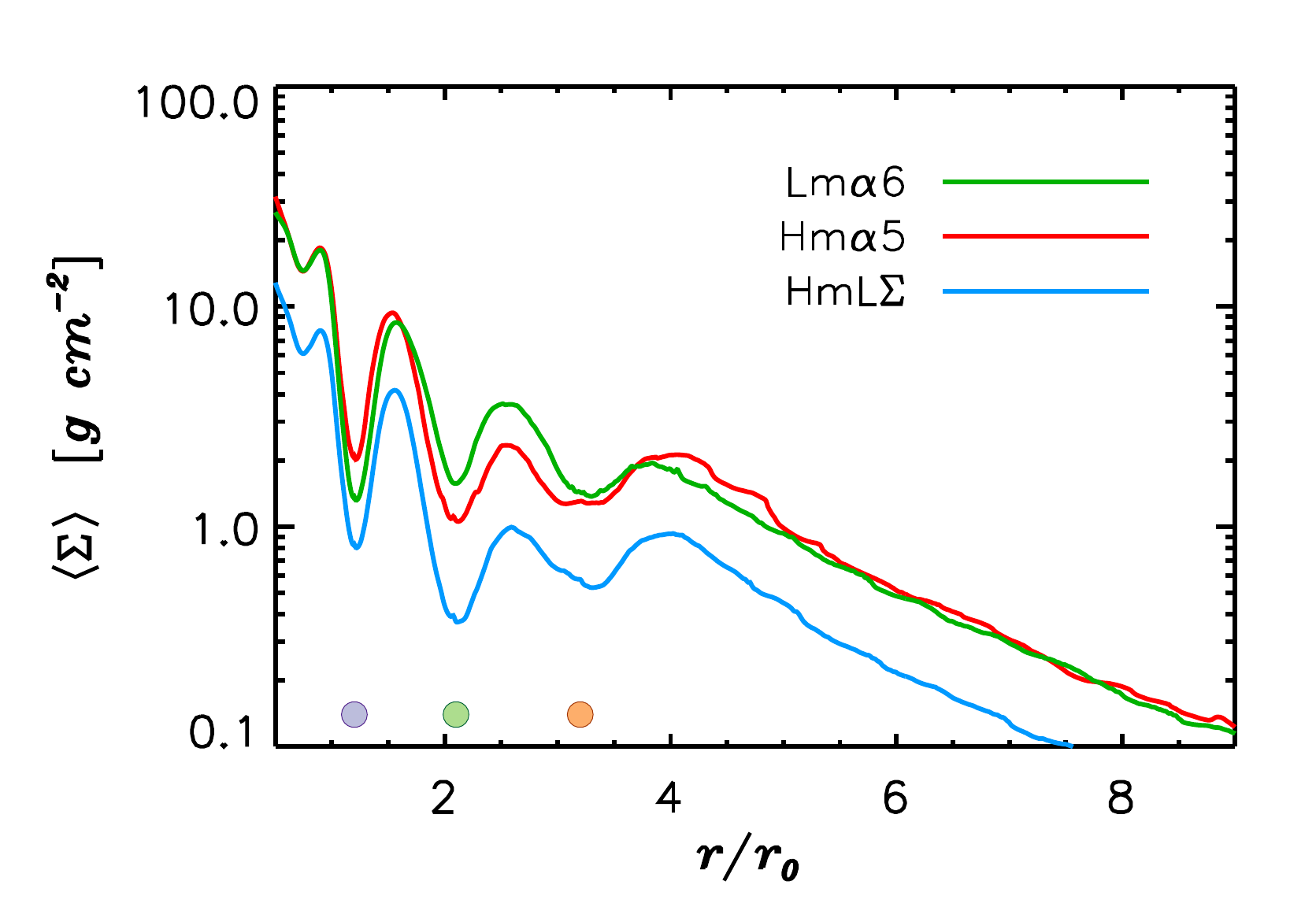}%
                             \includegraphics[clip]{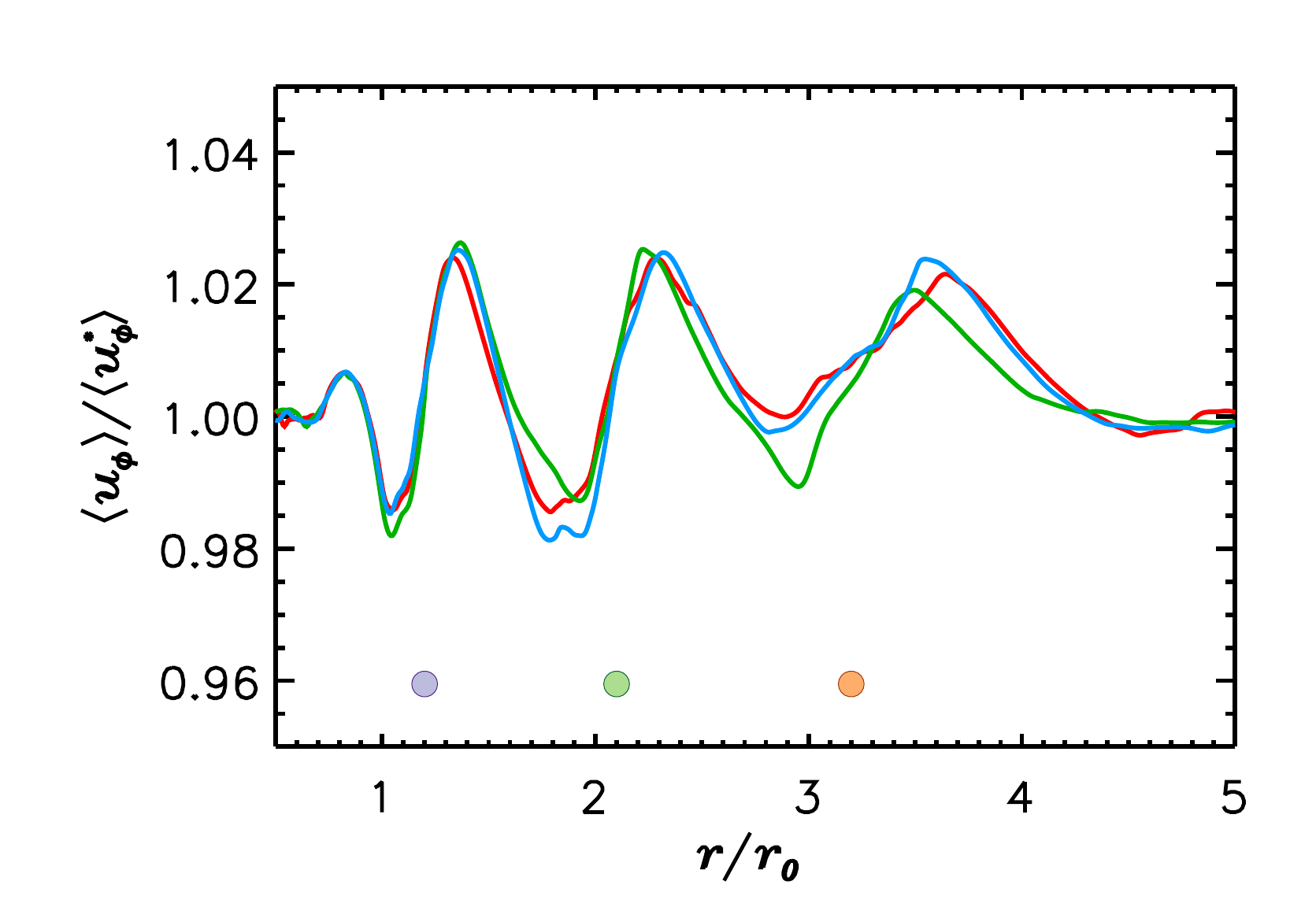}}
\caption{%
        Left:
        Surface density of the gas, averaged in azimuth around
        the star, as a function of the radial distance from the star 
        for the three cases listed in Table~\ref{tab:cases}.
        The symbols indicate the planets' orbital radius, 
        in units of the reference distance, $r_{0}$.
        Right:
        Ratio of the gas azimuthal velocity to its unperturbed value
        (i.e., without planets),
        averaged around the star for the density distributions
        displayed in the left panel.
        }
\label{fig:Sigma}
\end{figure*}

The gas density distribution along a planet's orbit, in a disc with finite
viscosity, remains unperturbed as long as viscous torques exceed the tidal 
torques generated by the planet's gravity. However, for low enough viscosity 
and/or large enough planet mass, tidal torques can overpower viscous torques.

Conservation of the gas radial momentum (and ignoring viscous stresses and
self-gravity) implies that the azimuthal velocity in the disc mid-plane
is such that
\begin{equation}
 u^{2}_{\phi}\approx v^{2}_{\mathrm{Kep}}%
 + \left(\frac{r}{\rho}\right)\frac{\partial p}{\partial r},
 \label{eq:uu}
\end{equation}
where the gas pressure is $p=c^{2}\rho$ and
$c=h v_{\mathrm{Kep}}$ is the gas sound speed.
If the disc is unperturbed (e.g., without planets) and $\rho$ is
a power-law of the radial distance $r$, the derivative term 
in Equation~\ref{eq:uu} is $\propto (h v_{\mathrm{Kep}})^{2}$.

The planets listed in Table~\ref{tab:cases} have normalized Hill radii
$\Rhill/r=\sqrt[3]{m_{p}/(3M_{\star})}$ comparable to, but somewhat smaller 
than, the disc aspect ratios at the planets' locations (which increase outward).
The effects of the tidal perturbations can be seen in Figure~\ref{fig:img}, 
which show $\Sigma(r,\phi)$ for the three models after $2600$ orbits at
$r_{0}$. The surface density averaged azimuth is displayed in the left 
panel of Figure~\ref{fig:Sigma}. 
The curves show gaps in the gas distribution with depths comparable to those 
in Figure~7 of \citet{liu2018}, obtained from two-dimensional ($r$--$\phi$)
simulations.

In a disc that is tidally perturbed by planets, gap formation alters
the density gradient
$\partial\rho/\partial r$ around the gap edges, and hence the gas 
pressure gradient, so that $u_{\phi}$ also becomes perturbed and
can rise above or drop below its unperturbed value, $u^{*}_{\phi}$,
depending on the sign of $\partial p/\partial r$. 
The right panel of Figure~\ref{fig:Sigma} shows the ratio
$u_{\phi}/u^{*}_{\phi}$, averaged in azimuth around the star. 
Deviations of the gas rotation profile from a Keplerian pattern 
have been reported for \HD\ by several groups 
\citep[e.g.,][]{teague2018,pinte2018,rab2020},
based on observations of CO spectral features.
In particular, \citet{teague2018} obtained fractional deviations
of a few percent in the region occupied by the second and third
planet, between $\approx 80$ and $\approx 200\,\AU$, in good 
accord with the results presented in the Figure (note that 
we refer to the unperturbed velocity $u^{*}_{\phi}$ rather than
to the Keplerian velocity $v_{\mathrm{Kep}}$).
\citet{rab2020} also performed a detail analysis of the rotational
velocity deviations of the ring region and arrived at very similar
conclusions (see total fractional variations in their Figure~7).
\citet{pinte2018} reported deviations at $\approx 260\,\AU$
amounting to $\approx 15$\% of the local Keplerian velocity,
which they modelled as induced by a $2\,\mjup$ planet orbiting 
at that distance. Since a massive planet at said location is
absent in our setup, this feature is not produced. However,
the models in Figure~\ref{fig:Sigma} confirm that local 
deviations from Keplerian rotation of the gas velocity 
in the proximity of the second and third planet
(whose masses here are between $\approx 0.5$ and $\approx 1\,\mjup$) 
can reach levels of $10$\%.

Solid grains have a tendency to concentrate in regions where
$\langle u_{\phi}/u^{*}_{\phi} \rangle>1$, although more localised 
features in the gas distribution, such as vortices, can alter the process
A clear example is presented in the Appendix, Figure~\ref{fig:sp_dist_comp},
which compares dust distributions for disc models with different levels 
of gas turbulence.
Note that in the cases displayed in Figure~\ref{fig:Sigma}, the local 
maxima of $\langle u_{\phi}/u^{*}_{\phi} \rangle$ occur somewhat inward 
of the local maxima in azimuth-averaged surface density.

\subsection{Vortex Formation and Dust Distribution}
\label{sec:VF}

\begin{figure}
\centering%
\resizebox{1.02\columnwidth}{!}{\includegraphics[trim={0.1cm 0cm 0cm 0cm},clip]{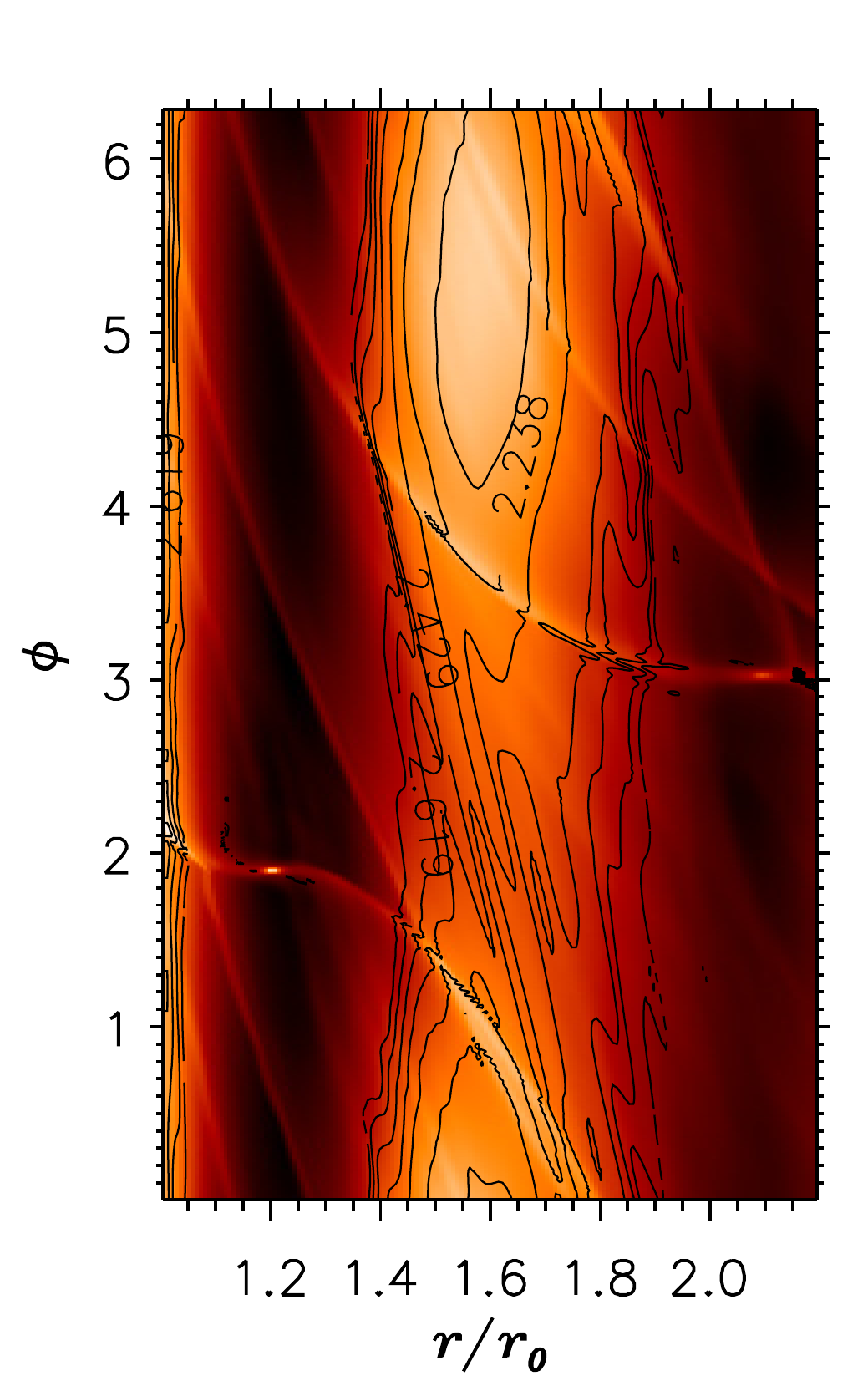}%
                            \includegraphics[trim={0.3cm 0cm 0cm 0cm},clip]{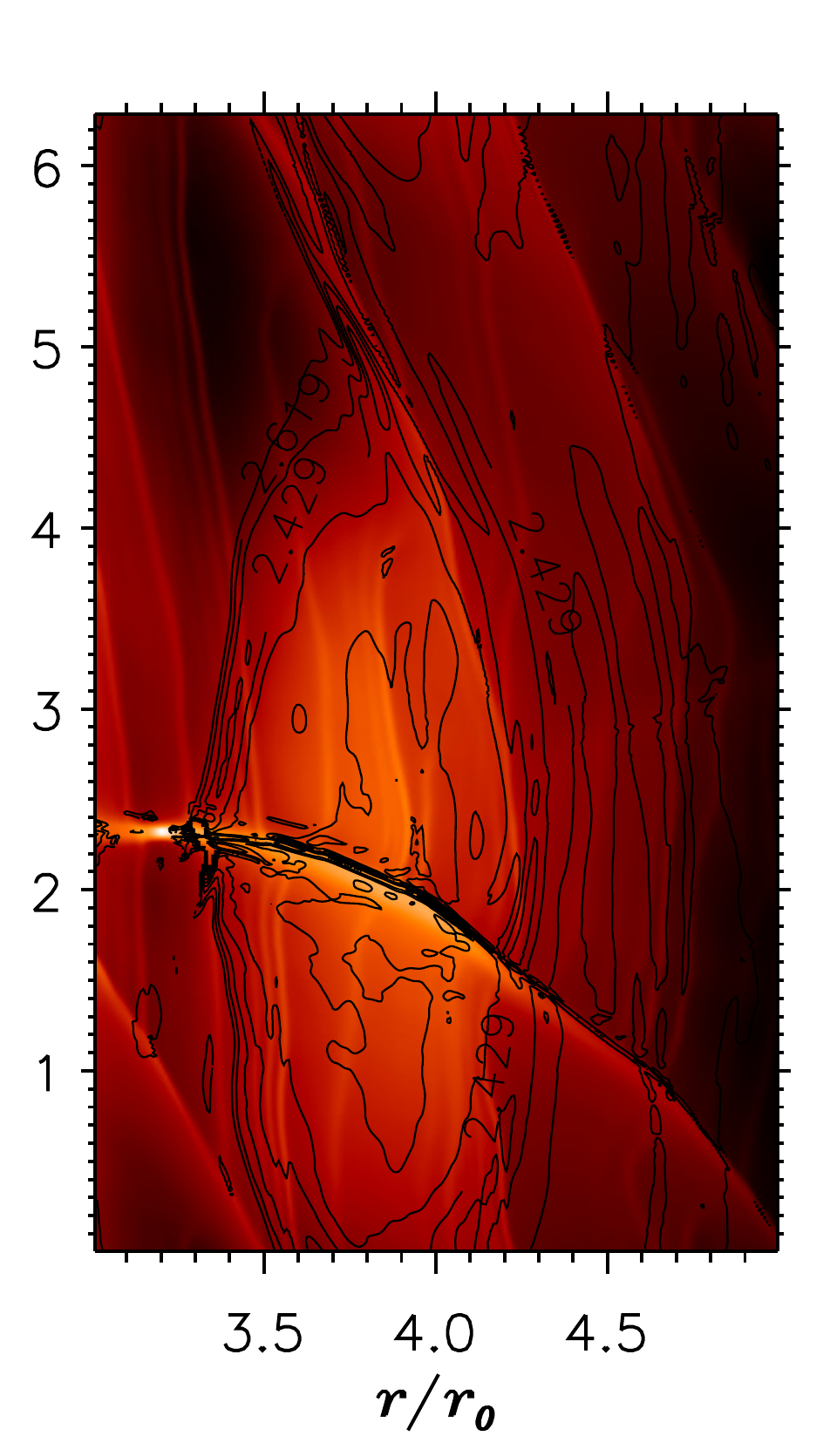}}
\caption{%
        Surface density (colour scale) exterior to the inner (left)
        and outer planet (right) in model \lmla. The black contours
        represent the potential vorticity of the gas, 
        Equation~\ref{eq:zeta}, and indicate the presence of
        vortices. In the left panel, the vortex spans an azimuthal
        angle of about $\pi$, and somewhat larger in the right panel
        (traversed by a spiral wave through the middle).
        }
\label{fig:vort_a16}
\end{figure}

\begin{figure*}
\centering%
\resizebox{2.05\columnwidth}{!}{\includegraphics[clip]{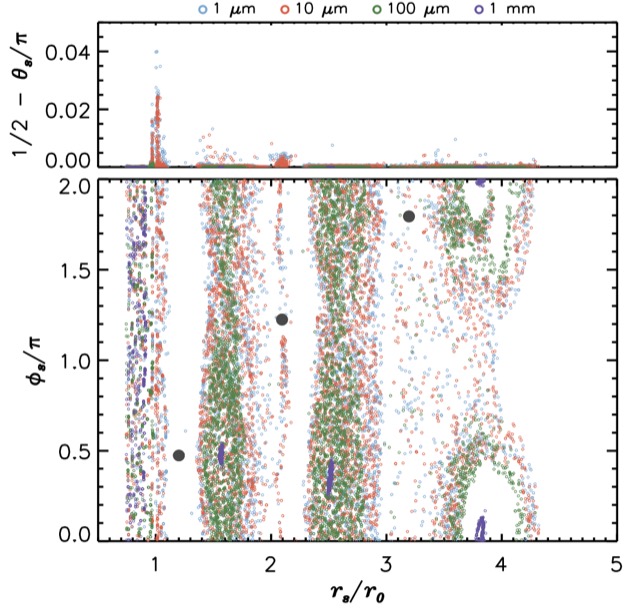}%
                             \includegraphics[clip]{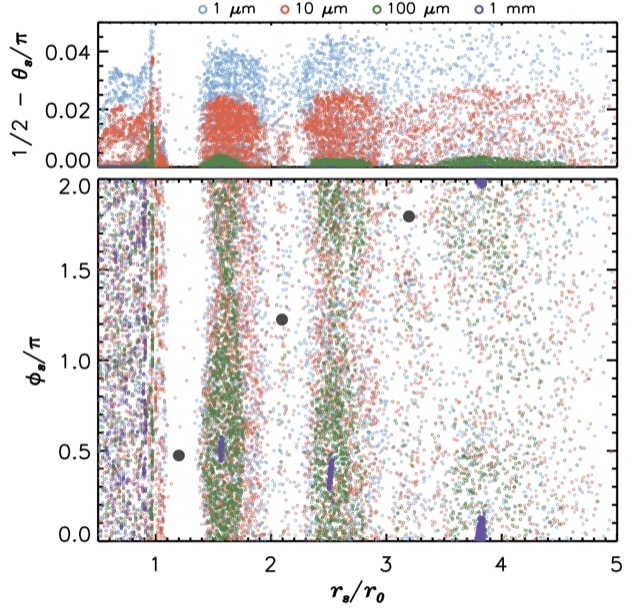}}
\caption{%
        Distribution of solid particles $500$ orbits after injection 
        ($t=3000$ orbits at $r=r_{0}$) in model \lmla. Top and bottom
        panels show the particles' positions projected in the
        $r$-$\theta$ and $r$-$\phi$ planes (in the top panels the
        angle is relative to the mid-plane, $\theta=\pi/2$, increasing
        northward). Different colours refer to different particle
        radii, as indicated in the legend. The grey solid circles
        represent the planets.
        Left panels refer to the simulation in which dust is released
        in the disc mid-plane whereas right panels refer to the
        model in which dust is produced via planetesimals' collisions.
        (\textit{High-resolution versions of this figure can be found
        \href{https://doi.org/10.6084/m9.figshare.16926040.v1}{here}}.)
        }
\label{fig:sp_dist_a16}
\end{figure*}

Rossby waves in thin Keplerian discs 
\citep[see, e.g.,][]{lovelace1999,tagger2001} can enter a non-linear 
regime and promote the formation of vortices \citep[see, e.g.,][]{li2001}
which, if persistent, may confine small solids
\citep[e.g.,][]{barge1995,bracco1999}. 
In the case of vortices arsing from planet-induced perturbations in
two-dimensional discs, \citet{fu2014} found that low levels of 
turbulent viscosity are required, corresponding to $\alpha\lesssim10^{-4}$,
for disc aspect ratios comparable to those applied here.
Consistent with those findings, the three-dimensional models in
Table~\ref{tab:cases} all show the emergence of vortices exterior
to the planets' orbits, as displayed in Figure~\ref{fig:vort_a16}.
The potential vorticity of the flow,
\begin{equation}
 \zeta=(\nabla\,\mathbf{\times}\,\mathbf{u})\cdot \mathbf{\hat{z}}/\Sigma
 \label{eq:zeta}
\end{equation}
($\mathbf{u}$ is the gas velocity in an inertial frame),
is displayed as contour levels, overlapped to the surface density
(colour scale), exterior to the orbits of the inner (left panel) and
outer planet (right panel). The vortices span an angle $\gtrsim \pi$
in azimuth.
In three dimensions, vortices are expected to also have a vertical
structure \citep[e.g.,][]{meheut2012}.
The presence of a similar feature exterior 
to the second planet, based on the contours of $\zeta$, 
is less evident due to the interference of spiral waves 
propagating through the region.

\begin{figure*}
\centering%
\resizebox{2.07\columnwidth}{!}{\includegraphics[clip]{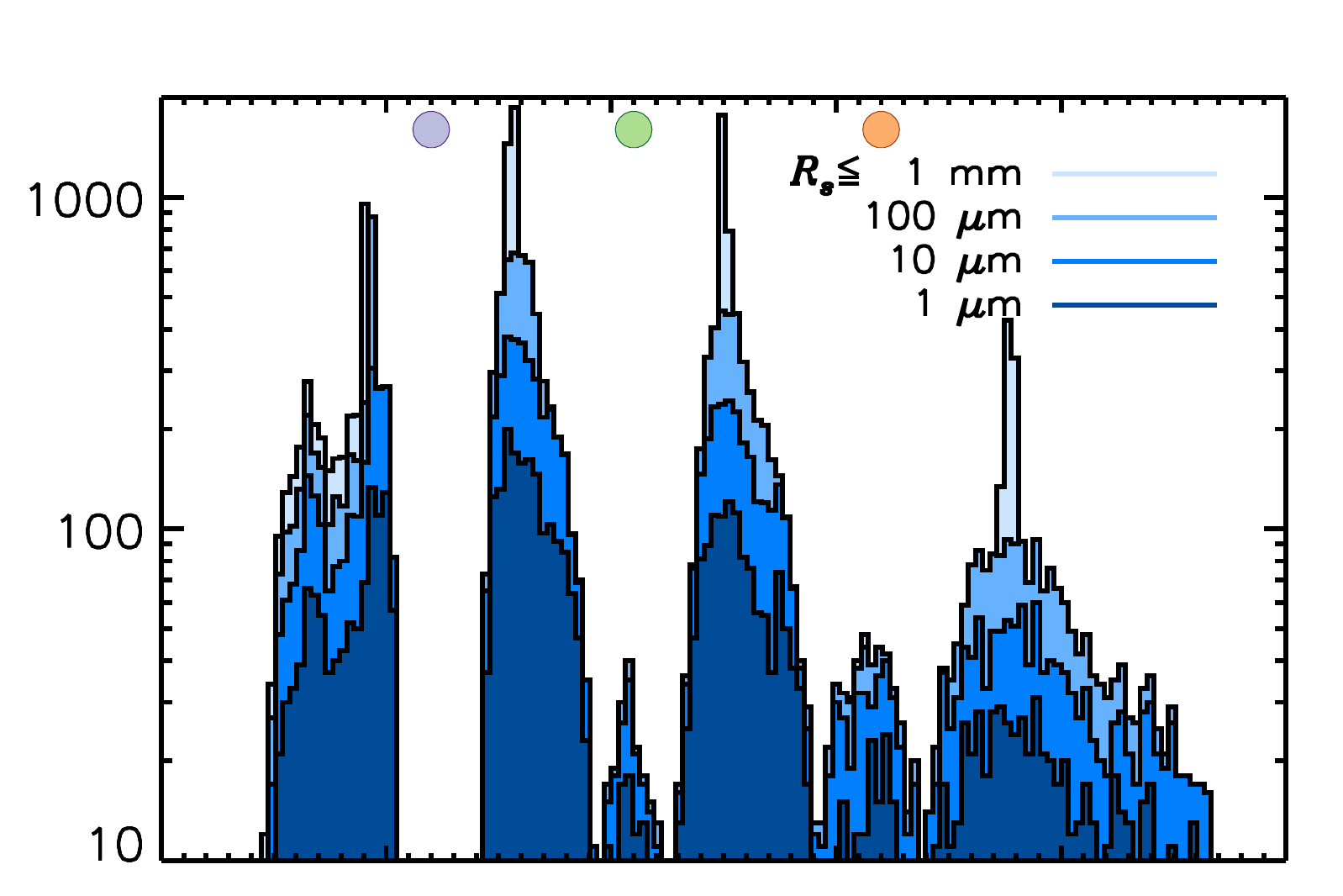}%
                             \includegraphics[clip]{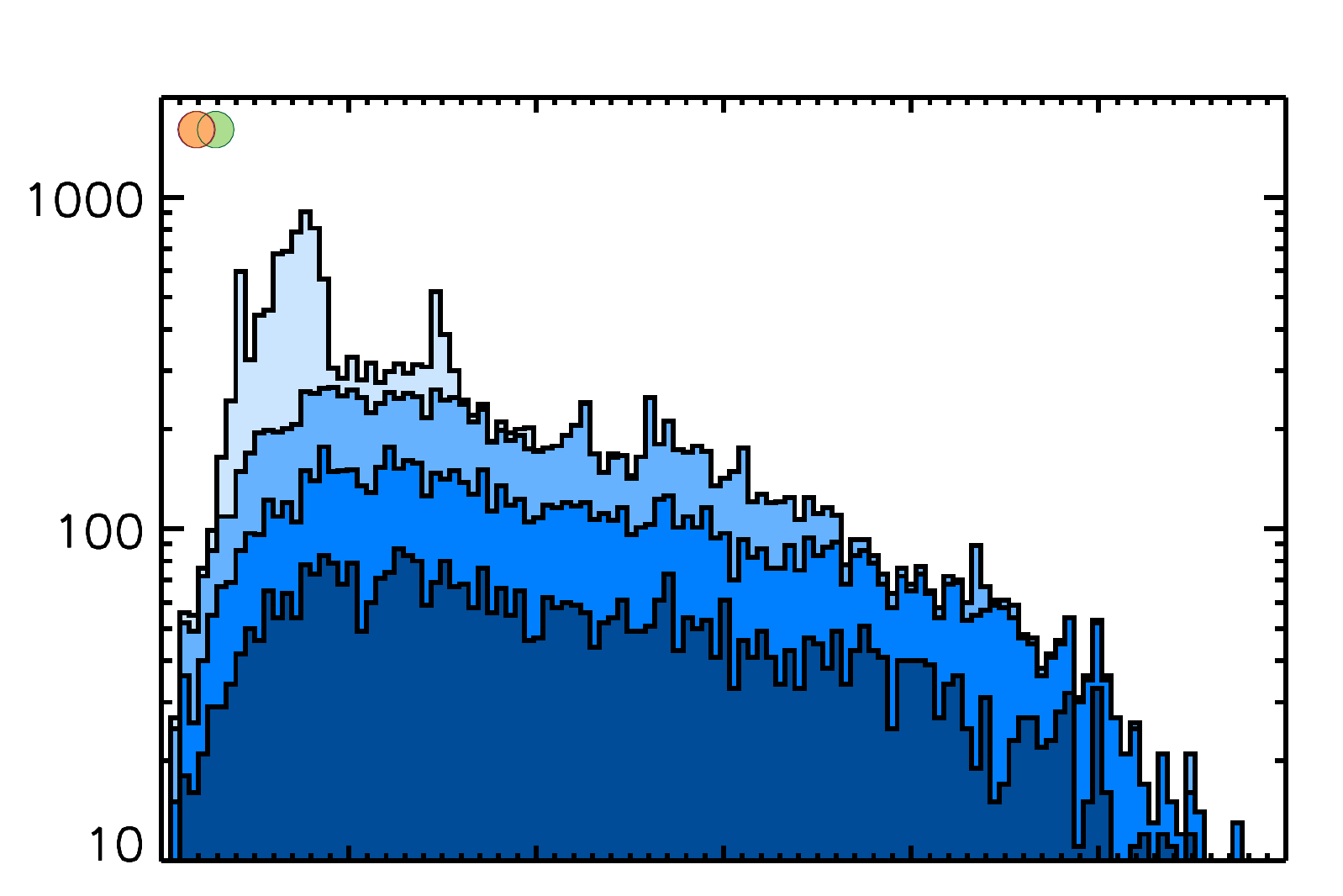}%
                             \includegraphics[clip]{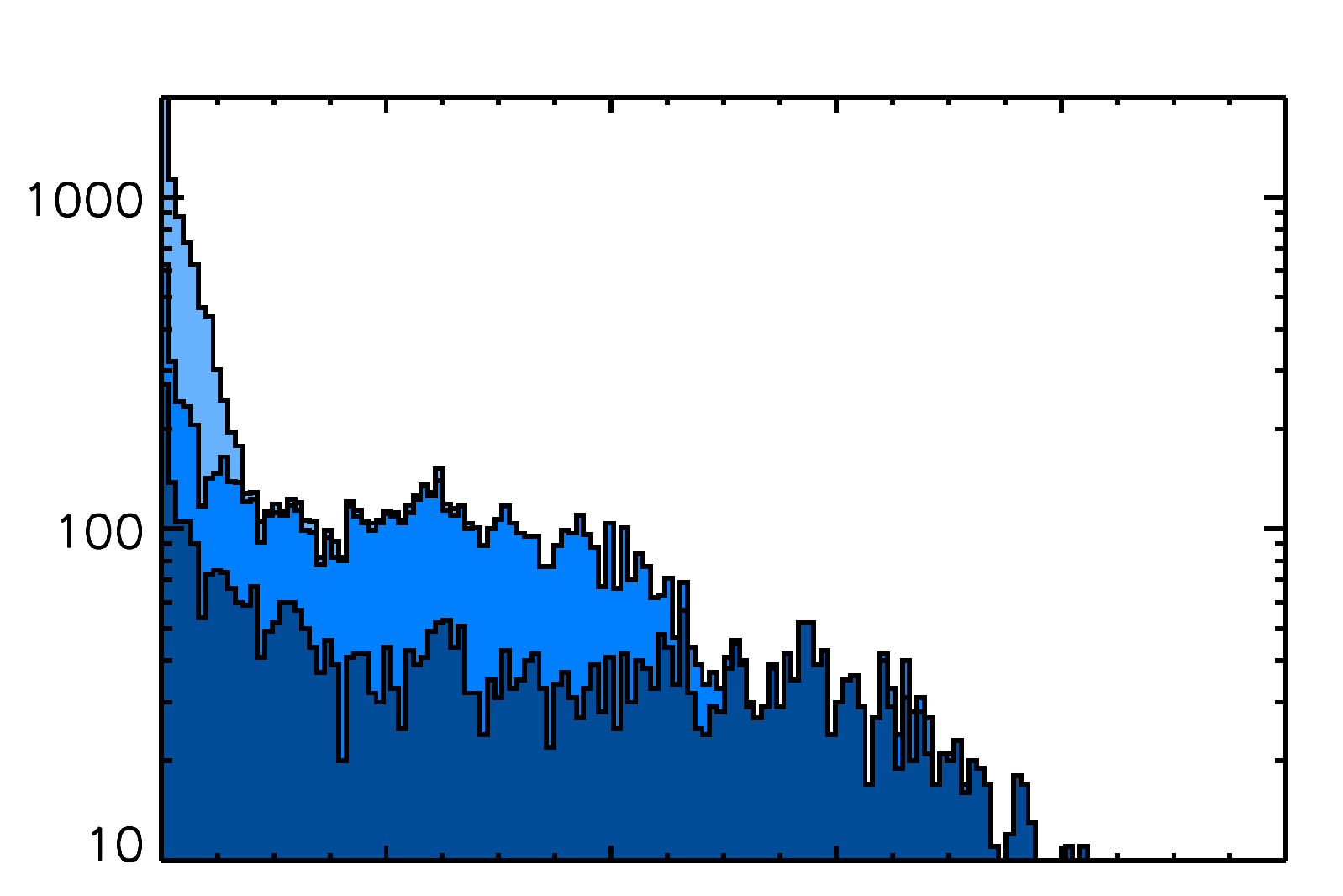}}
\resizebox{2.07\columnwidth}{!}{\includegraphics[clip]{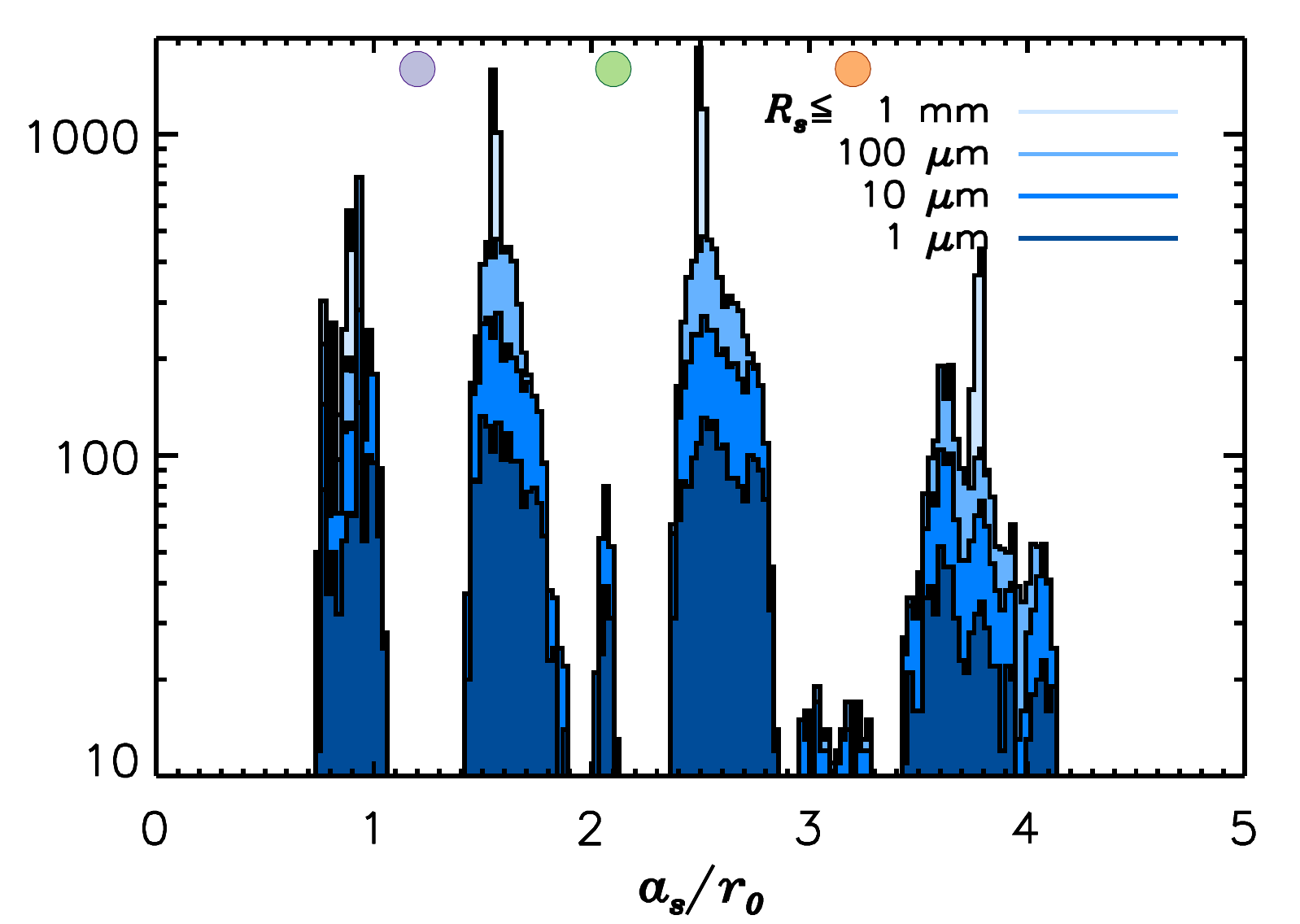}%
                             \includegraphics[clip]{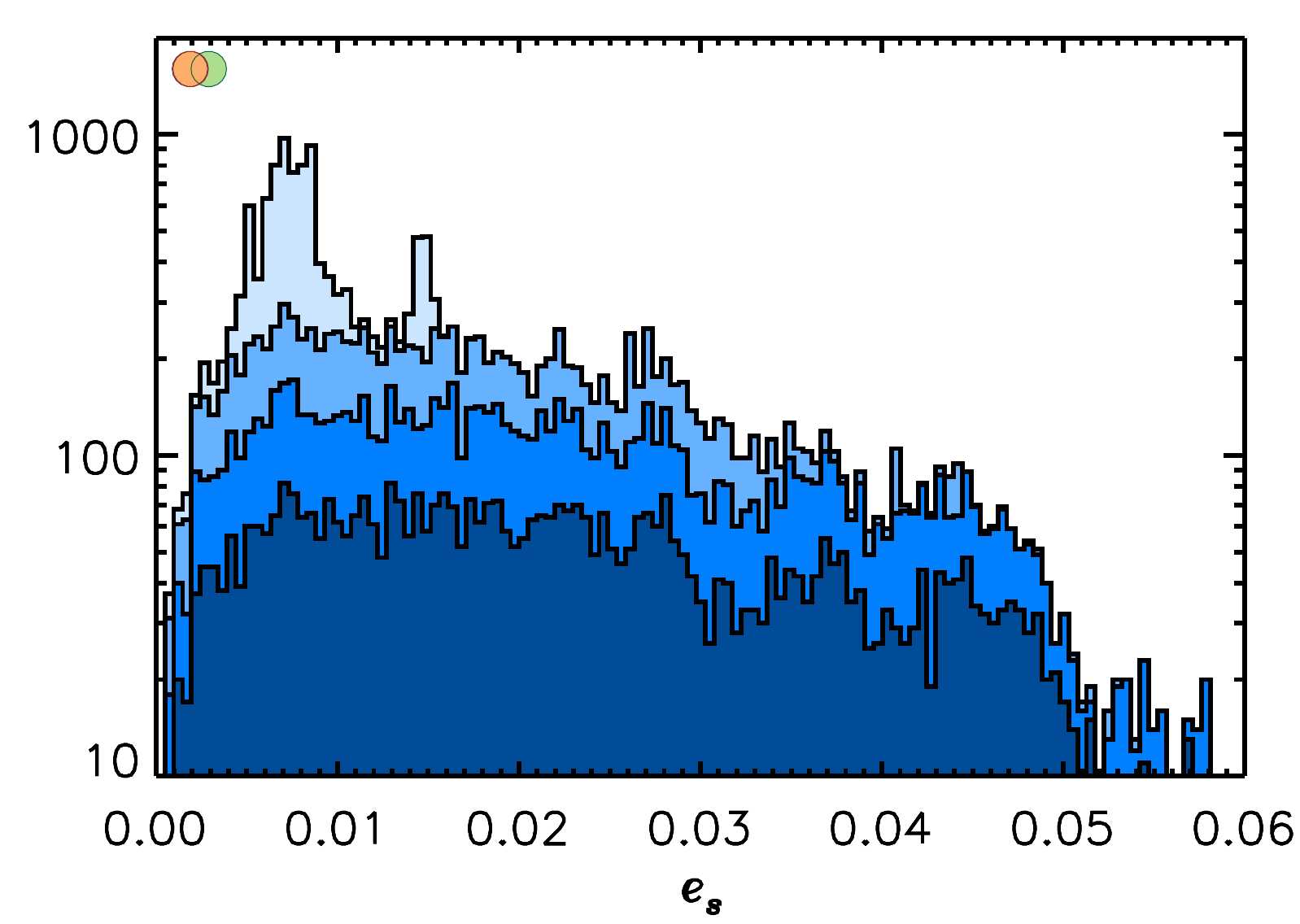}%
                             \includegraphics[clip]{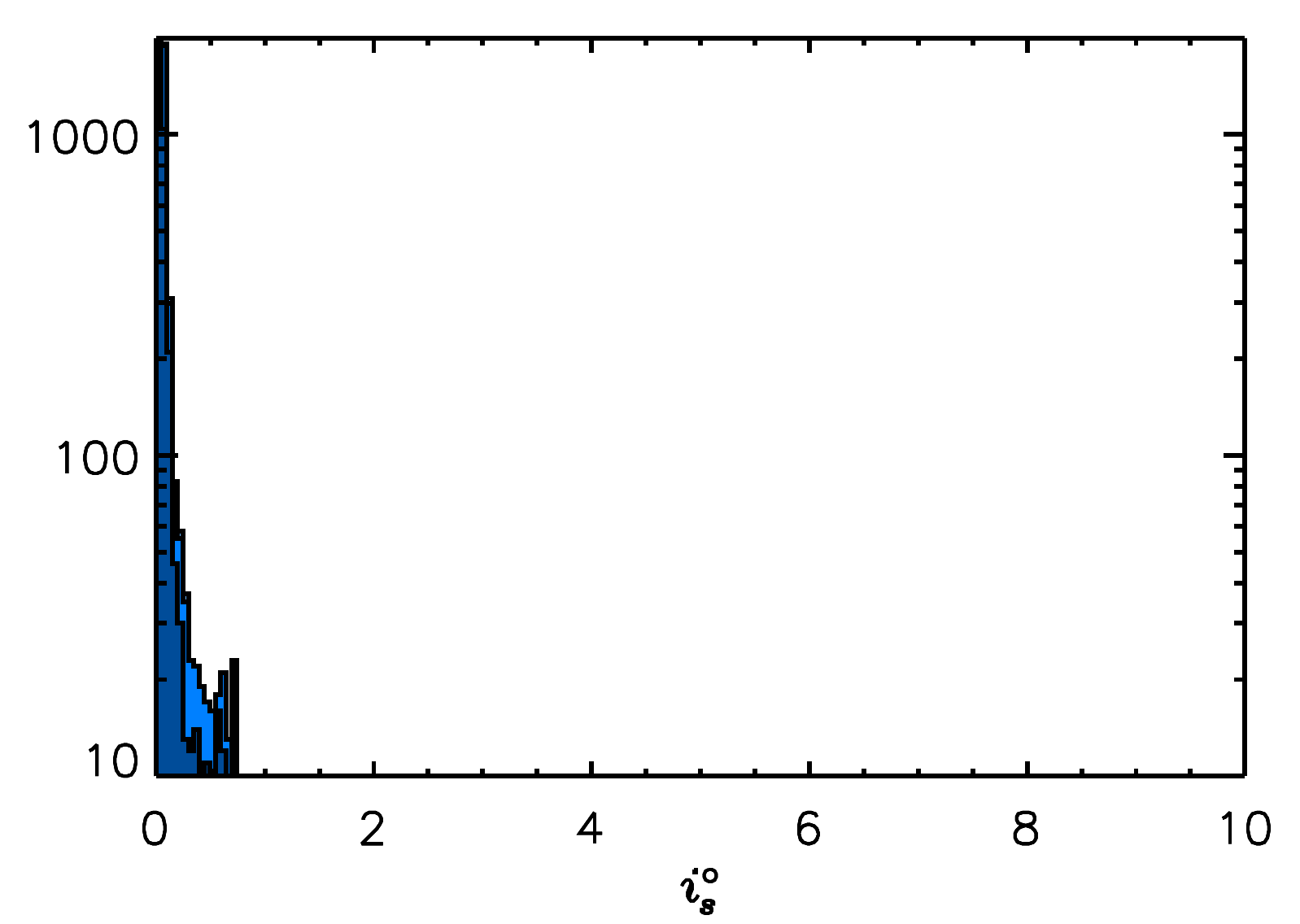}}
\caption{%
        Histograms of the particles' distribution at $t=3000$ orbits,
        $500$ orbits after injection, for models \lmla. Top/bottom
        panels show models with collision-produced/first-generation  
        dust.
        From left to right, the histograms display the distributions of
        semi-major axis, orbital eccentricity, and inclination. 
        Different shades of blue refer to different ranges of particle 
        radii, as indicated in the legend (lightest shade for the
        total distribution and darkest shade for the smallest particles). 
        The coloured circles represent the planets semi-major axis
        (left panels) and orbital eccentricity (centre panels).
        }
\label{fig:sp_hist_a16}
\end{figure*}

Figure~\ref{fig:sp_dist_a16} shows radius and azimuth (projected) 
positions of the particles in the lower panels and the radial and 
meridional (projected) positions in the upper panels, over $3\,H$,
for model \lmla. 
Particles are colour-coded by size. The dynamical behaviour of
the particles orbiting exterior to the planets' orbits are consistent
with the presence of vortical motions in the gas.
The largest solids, $R_{s}=100\,\mu\mathrm{m}$--$1\,\mathrm{mm}$,
which move relative to the gas by drag on the shortest timescales,
tend to be concentrated toward the centres of the vortices.
Over the evolution of these models, the effect appears
most evident for $1\,\mathrm{mm}$ solids, which are collected
in regions with high concentrations relative to their surroundings,
but smaller particles are affected as well, especially when close
to the mid-plane (compare particle distributions around the outermost
vortex in the left and right panel of Figure~\ref{fig:sp_dist_a16}).
Note that the motion of dust clearly indicate the presence of
a vortex also between the two outermost planets.
The confinement of dust in radius and azimuth, for various sizes,
can be evaluated more quantitatively in the histogram of
Figure~\ref{fig:sp_hist_a16} (left panels), in which different
colours represent populations of different radii.
The $1\,\mathrm{mm}$ solids show sharp peaks around the locations
of the vortex centres. By the end of the simulation, the peak 
concentration of these particles is between $10$ and $40$ times
as large as in the initial concentration at those locations.
The distributions of orbital eccentricity of the particles
(centre panels) are governed by the eccentric motion of the gas
\citep[driven by tidal interactions with the planets, e.g.,][]{ogilvie2003,gennaro2006},
which dictates values significantly
larger than those of the planets (see symbols). 
The eccentricities of collisionally-produced dust decay relatively
quickly, and therefore would not be useful to probe the origin
of the two dust species (compare top and bottom panels).

\begin{figure*}
\centering%
\resizebox{2.05\columnwidth}{!}{\includegraphics[clip]{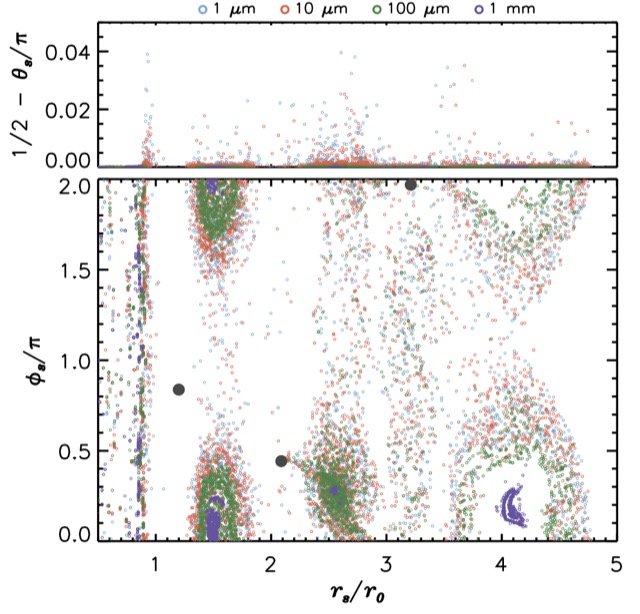}%
                             \includegraphics[clip]{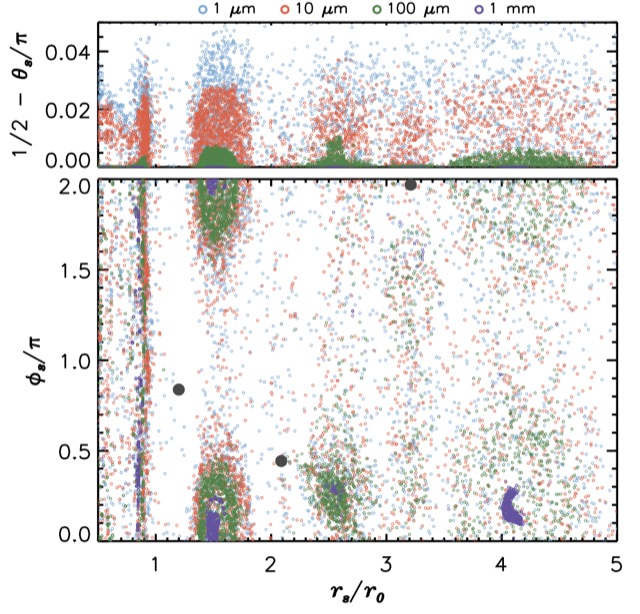}}
\caption{%
        As Figure~\ref{fig:sp_dist_a16}, but the solids' distributions
        refer to model \hmha.
        Left panels show results from the simulation in which dust 
        is released in the disc mid-plane whereas right panels show
        results from the simulation in which dust is produced by 
        planetesimals' collisions.
        (\textit{High-resolution versions of this figure can be found
        \href{https://doi.org/10.6084/m9.figshare.16926040.v1}{here}}.)
        }
\label{fig:sp_dist_hma15}
\end{figure*}

A behaviour similar to that of model \lmla\ for the confinement
and distribution of dust is observed in the other model configurations
(see Figures~\ref{fig:sp_dist_hma15} and \ref{fig:sp_dist_dhma15}).
The cases better show that also $\sim 1\,\mu\mathrm{m}$ grains are
collected within the vortices over the duration of the simulations
(see left panels).
Experiments conducted at higher viscosity, $\alpha=10^{-3}$, for both
configuration of the planet masses, indicate no such features exist
(an example is presented in Appendix~\ref{sec:DSP}, Figure~\ref{fig:sp_dist_comp}).

\begin{figure*}
\centering%
\resizebox{2.07\columnwidth}{!}{\includegraphics[clip]{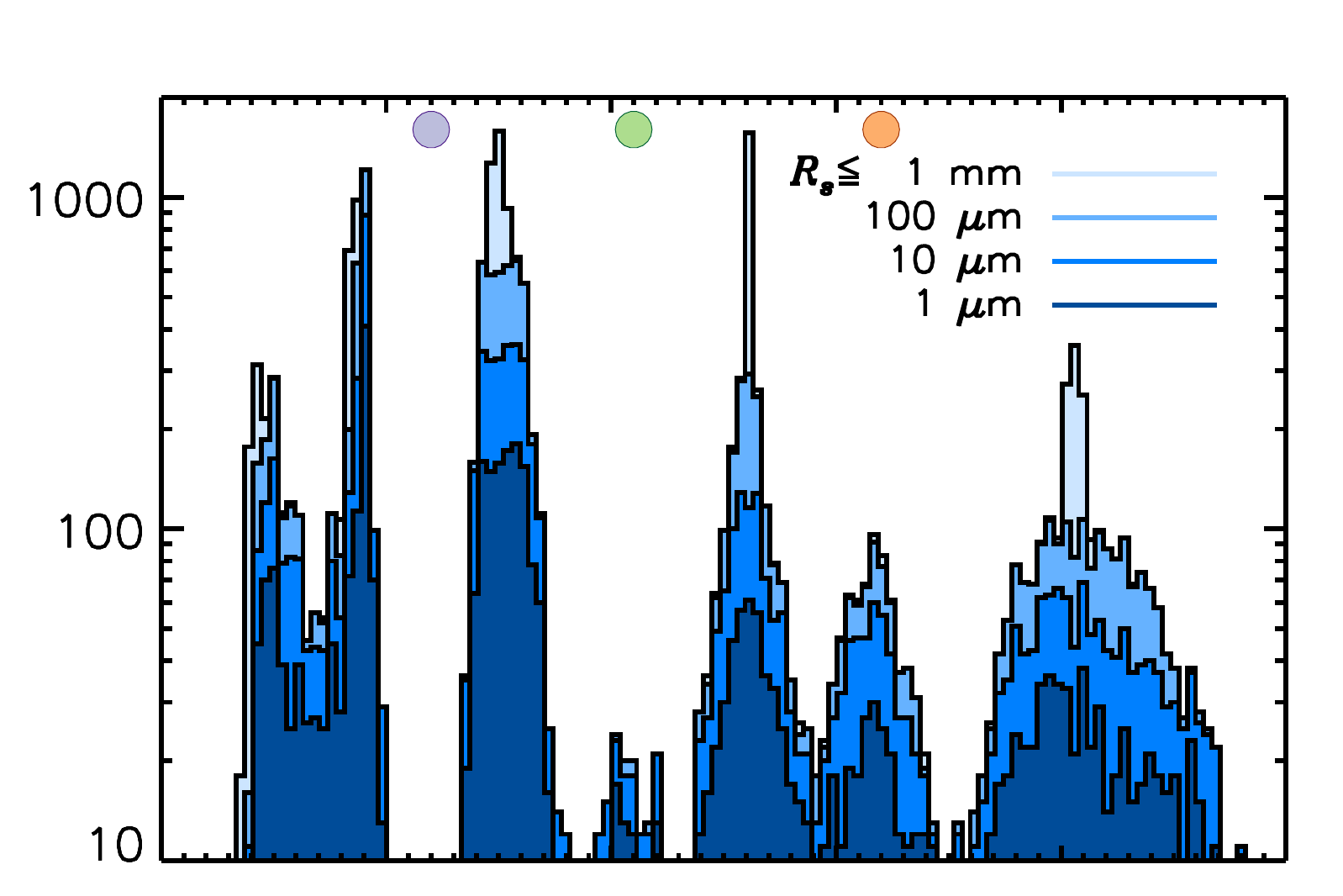}%
                             \includegraphics[clip]{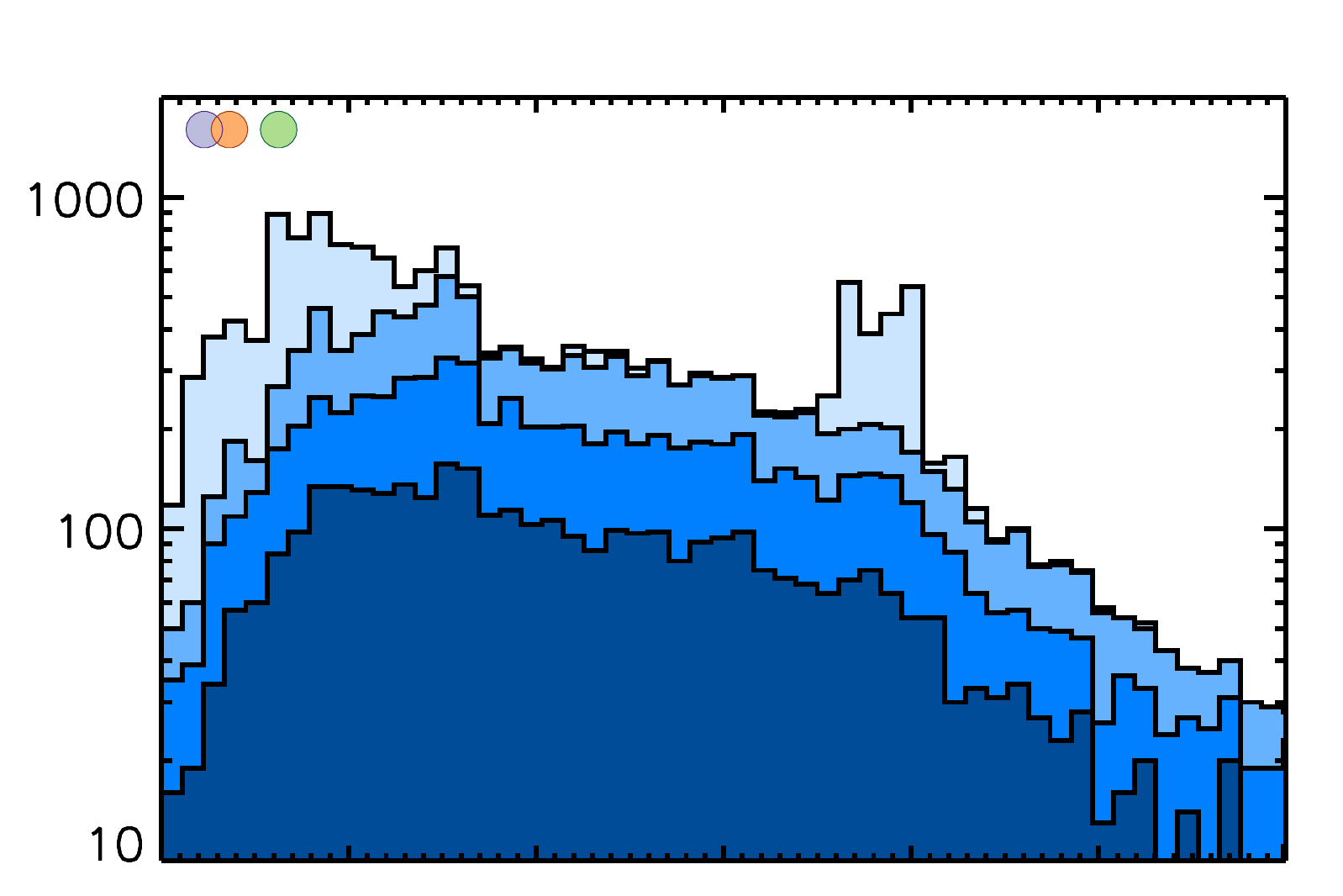}%
                             \includegraphics[clip]{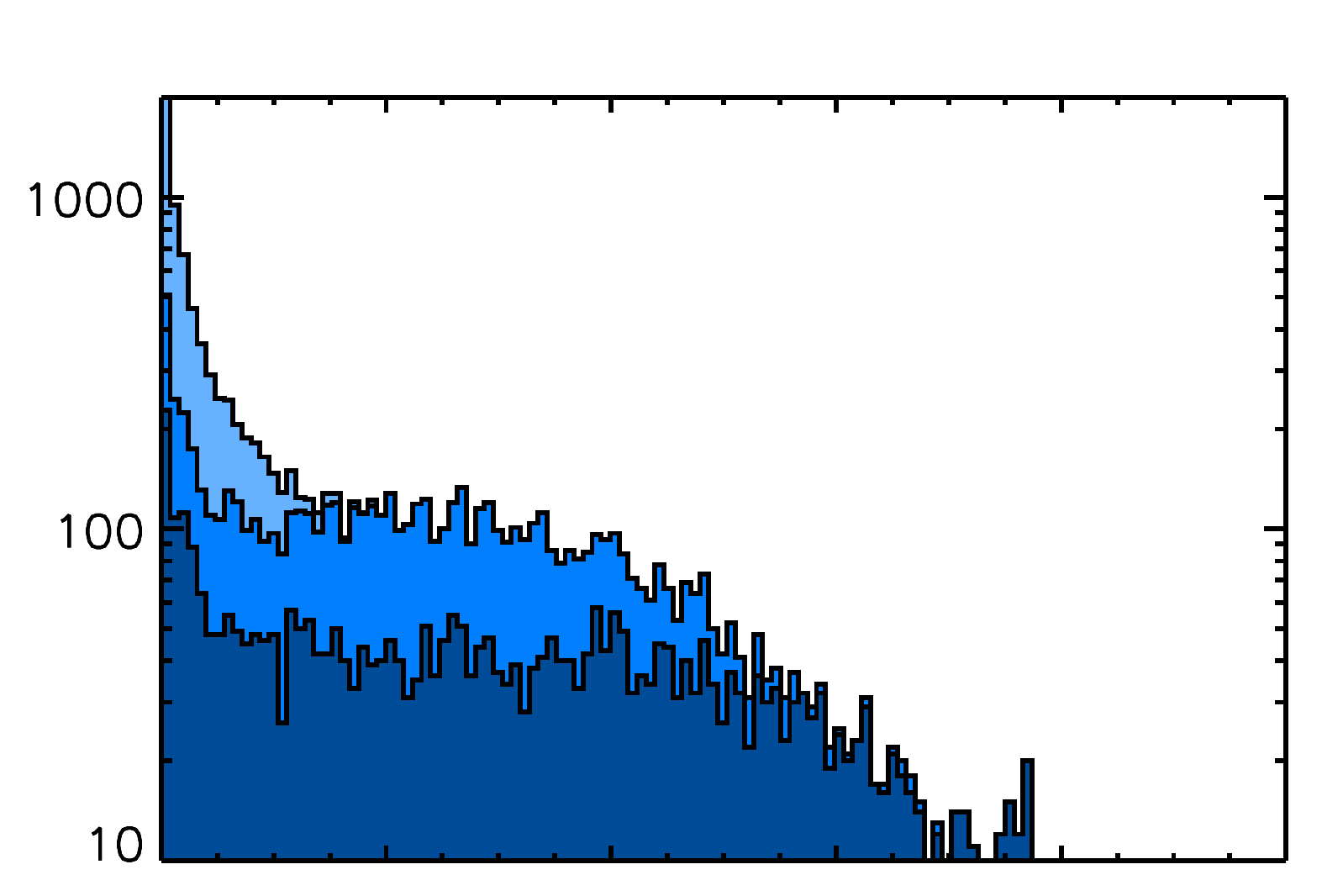}}
\resizebox{2.07\columnwidth}{!}{\includegraphics[clip]{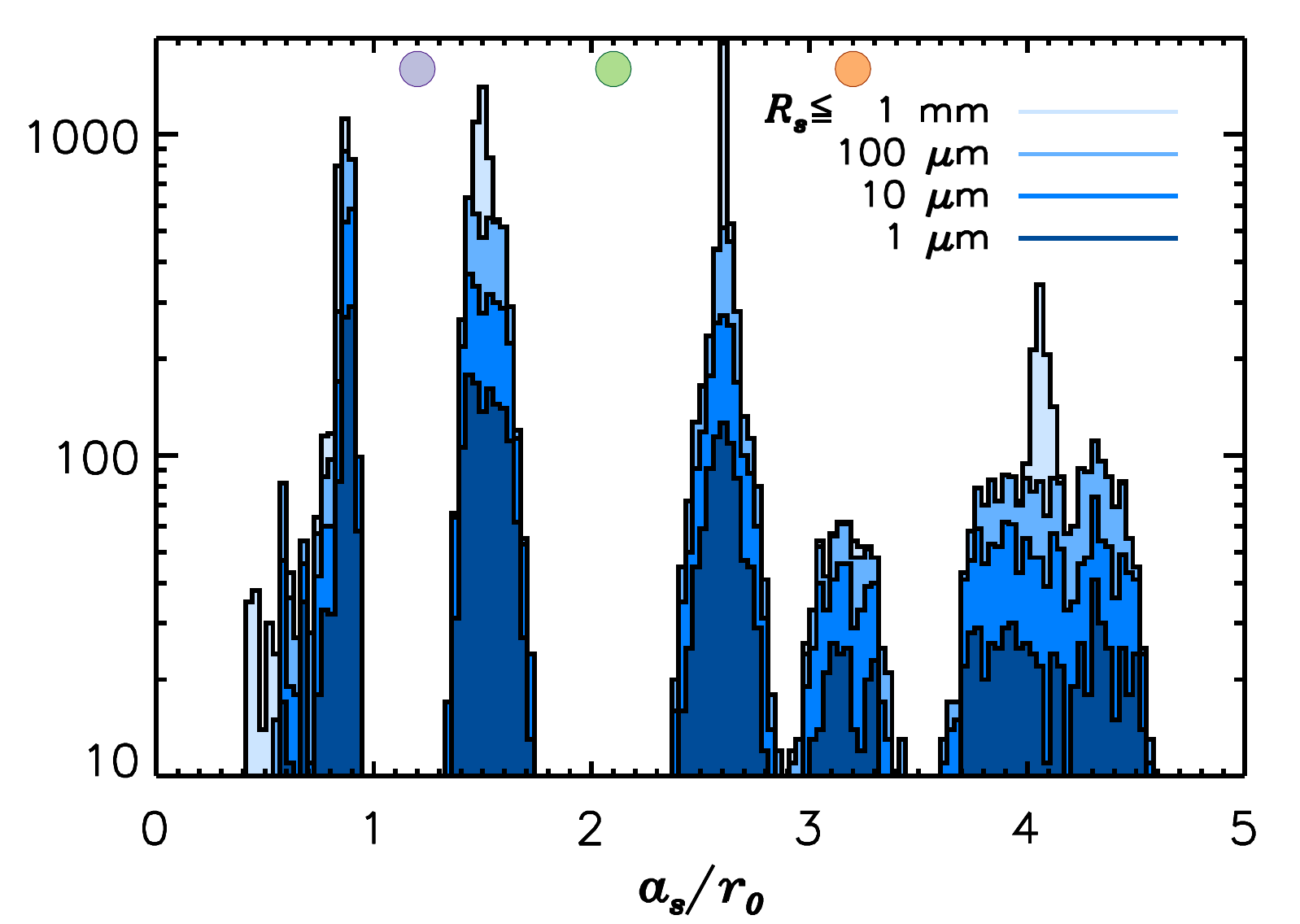}%
                             \includegraphics[clip]{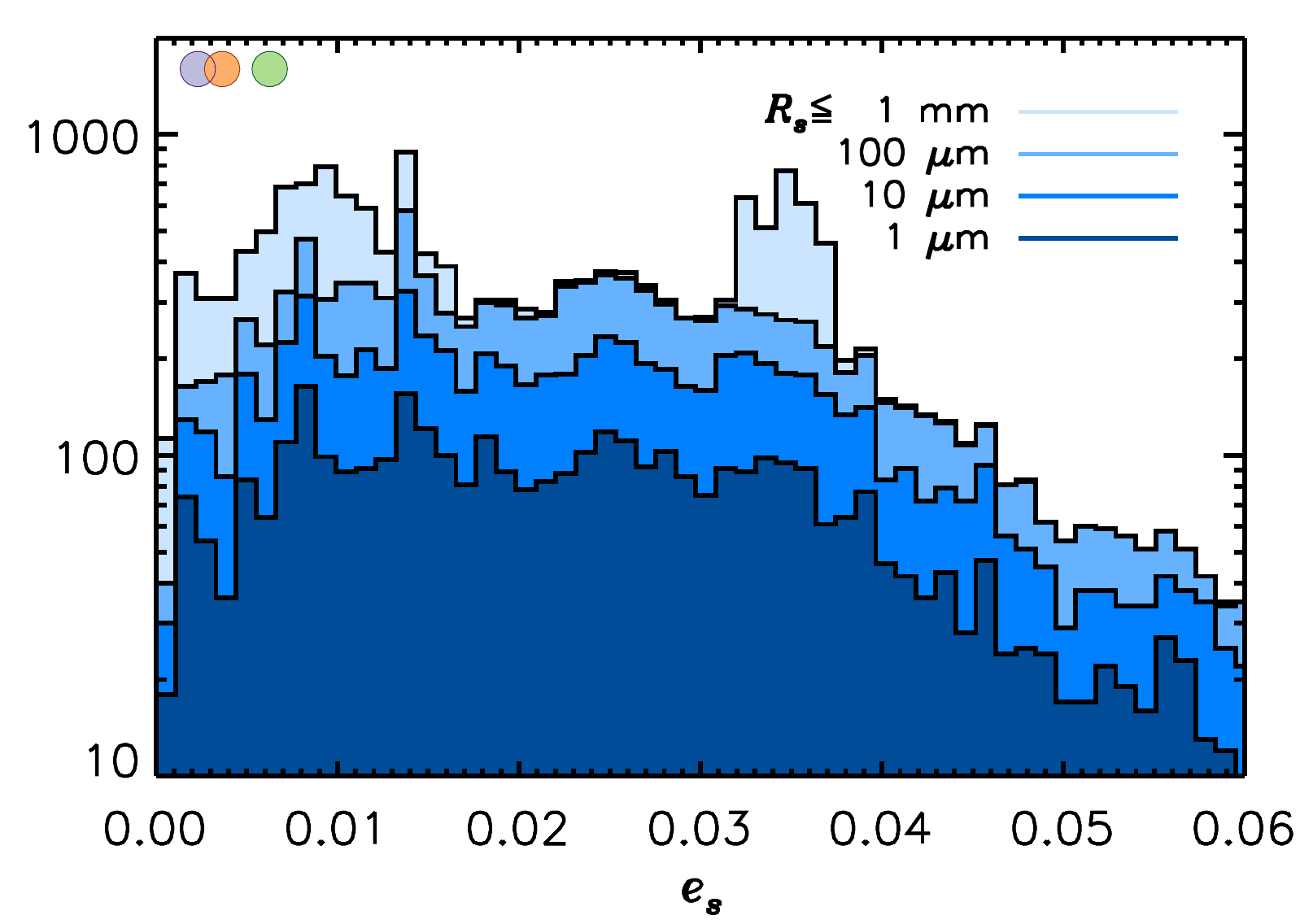}%
                             \includegraphics[clip]{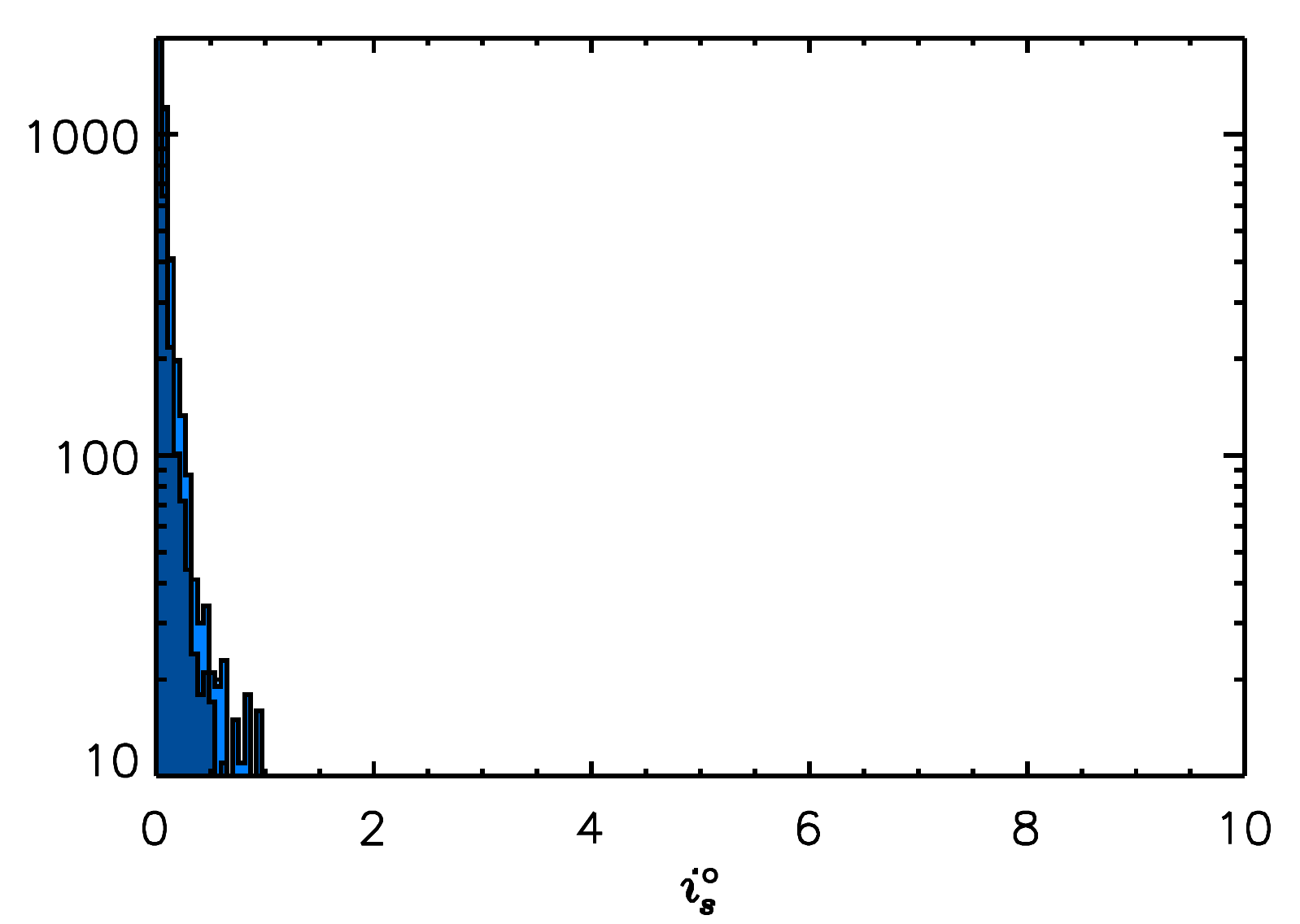}}
\caption{%
        As Figure~\ref{fig:sp_hist_a16}, but for models \hmha. 
        Top and bottom panels show results for
        collision-produced and first-generation dust, respectively.
        Different shades of blue render different particle radii, 
        as indicated. 
        }
\label{fig:sp_hist_hma15}
\end{figure*}

\begin{figure*}
\centering%
\resizebox{2.05\columnwidth}{!}{\includegraphics[clip]{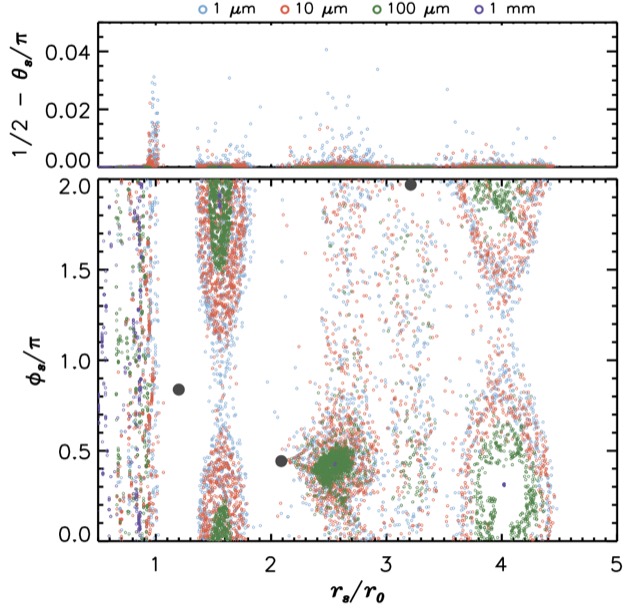}%
                             \includegraphics[clip]{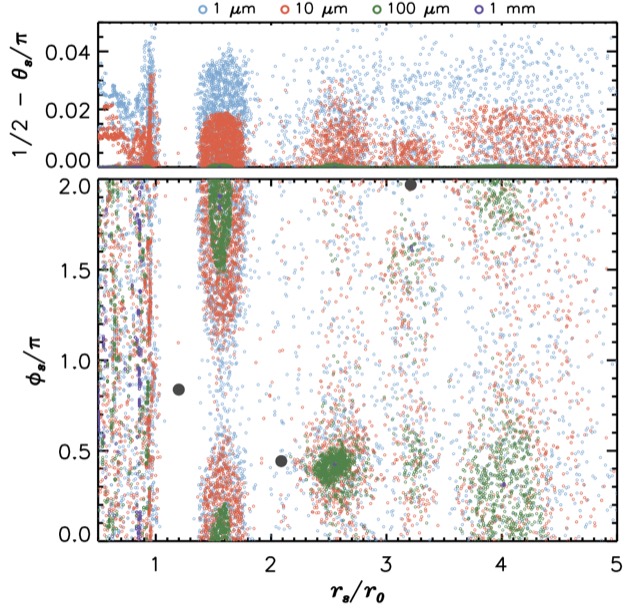}}
\caption{%
        As Figure~\ref{fig:sp_dist_a16}, but the solids' distributions
        refer to model \hmls.
        Left and right panels show, respectively, results from the models
        in which dust is released in the disc mid-plane and in which dust 
        is assumed to arise from planetesimals' collisions.
        (\textit{High-resolution versions of this figure can be found
        \href{https://doi.org/10.6084/m9.figshare.16926040.v1}{here}}.)
        }
\label{fig:sp_dist_dhma15}
\end{figure*}

\begin{figure*}
\centering%
\resizebox{2.07\columnwidth}{!}{\includegraphics[clip]{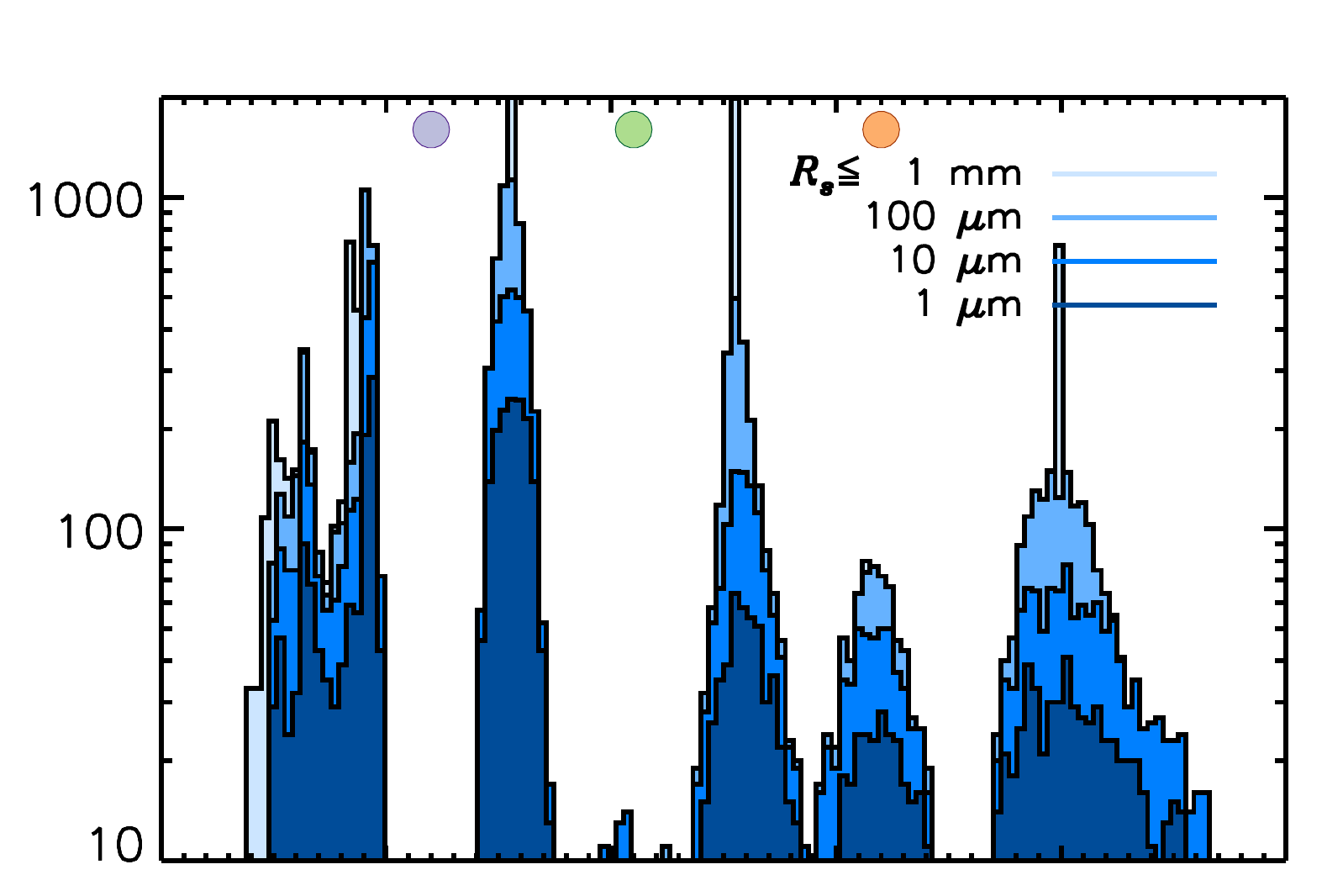}%
                             \includegraphics[clip]{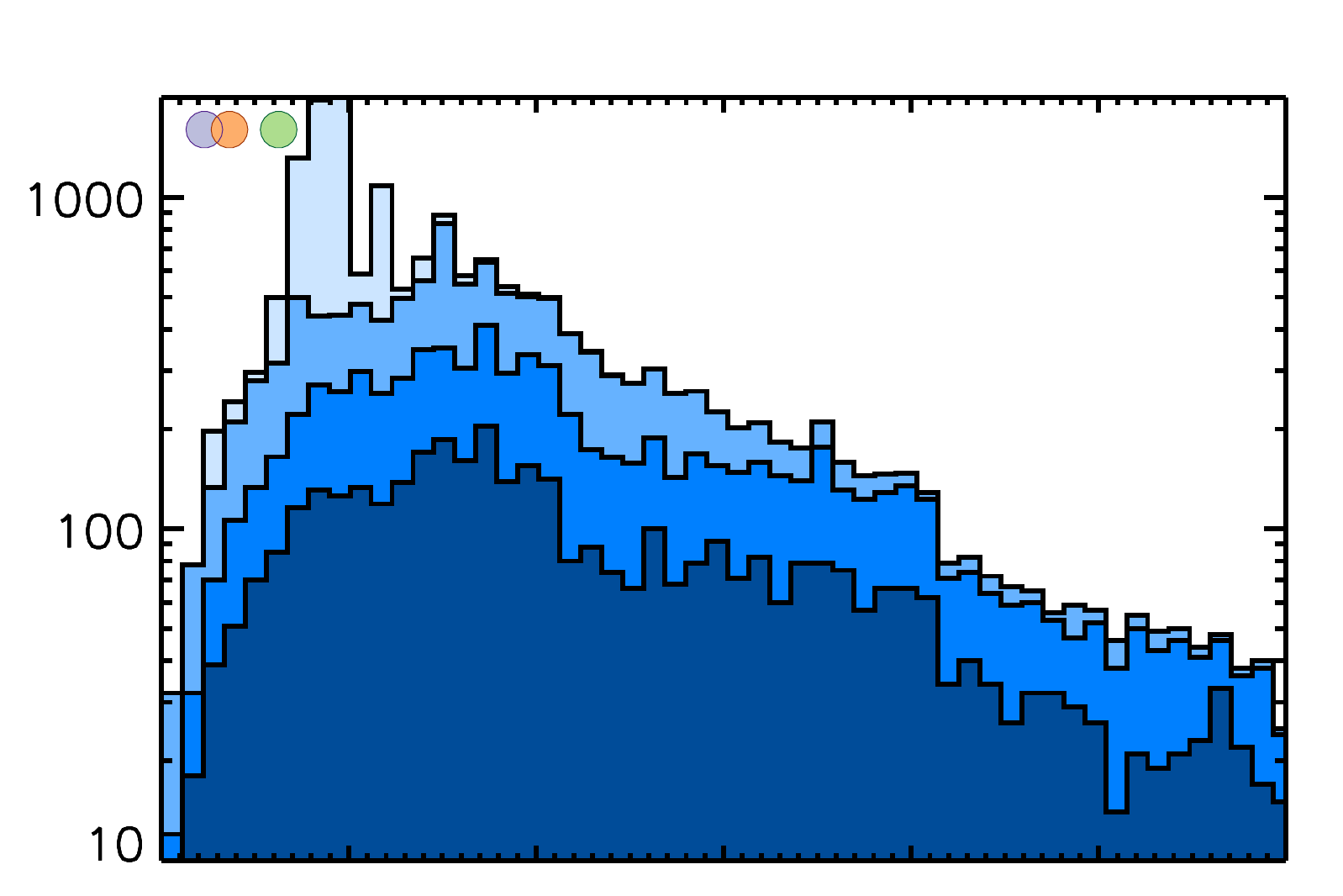}%
                             \includegraphics[clip]{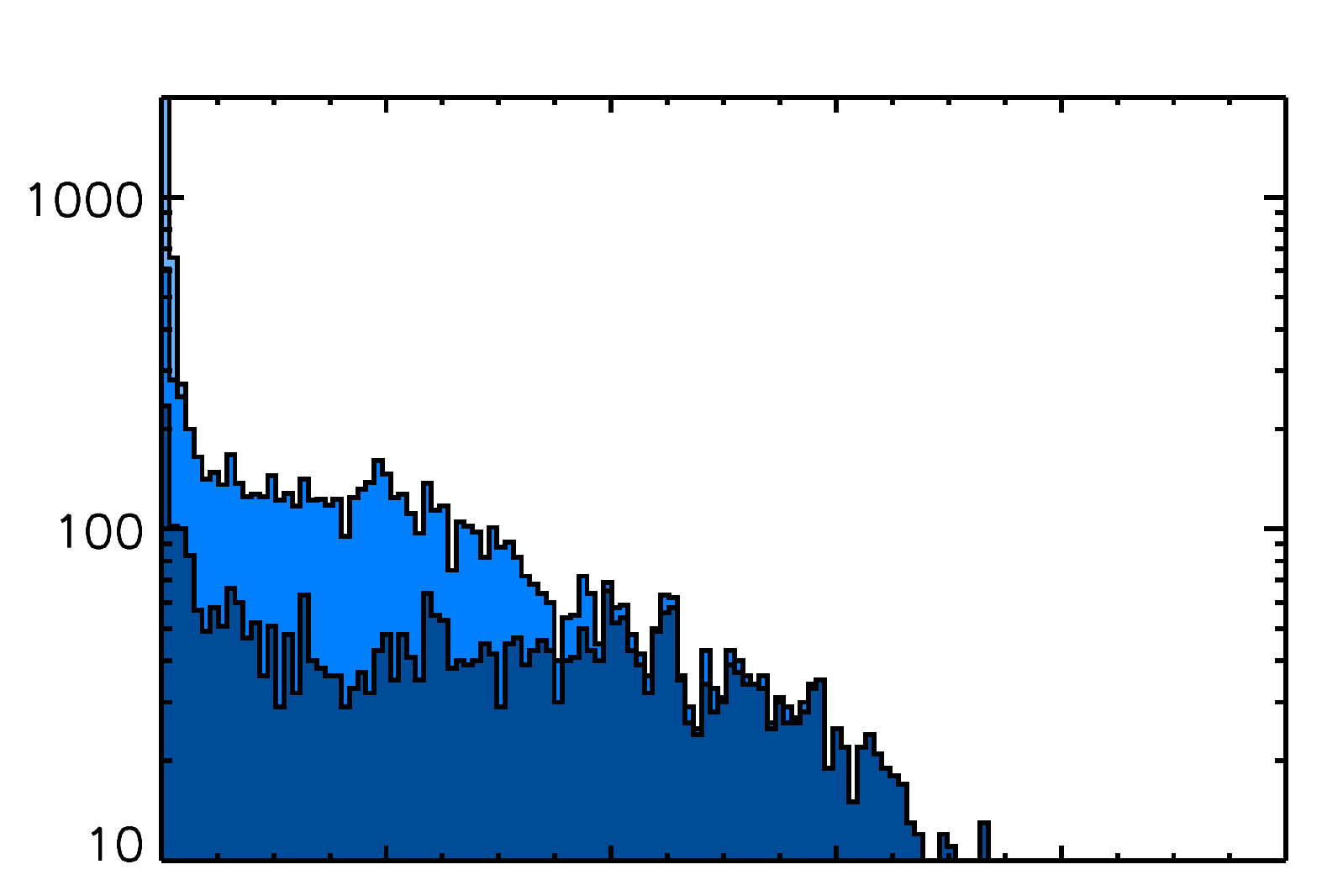}}
\resizebox{2.07\columnwidth}{!}{\includegraphics[clip]{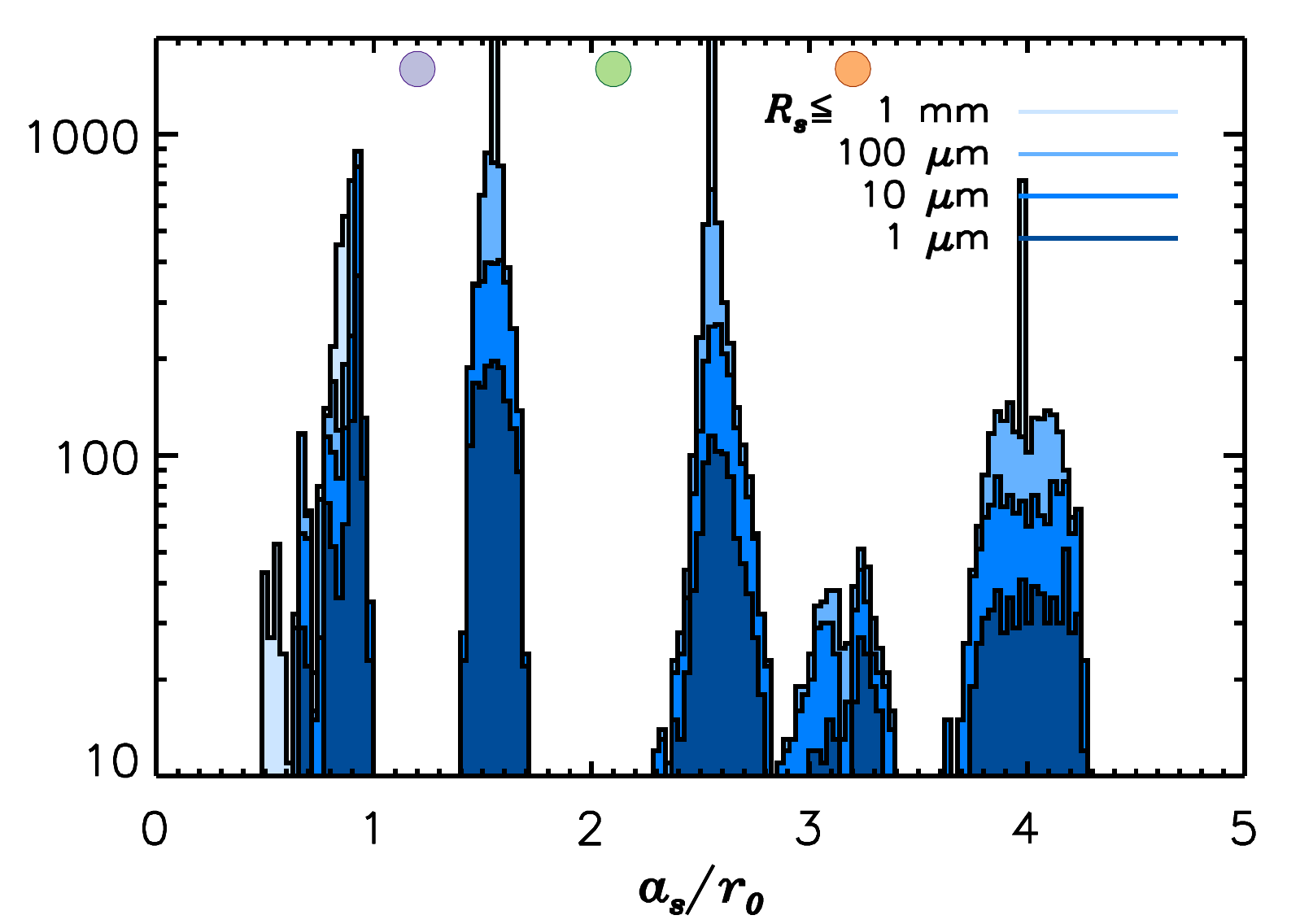}%
                             \includegraphics[clip]{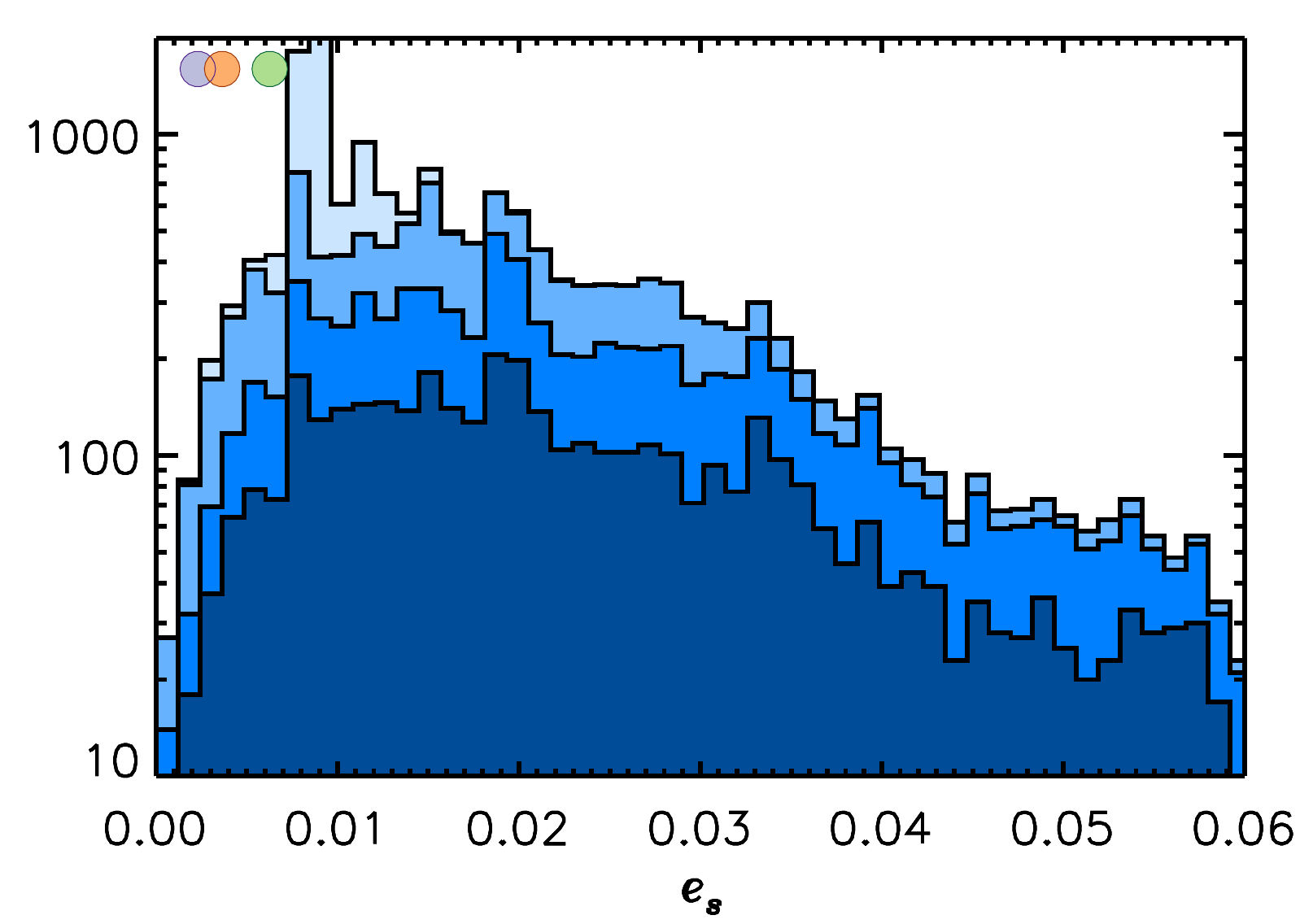}%
                             \includegraphics[clip]{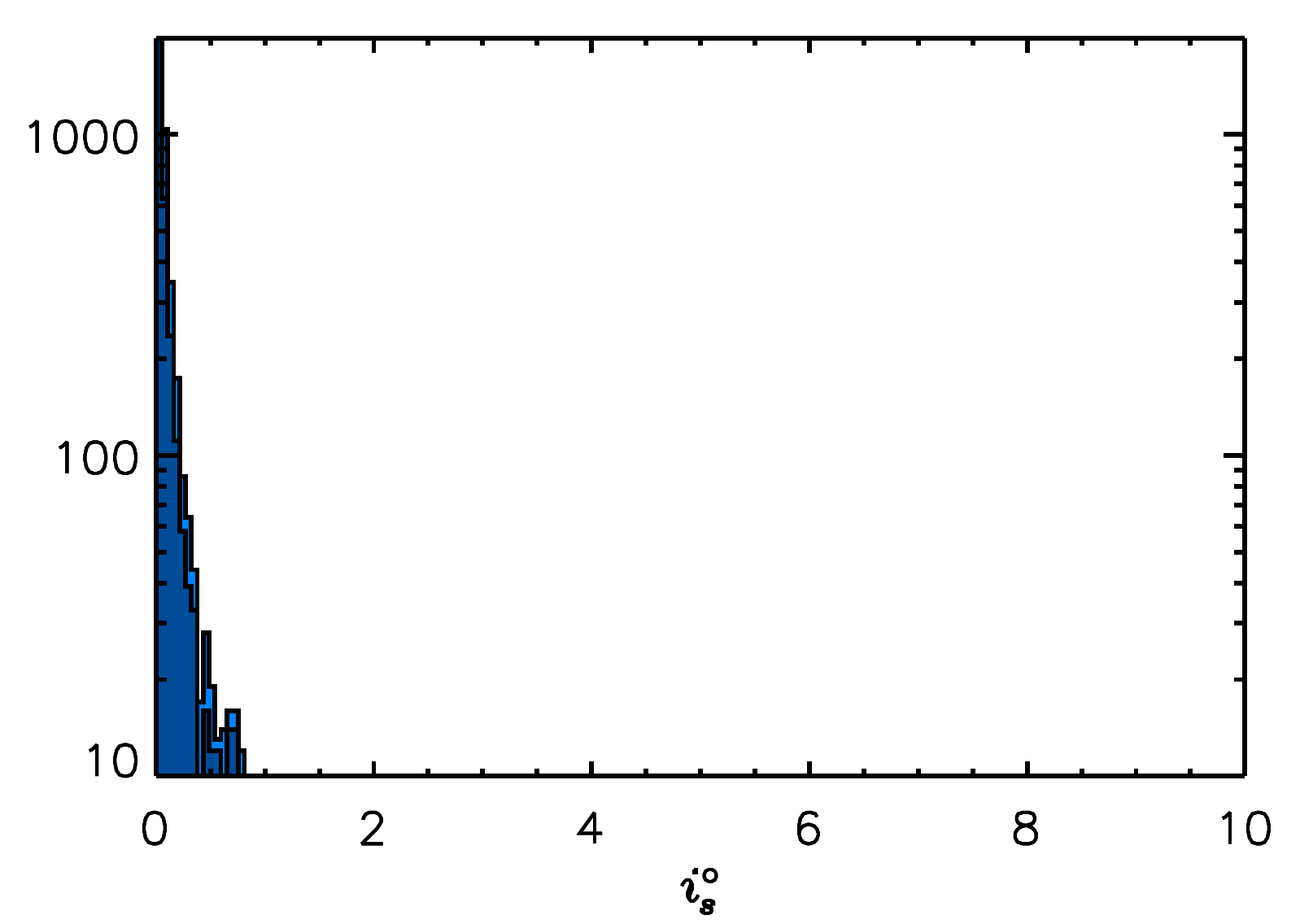}}
\caption{%
        As Figure~\ref{fig:sp_hist_a16}, but for models \hmls. 
        Top/bottom panels show simulations with
        collision-produced/first-generation dust.
        Different shades of blue render different particle radii, 
        as indicated. 
        }
\label{fig:sp_hist_dhma15}
\end{figure*}

Although the long-term evolution of vortices was not monitored in
detail, these features did persist for the duration of the calculations.
We do not speculate on the fate of the solids' concentrations, which
is not simulated in the models (since particles are treated as single
physical units, not as super-particles), although above some threshold
they may collapse into larger solids 
\citep[e.g.,][and references therein]{heng2010,cuzzi2010}.
It is expected that both first- and second-generation dust orbiting
close to the disc mid-plane would be affected by vortices in a similar 
manner, and therefore they would not provide an indicator to distinguish 
the two populations.
Aside from the presence of these flow features, which requires
particular conditions (e.g., low $\alpha$ and low gas temperatures),
the conclusion stemming from the mid-plane gas dynamics is more general:
it may affect dust differently based on their size but not on their
origin, as can also be concluded by comparing the semi-major
axis and eccentricity distributions in the upper and lower
panels in Figures~\ref{fig:sp_hist_hma15} and \ref{fig:sp_hist_dhma15}.
Some differences can be seen in the presence of dust within the
gas gaps generated by the planets, which take longer to be cleared
when grains are initially distributed off the mid-plane (see, e.g.,
left panels of Figure~\ref{fig:sp_hist_a16}), an effect connected
to their settling timescale. 
The extended clearing time in gaps, added to the continued production,
of second-generation dust could help explain some observations
\citep[e.g.,][]{isella2016} and also affect the estimates of turbulence
levels in the gas.
Note that the right panels of Figures~\ref{fig:sp_dist_a16},
\ref{fig:sp_dist_hma15}, and \ref{fig:sp_dist_dhma15} show that
particles approach the star closer than those in the left panels.
This is because of the smaller perihelia (larger $e_{s}$ on average)
of the initial distribution of second-generation particles (see
top panels of Figure~\ref{fig:sp_hist_comp}).

\begin{figure}
\centering%
\resizebox{1.05\columnwidth}{!}{\includegraphics[clip]{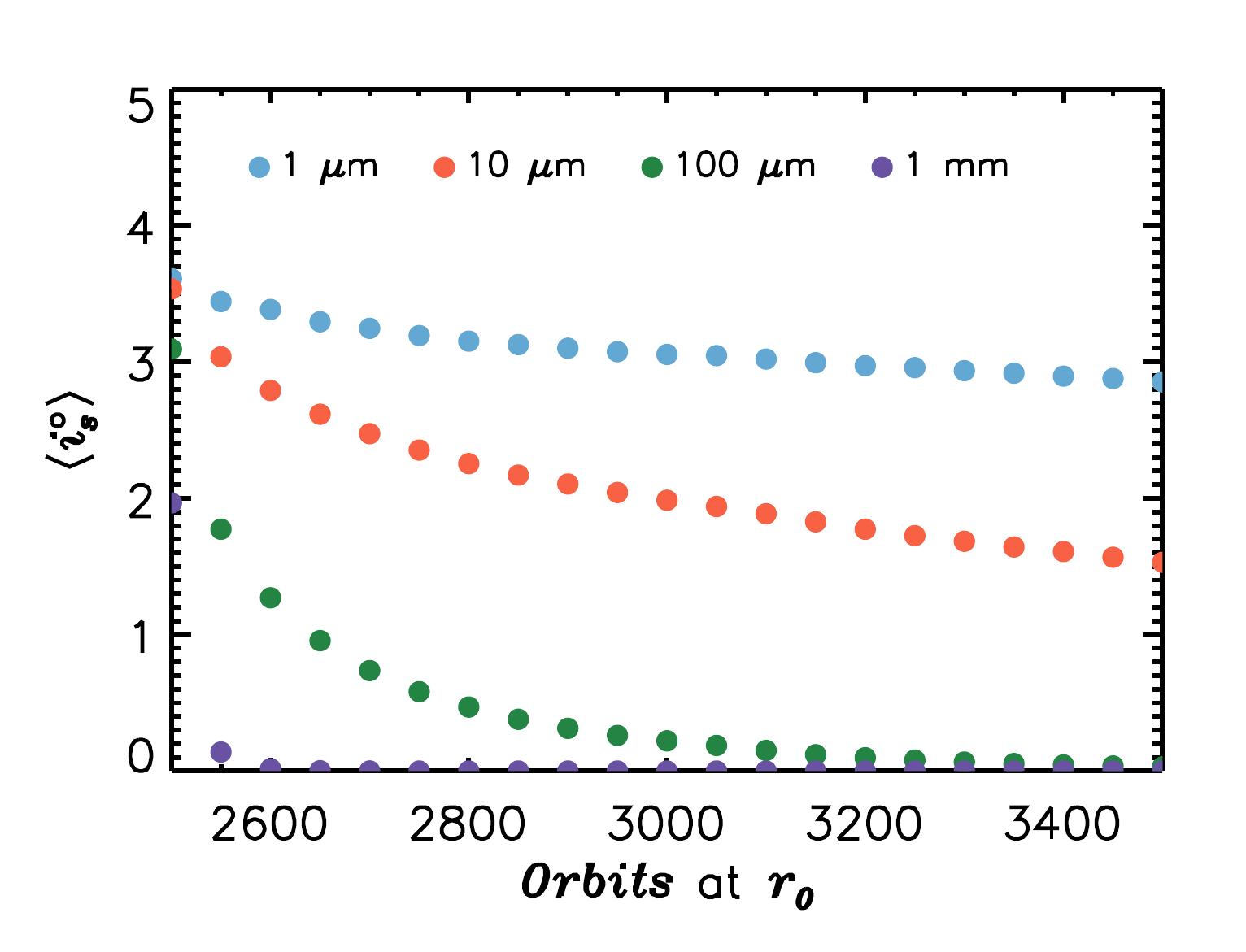}}
\caption{%
        Average orbital inclination of dust grains, for different
        radii (as indicated), as a function of time for model
        \lmla. The dust is released in the disc at $t\approx 2500$
        orbits (the time refers to the reference radius $r_{0}$).
        The average includes solids at all orbital radii.
        }
\label{fig:is_a16}
\end{figure}

Removal of dust by sublimation is negligible in these models.
Since the inner boundary of the computational domain only extends
inward to $4$--$5\,\AU$ (depending on $r_{0}$), ablation of ice
by thermal radiation is ineffective \citep[e.g.,][]{gennaro2015}.
Nonetheless, a small fraction the modelled solids' mass
($0.1$--$1$\%) evaporates. This is likely due to frictional 
heating undergone by some particles in the proximity of the planets,
where both $\rho$ and relative velocities (between gas and solids)
are enhanced (in fact, the loss is greatest in model \hmha).
Accretion on the planets removes up to a few percent of the modelled
solids' mass. Both removal processes affect the two dust populations
to a similar extent, implying that the largest grains (which orbit
closest to the mid-plane) are mainly involved.

\subsection{Dust Vertical Settling}
\label{sec:DVS}

The largest dynamical differences between the two scenarios representing
first-and second-generation dust are, as expected, found in the vertical
distributions of the smallest particles, $R_{s}=1$--$10\,\mu\mathrm{m}$,
which take the longest to settle toward the mid-plane of the disc 
compared to larger grains (see upper panels of 
Figures~\ref{fig:sp_dist_a16},
\ref{fig:sp_dist_hma15} and \ref{fig:sp_dist_dhma15}). 
After a period of a few hundreds of orbits,
$R_{s}=100\,\mu\mathrm{m}$--$1\,\mathrm{mm}$ particles are mostly
confined in (or in the proximity of) the mid-plane. 
These and larger grains would quickly mix with primordial 
mid-plane dust.

In a disc unperturbed by planets, assuming that the (spherical) grains 
are within the Epstein regime of gas drag (i.e., $R_{s}$ is much 
smaller than the mean-free path of gas atoms/molecules) and 
their velocity relative to the gas remains subsonic, 
the stopping time of the particles (i.e., their relative velocity
divided by the drag acceleration) is
\begin{equation}
 \tau_{\mathrm{stop}}=%
 \left(\frac{\rho_{s}}{\rho}\right)%
 \left(\frac{R_{s}}{c}\right),
 \label{eq:tstop}
\end{equation}
in which $\rho$ is the gas density (and $c$ the sound speed).
The stopping time increases as the particle size increases and
the gas density reduces.
The dust settling timescale, $\tau_{\mathrm{set}}$, is given 
by the distance over the mid-plane, $z$, divided by the settling
velocity,
$|v_{\mathrm{set}}|=z\Omega^{2}_{\mathrm{K}}\tau_{\mathrm{stop}}$
\citep[see, e.g.,][]{dullemond2004}, so that
\begin{equation}
 \Omega_{\mathrm{K}}\tau_{\mathrm{set}}=%
 \frac{1}{\Omega_{\mathrm{K}}\tau_{\mathrm{stop}}}.
 \label{eq:tset}
\end{equation}
Therefore, as $\tau_{\mathrm{stop}}$ approaches zero, the settling
timescale diverges since dust tends to move as a parcel of gas.
At a given radius in the disc, expanding the gas density
in Equation~(\ref{eq:tstop}), this timescale is
\begin{equation}
 \Omega_{\mathrm{K}}\tau_{\mathrm{set}}\propto%
 \frac{1}{R_{s}}%
 \left(\frac{\Sigma}{\rho_{s}}\right)%
 \exp{\left[-\left(\frac{\cot{\theta}}{h}\right)^2\right]},
 \label{eq:tset_red}
\end{equation}
where $\cot{\theta}=z/r$ ($r$ is the cylindrical radius).
Dust grains decay very quickly far above the disc mid-plane
because of low gas density (hence, long $\tau_{\mathrm{stop}}$)
but $\tau_{\mathrm{set}}$ becomes nearly independent of height 
below the gas pressure scale-height, $H$.

The expectation is that the settling timescale locally drops
as the particle size increases, which is in agreement with
the dust inclination distributions in the right panels of
Figures~\ref{fig:sp_hist_a16}, \ref{fig:sp_hist_hma15} and
\ref{fig:sp_hist_dhma15}.
It should be noted, however, that Equations~(\ref{eq:tstop}),
(\ref{eq:tset}),
and (\ref{eq:tset_red})
are of limited application in the interpretation of the simulation
results since settling grains of a given size may be outside of 
the Epstein drag regime as gas density increases toward the mid-plane.
In addition, since second-generation dust evolves in a disc
perturbed by planets, the sedimentation velocity is also affected
by their presence (directly through their gravity and indirectly 
through perturbations induced in the gas flow).
Therefore, the settling timescale can be different, or somewhat different,
from that predicted by the analytical estimate (in the applicable
size range) provided by Equation~(\ref{eq:tset}).
In fact, the top-left panels of Figures~\ref{fig:sp_dist_a16},
\ref{fig:sp_dist_hma15} and \ref{fig:sp_dist_dhma15} all indicate
that the planets cause some amount of vertical stirring on the 
mid-plane dust, mostly confined to radial locations in the proximity
of the planets' orbits.
Since the effect appears larger on the small grains, it is
mainly associated to the vertical motion of the gas induced by 
the planets.
From the inclination distributions in the bottom panels of
Figures~\ref{fig:sp_hist_a16}, \ref{fig:sp_hist_hma15} and
\ref{fig:sp_hist_dhma15}, the bulk vertical stirring (including
particles of all sizes) in the disc region perturbed by the planets
can be estimated to not exceed $\approx 25$\% of gas scale-height.
This estimate depends on the planet masses (and gas temperature)
considered herein and would be enhanced if more massive planets 
were perturbing the gas.
Obviously, this effect would equally impact first- and second-generation 
dust.
Nonetheless, vertical stirring of dust by planets could potentially
influence the interpretation of observations constraining the gas-to-dust scale-heights 
\citep[see, e.g.,][]{ohashi2019,doi2021}.

\subsection{Dust Sedimentation and Replenishment}
\label{sec:DSR}

The sedimentation behaviour of dust is more clearly observed in
Figure~\ref{fig:is_a16}, which presents the average orbital
inclination of the grains versus time, in model \lmla, for 
the various particle sizes.
A very similar settling behaviour is found in model \hmha.
The averages are computed as mean values of the entire population
of solids (binned by grain radius), including all orbital radii 
($0.5\,r_{0}\lesssim a_{s}\lesssim 4.5\,r_{0}$).
Clearly, the presence of small dust off the disc mid-plane would
depend on the balance between sedimentation and replenishment by
collisional comminution of planetesimals. If the latter process
occurred on a timescale comparable to (or shorter than) those
involved in Figure~\ref{fig:is_a16}, then dust would be distributed,
and possibly observable, over a range of heights.
It should be noted that turbulent stirring would not contribute
significantly to the dust vertical distribution because of 
the low values of the parameter $\alpha$ employed herein
\citep[e.g.,][]{dubrulle1995}.
These low turbulence values, however, appear to be consistent 
with the observed structures of the dust in the disc mid-plane
\citep[e.g.,][]{isella2016,liu2018}.
Weak turbulence is also supported by observations of CO emission 
lines  \citep{flaherty2015}.

Figure~\ref{fig:is_a16} shows, as expected, that the initial decline 
of the dust inclination is rapid at heights $\gtrsim H$ but it
slows down as grains approach the mid-plane.
The mean settling timescales, evaluated over the last $\approx 800$
orbits of evolution from the data plotted in the Figure,
are $\approx 6700$ and $\approx 1700$ orbits (at $r_{0}$) for $R_{s}=1$
and $10\,\mu\mathrm{m}$, respectively. These timescales are determined
by fitting an exponential curve to the average dust inclination, 
so that 
$d\langle i_{s}\rangle/dt=-\langle i_{s}\rangle/\tau_{\mathrm{mod}}$
and $\tau_{\mathrm{mod}}$ is the model estimate of the settling 
timescale.
As anticipated above, $\tau_{\mathrm{mod}}$ represents a bulk value
that includes particles at all radial distances in the distribution
(see, e.g., Figures~\ref{fig:sp_dist_a16} and \ref{fig:sp_hist_a16}), 
and therefore accounts for variations associated to the varying gas 
density.
Larger grains settle, on average, on timescales shorter than $250$
orbits. Model \hmls, with lower gas density (hence longer stopping
times), provides shorter settling timescales, 
between $\tau_{\mathrm{mod}}\approx 5100$
($R_{s}=1\,\mu\mathrm{m}$) and $1000$ ($R_{s}=10\,\mu\mathrm{m}$) 
orbits.

According to the calculations of \citet{turrini2019}, collisions
among planetesimals, whose orbital dynamics is excited by the planets'
gravity, would produce and distribute dust over many gas
scale-heights. In the presence of fully formed (Jupiter-mass) planets,
that study estimated average dust production timescales by collisions
of $6200$--$6600$ orbits (at $r_{0}$).
These would be average values over the planetary region, but 
production rates also depend on orbital distance and decline
with $r$. Sedimentation would bring
this dust down to heights $\approx H$ on relatively short times,
as models indicate,
but further settling of $\approx 1\,\mu\mathrm{m}$ (and smaller) 
grains would occur on timescales comparable with dust resupply 
by collisions.
Henceforth, a population of small dust distributed above the mid-plane,
up to $\approx H$, would be expected if planetesimal collisions
were ongoing.

The production of second-generation dust would depend on several
quantities, such as number density of planetesimals, their size
distribution, and the interior strength of the bodies. Collision
rates would also depend on the planets' masses. Higher/lower gas
densities in the disc could affect dust dynamics, but much less
planetesimal collisions. Nonetheless, if the parameters applied
here are representative of \HD\ system, based on the settling
timescales of small grains, resupply of dust should occur over
timescales $\lesssim 1.5\,\mathrm{Myr}$ in order for second-generation
dust to be observable above the disc mid-plane. 
Over longer collisional timescales, this dust population would 
likely be completely mixed with primordial dust.
The admixture of the two populations, whose orbits would be
largely confined to the disc mid-plane, would show similar
dynamical features (including some vertical stirring by 
the planets) and would not offer distinctive dynamical
signatures for separating the two species.
A lack of indicative features would also be expected if collisions 
were unable to produce significant quantities (from an observational
point of view) of $\lesssim 10\,\mu\mathrm{m}$ grains, although 
the size limit would depend on the thermodynamic properties of the gas.

Multi-wavelength observations of \HD\ \citep{muro-arena2018} indicate
that although three rings are detectable in millimetre thermal 
emission, only the innermost ring is detectable in polarised scattered 
light in the near infrared. These observations suggest that 
the surface of the disc around the two outer planets may lack small 
dust grains ($R_{s}\lesssim 1\,\mu\mathrm{m}$).
\citet{muro-arena2018} showed that ``enhanced'' settling can reproduce
observations if the applied turbulence parameter in the region is 
$\alpha\sim 10^{-5}$, as applied herein.
These observations are compatible with (or do not exclude) the presence
of collisionally-produced dust in the system since, as mentioned above,
the collision frequency of planetesimals reduces as their orbital 
distance increases. 
Hence, the supply rate of second-generation dust may fall beneath 
the removal rate by sedimentation in the region of the outer rings.

\section{Conclusions}
\label{sec:C}

\HD\ is probably one of the most thoroughly studied protoplanetary disc
because of its size and complex structure, characterised by well-defined
rings and gaps, which possibly originate from either forming or
recently-formed giant planets.
Past work \citep{turrini2019} suggests that orbital excitation of
leftover planetesimals by nearby giant planets leads to collisions
and to the production of second-generation dust. This process would point
to a resurgence of particulate material in circumstellar discs that harbour
massive planets, once these planets are nearly fully grown (i.e., at ages 
$\gtrsim 1\,\mathrm{Myr}$). 

The goal of the paper is to test this scenario and investigate the evolution
of second-generation dust in the system, under the assumption that 
said dust is produced by collisional comminution of planetesimals whose
orbits have been excited by the embedded planets.
Contrary to primordial dust, which is old enough to have settled on
near-circular orbits in the disc mid-plane and to have been cleared out
of the gaps carved by the planets in the gas distribution, second-generation
dust is continuously produced and is more dynamically active, being delivered
by collisions to highly eccentric and inclined trajectories. 

We present hydrodynamic calculations of the \HD\ system, simulating
both gas and dust components, and applying different initial conditions
to the grains, depending on whether they are intended to represent first-
or second-generation dust populations.
We also consider different approximations for the state of the gaseous
disc, based on prior modelling work that tried to reproduce some
observational features.
Contrary to prior studies of this system that only modelled the mid-plane
distribution of dust in two dimensions \citep[e.g.,][]{isella2016,liu2018},
we constructed three-dimensional disc models in order to examine 
the vertical distribution and settling dynamics of grains of various 
sizes. These are the first 3-D hydrodynamic models of the system 
(to our knowledge) that include dust and show some important features, 
such as the formation of vortices induced by Rossby
waves \citep[caused by the low levels of turbulence viscosity predicted 
for the system, e.g.,][]{liu2018}.
These vortices are effective in collecting dust in the disc mid-plane
and may represent sites for additional planetesimal formation
\citep[see, e.g.,][]{heng2010,cuzzi2010}, which would contribute 
to replenish the population of large bodies.

The most noticeable differences between the density distributions
of first- and second-generation dust appear in the vertical direction.
Second-generation dust is expected to have, on average, higher inclination,
with small grains taking longer to settle (than large grains) and therefore
extending farther above the mid-plane. 
Grains produced by collisions of planetesimals are also expected
to be injected on highly eccentric orbits which, however, are rapidly
circularised. The presence of small gains distributed in the vertical
direction is determined by a balance between the rate of sedimentation
and the rate of production by collisions (of larger bodies on inclined
orbits).
We find that the sedimentation rate of $\sim \mu\mathrm{m}$-size grains
is comparable to estimates of the production rate of second-generation
dust \citep{turrini2019}, therefore allowing this population to be 
persistent and potentially observable.
This possibility may reduce as the disc ages and gas disperses since 
settling timescale would become shorter (as gas densities decline).
Large, $\sim \mathrm{mm}$-size grains would instead settle too quickly 
to be observable. First-generation dust remains broadly confined to 
the mid-plane of the disc,
although subject to some amount of vertical stirring by the planets.

Production rates of second-generation dust are expected to depend
on some specific details (e.g., number density of planetesimals and
their size distribution). Assuming we applied representative ranges
of parameters for \HD, as long as planetesimal collisional timescales
are $\lesssim 1.5\,\mathrm{Myr}$, second-generation dust should be
distinguishable from primordial dust. If collisional timescales are
longer, the two populations would be likely completely mixed.
The admixture, largely confined to the disc mid-plane, would not 
offer distinctive dynamical signatures for separating the two
populations.
Coagulation and weathering processes would likely smooth
compositional differences.
Once gas is completely dissipated, collisions among planetesimals
may continue but the lifetime of second-generation dust would
be mainly dictated by Poynting–Robertson drag and planetesimal
scattering timescales.

The behaviour of small, second-generation dust may also aid in the
interpretation of some modelling results \citep[e.g.,][]{ohashi2019},
which show that beyond $70\,\AU$, outside the location of the innermost
planet, the dust scale-height is about twice as large as that of inner
disc regions. \citet{ohashi2019} suggested that an increase in gas
turbulence strength at those radial locations, and therefore enhanced
vertical stirring, may be responsible for these results.
However, collisionally-produced dust may also produce an enhanced
disc scale-height of solids in the proximity of the planet's orbits. 
Clearly, this latter interpretation would not rule out the possibility
of a radial variability of the turbulence strength, but it would rather
offer another, or contributing, explanation for the larger dust
scale-height observed in some sections of the disc.

Another finding of this work is that second-generation dust particles
tend to survive longer within the gaps carved by the planets in the gas
distribution. The effect is expected as long as the settling timescales
of the grains remain longer than the clearing timescales from the gap
regions. This process may offer an additional observational test for
the presence of second-generation dust particles.

Although derived for the \HD\ system, the general conclusion of this
study would be applicable to similar systems with embedded giant planets.
For example, collisionally-produced grains may be present in the HD~100546
system, which is suspected to harbour a planet at about $70\,\AU$
from the star, and for which there is observational evidence of
dust orbiting at considerable heights over the disc mid-plane
\citep{sissa2018}.

\section*{Acknowledgements}

We thank an anonymous reviewer whose comments helped improve
this paper.
Primary support for this work was provided by NASA’s Research
Opportunities in Space and Earth Science (proposals 80HQTR19T0071 and
80HQTR19T0086).
Computational resources supporting this work were provided by 
the NASA High-End Computing (HEC) Program through the NASA Advanced 
Supercomputing (NAS) Division at Ames Research Center.

\section*{Data Availability}

The data underlying the research results described in the article will 
be shared upon reasonable request to the authors.









\appendix

\section{Dust Distribution Sensitivity on Particle Number and Grid Resolution}
\label{sec:DSP}

\begin{figure*}
\centering%
\resizebox{1.9\columnwidth}{!}{\includegraphics[clip]{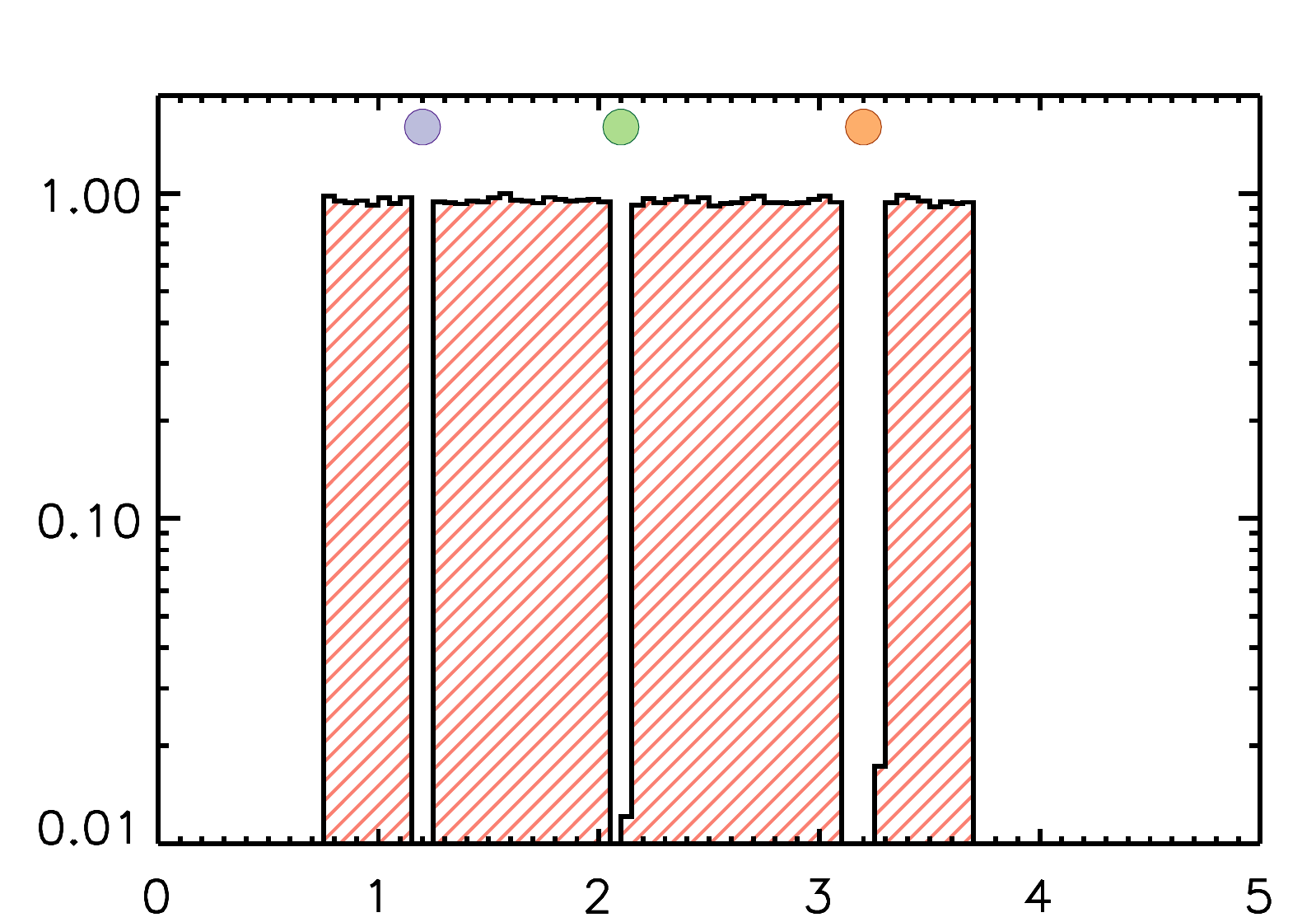}%
                             \includegraphics[clip]{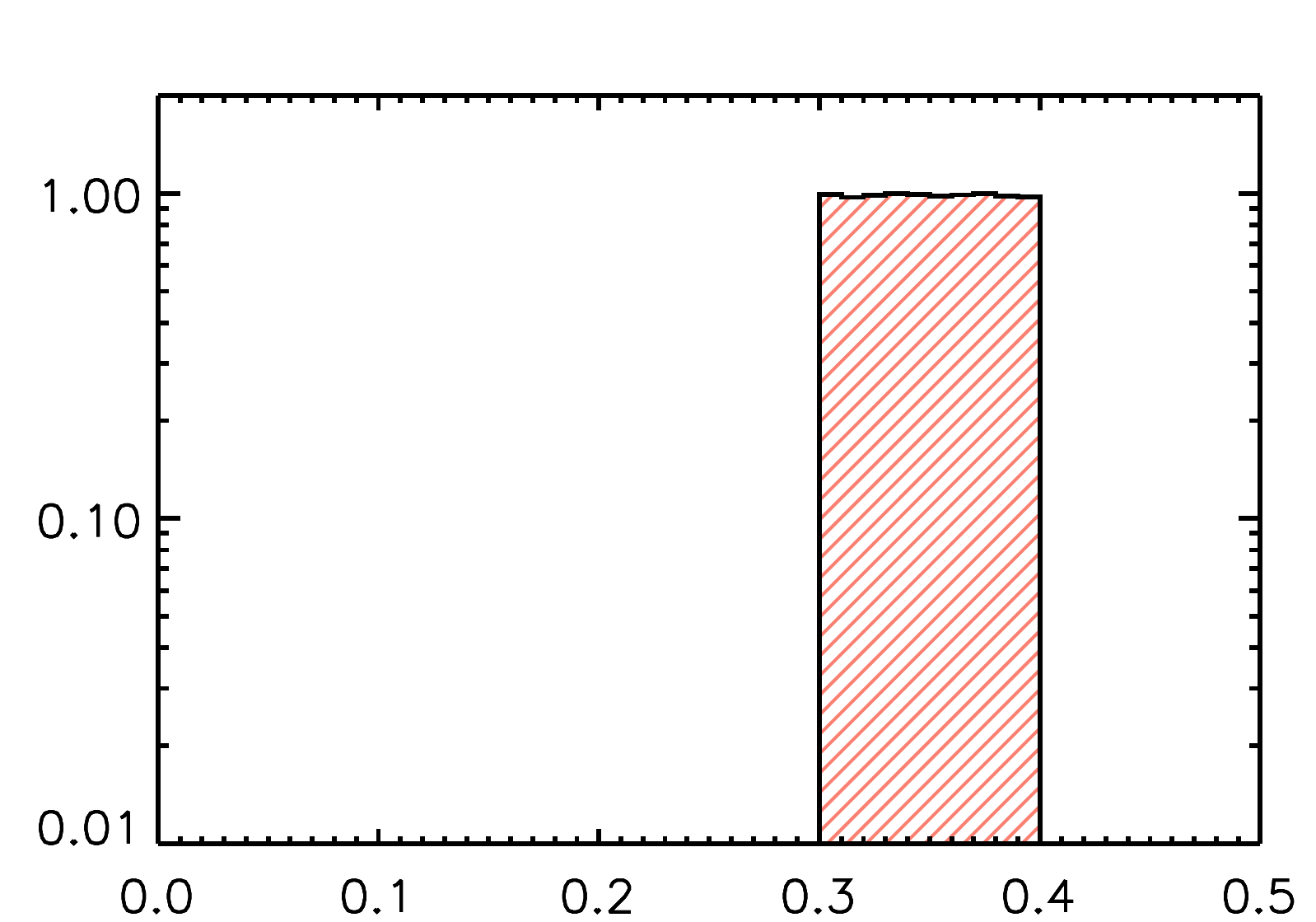}%
                             \includegraphics[clip]{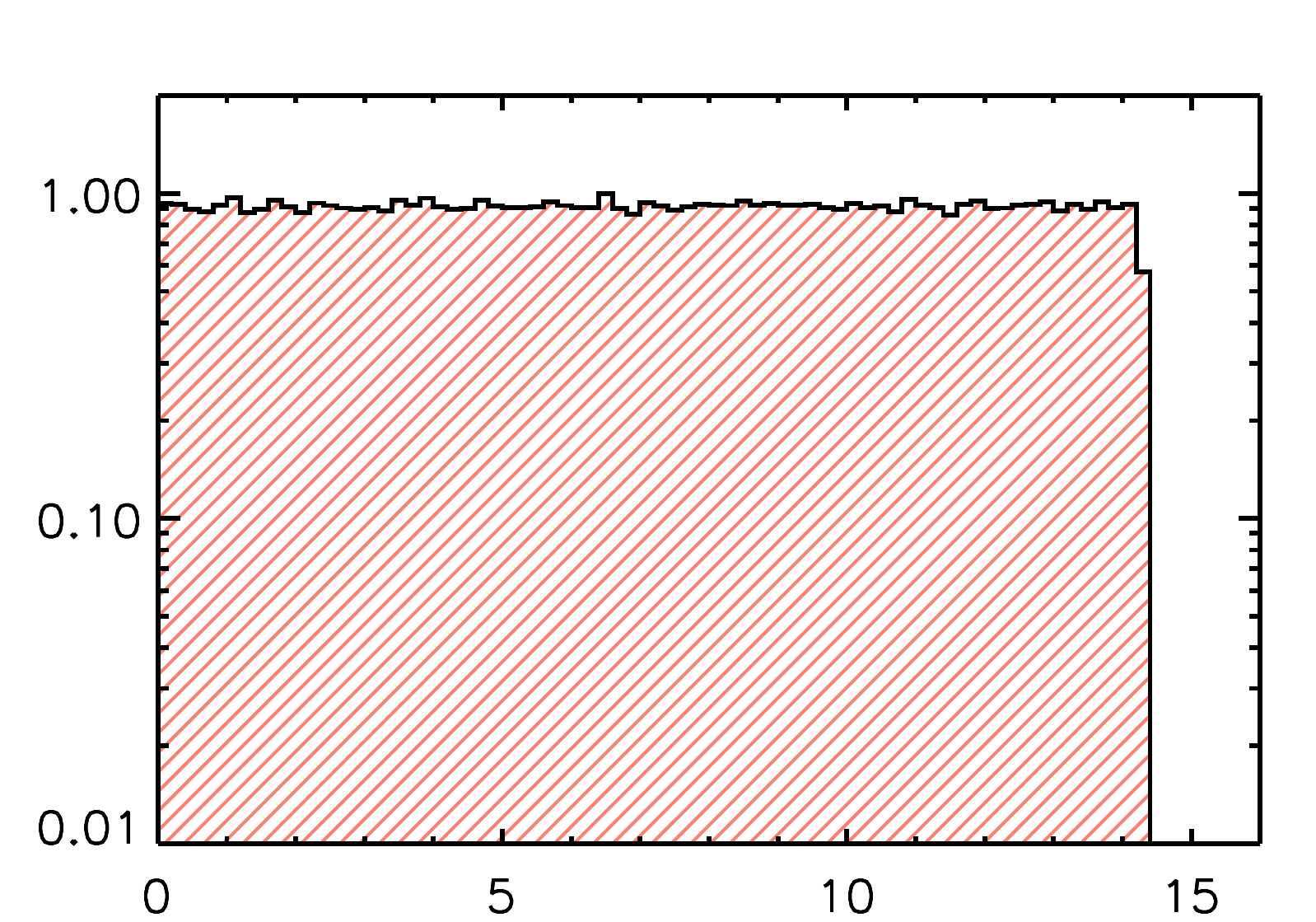}}
\resizebox{1.9\columnwidth}{!}{\includegraphics[clip]{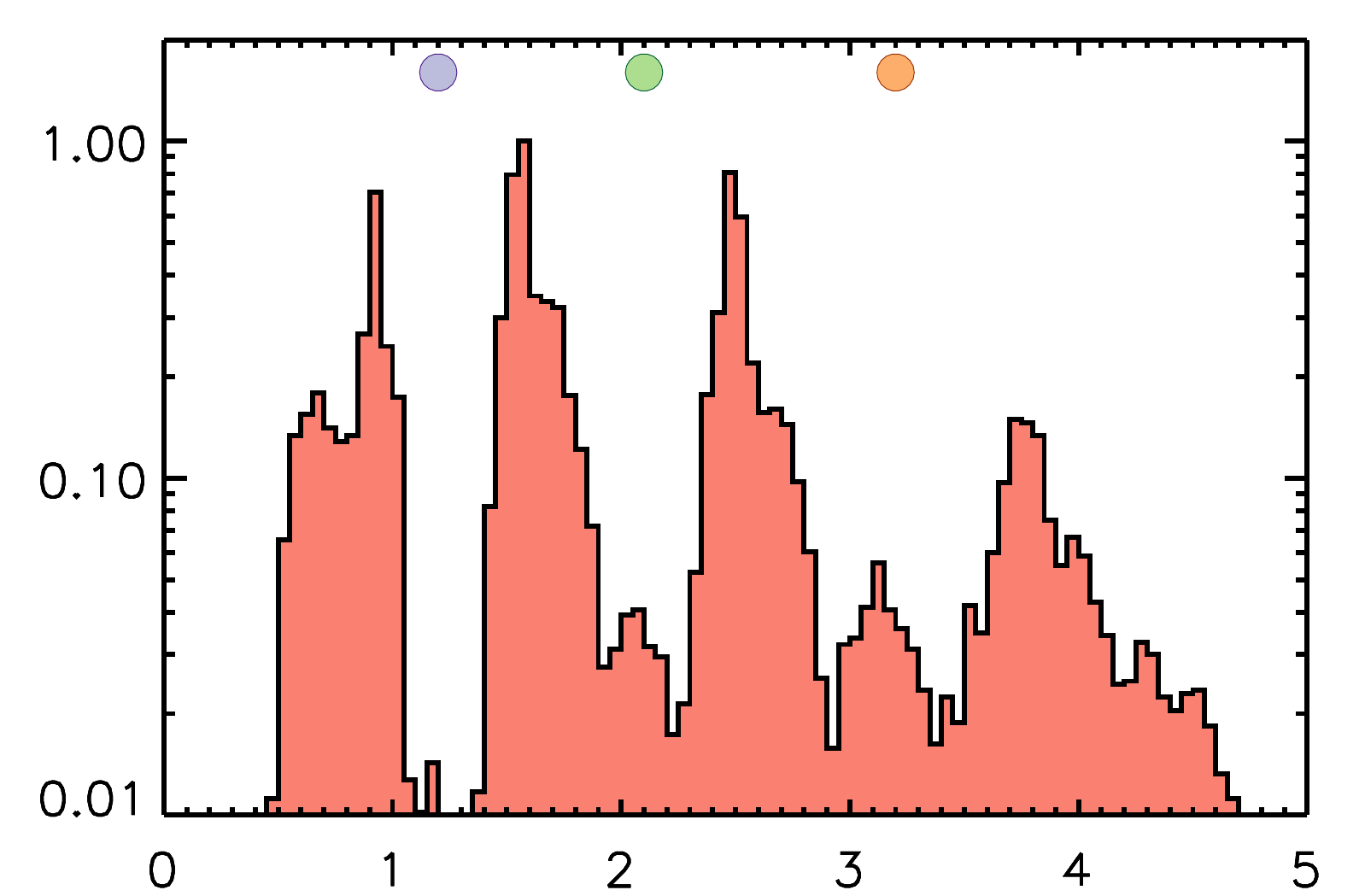}%
                             \includegraphics[clip]{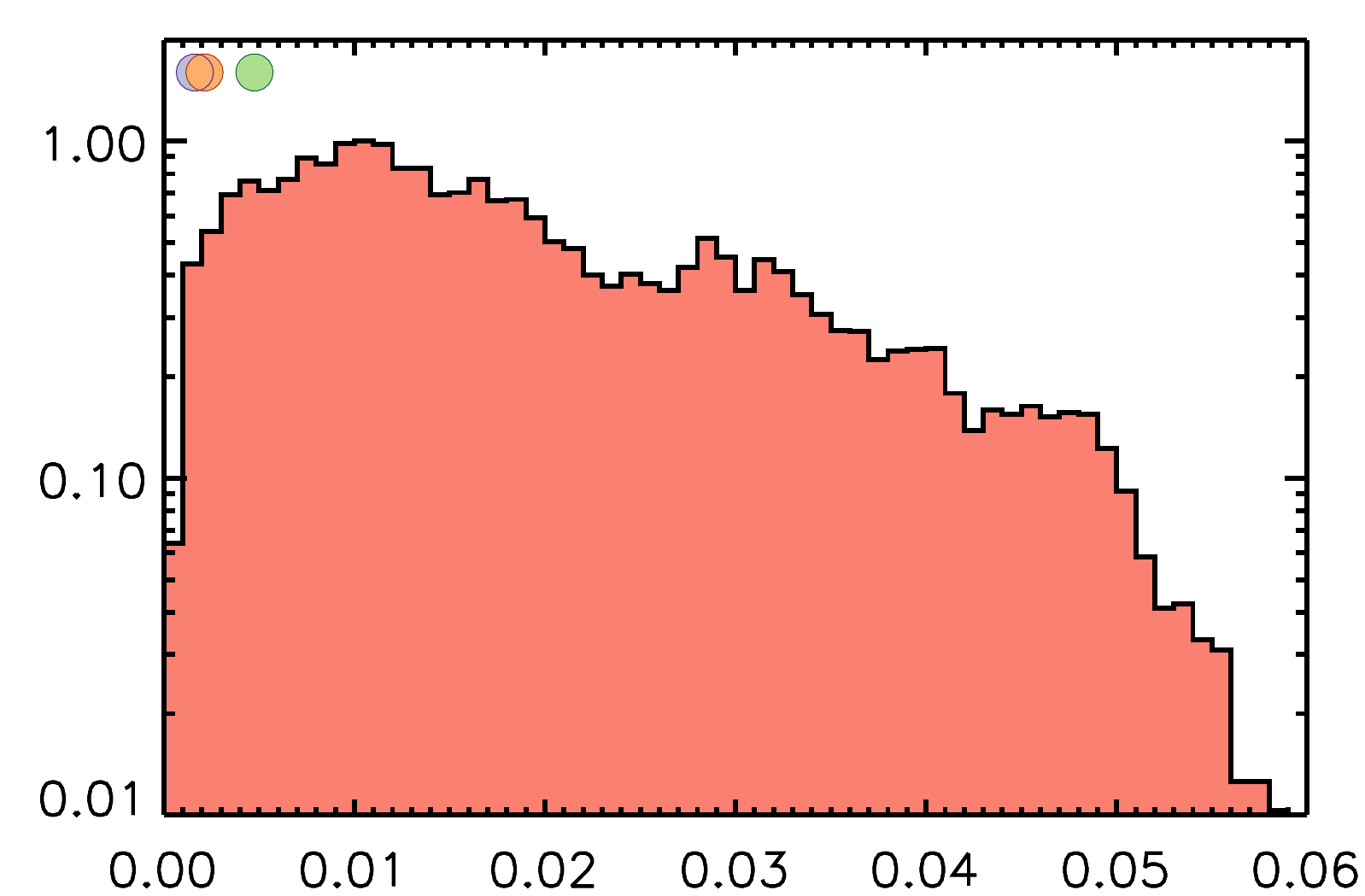}%
                             \includegraphics[clip]{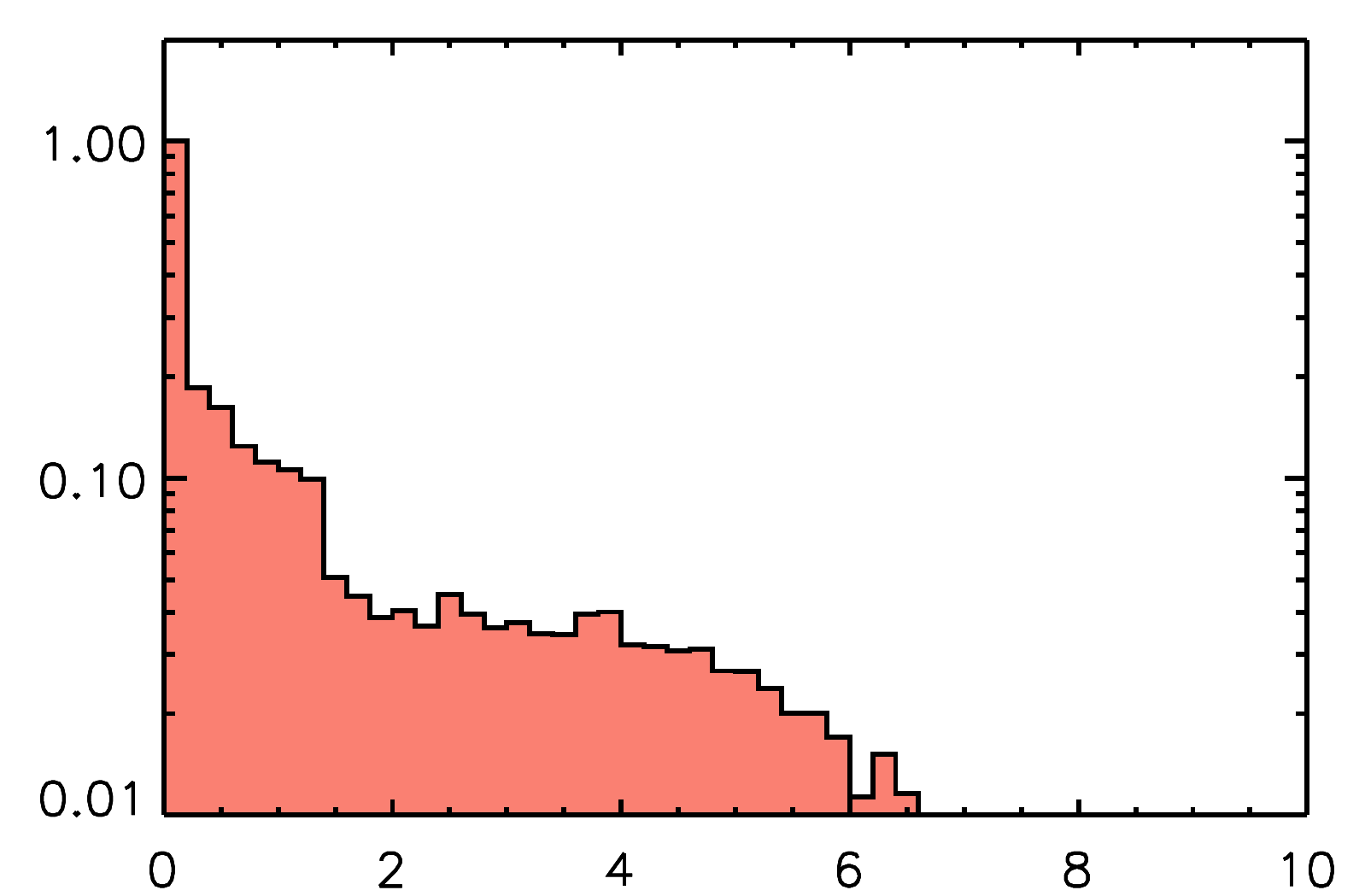}}
\resizebox{1.9\columnwidth}{!}{\includegraphics[clip]{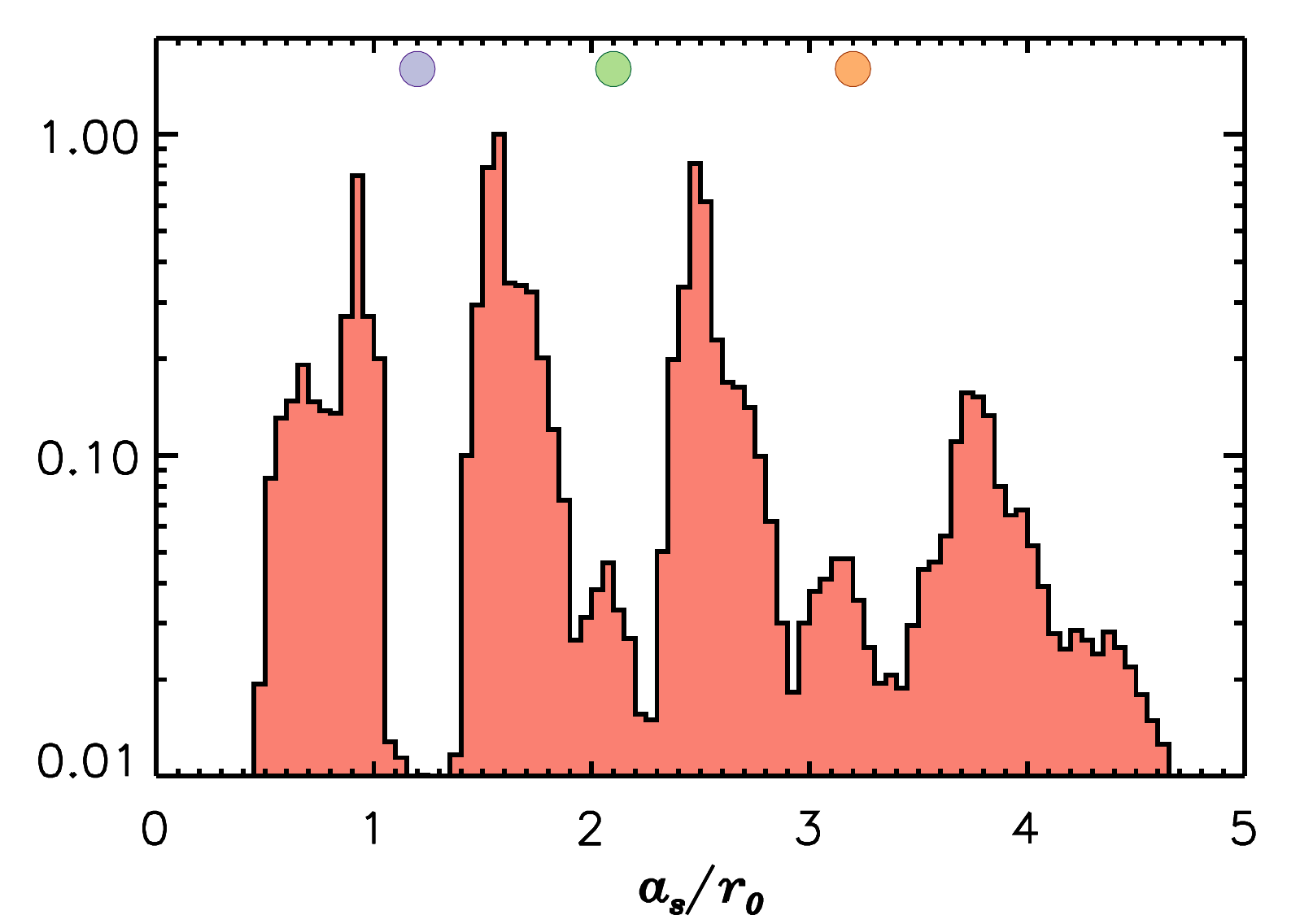}%
                             \includegraphics[clip]{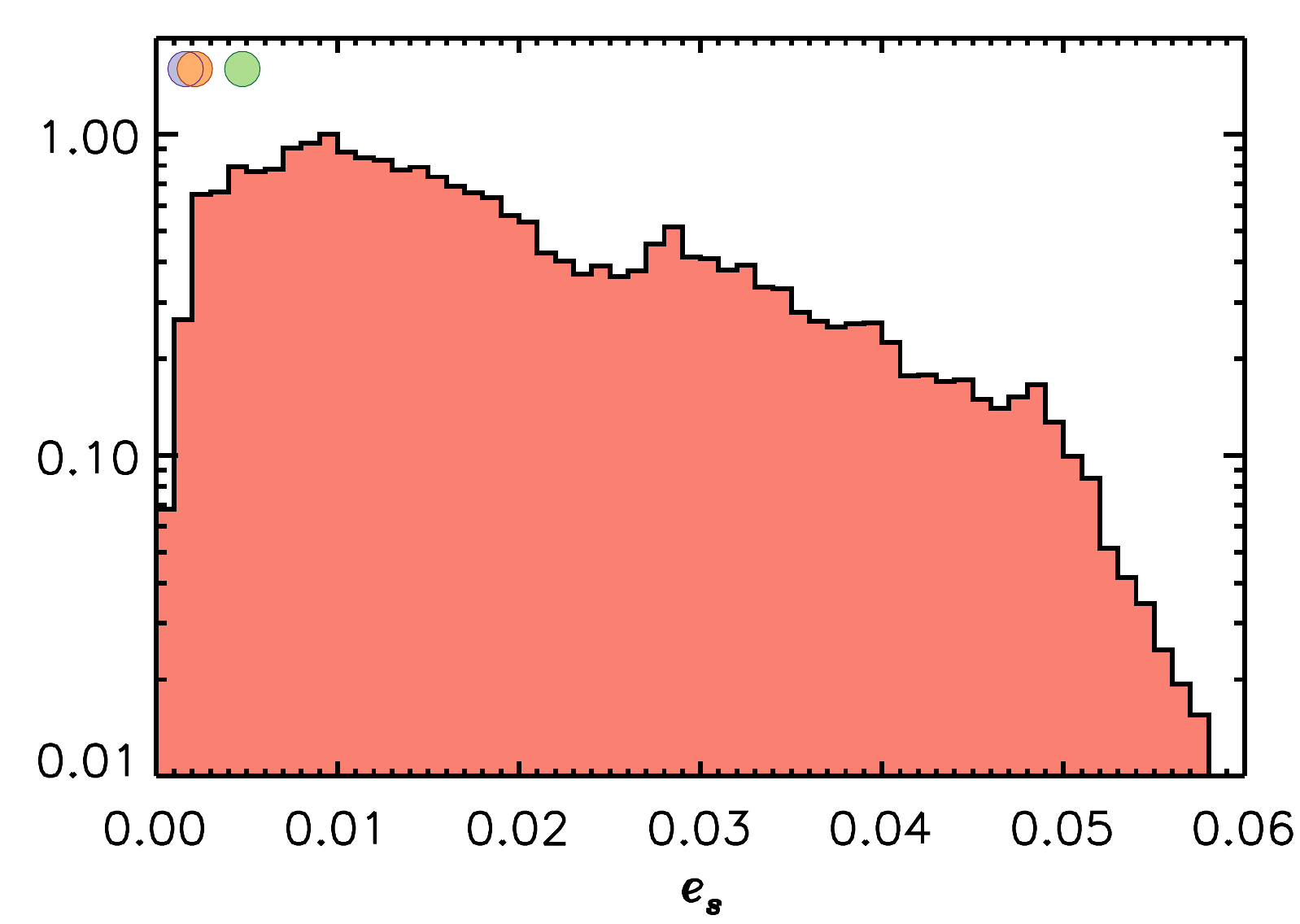}%
                             \includegraphics[clip]{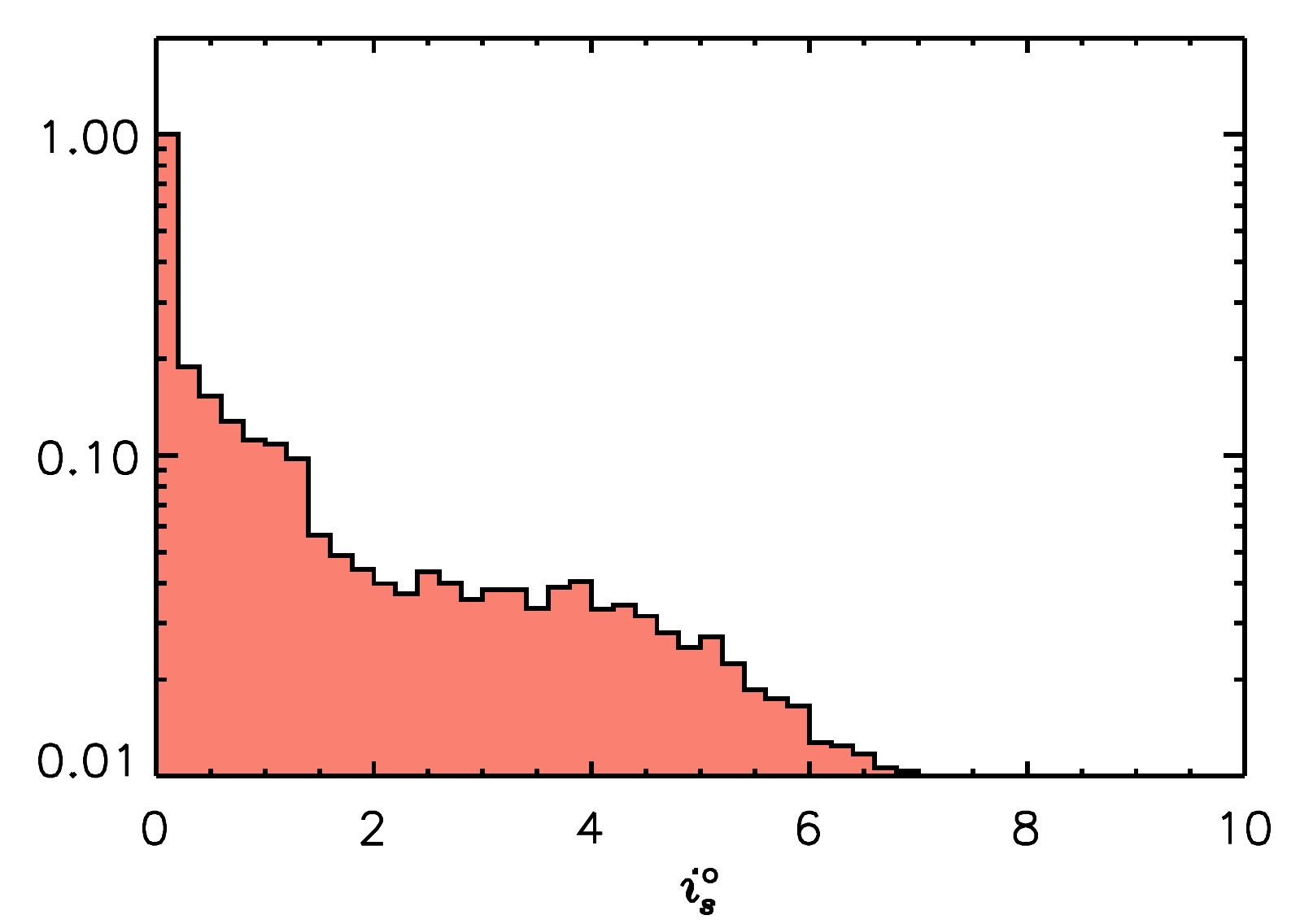}}
\caption{%
        Comparison of histograms with different number of particles
        from two versions of model \lmla. The evolution time is $t=2750$ 
        orbits (at $r=r_{0}$) and the solids have evolved for $250$ orbits.
        Top Row: Initial (normalized) dust distributions of
        semi-major axis (left), orbital eccentricity (centre), and 
        inclination (right) of second-generation dust.
        Middle row: Normalised distributions from the model with
        smaller number of particles.
        Bottom row: Normalised distributions from the model with
        larger number of particles.
        The coloured circles represent the planets' semi-major axis
        (left panels) and orbital eccentricity (centre panels).
        }
\label{fig:sp_hist_comp}
\end{figure*}

\begin{figure*}
\centering%
\resizebox{2.0\columnwidth}{!}{\includegraphics[clip]{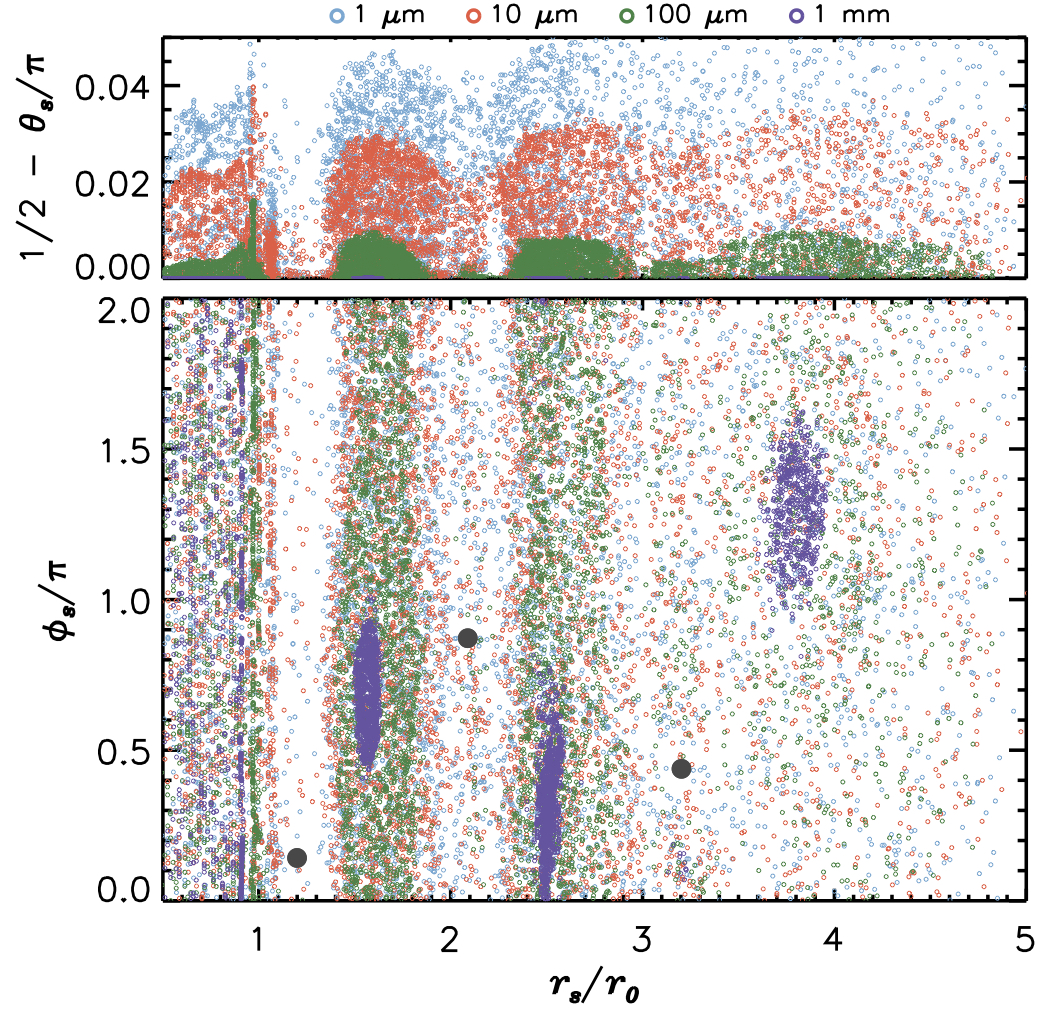}%
\includegraphics[clip]{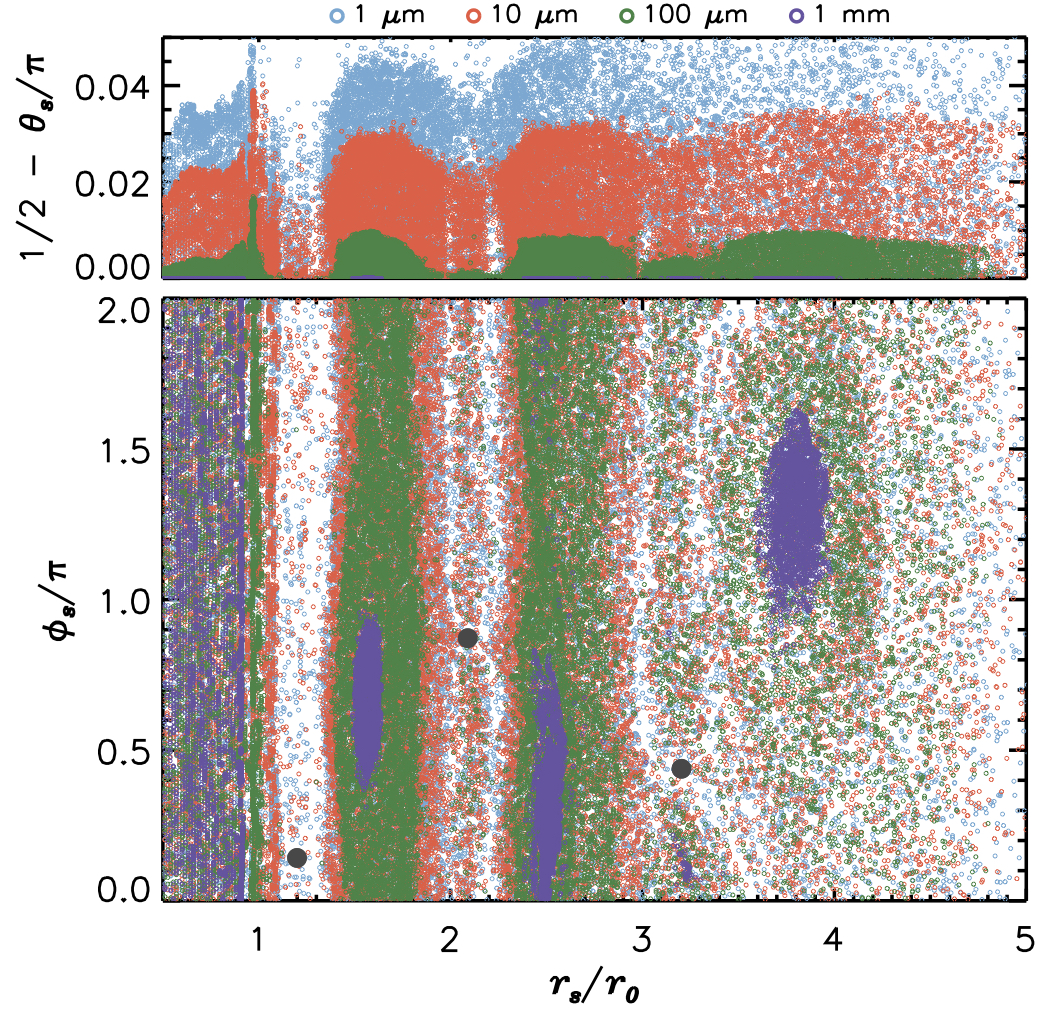}%
\includegraphics[clip]{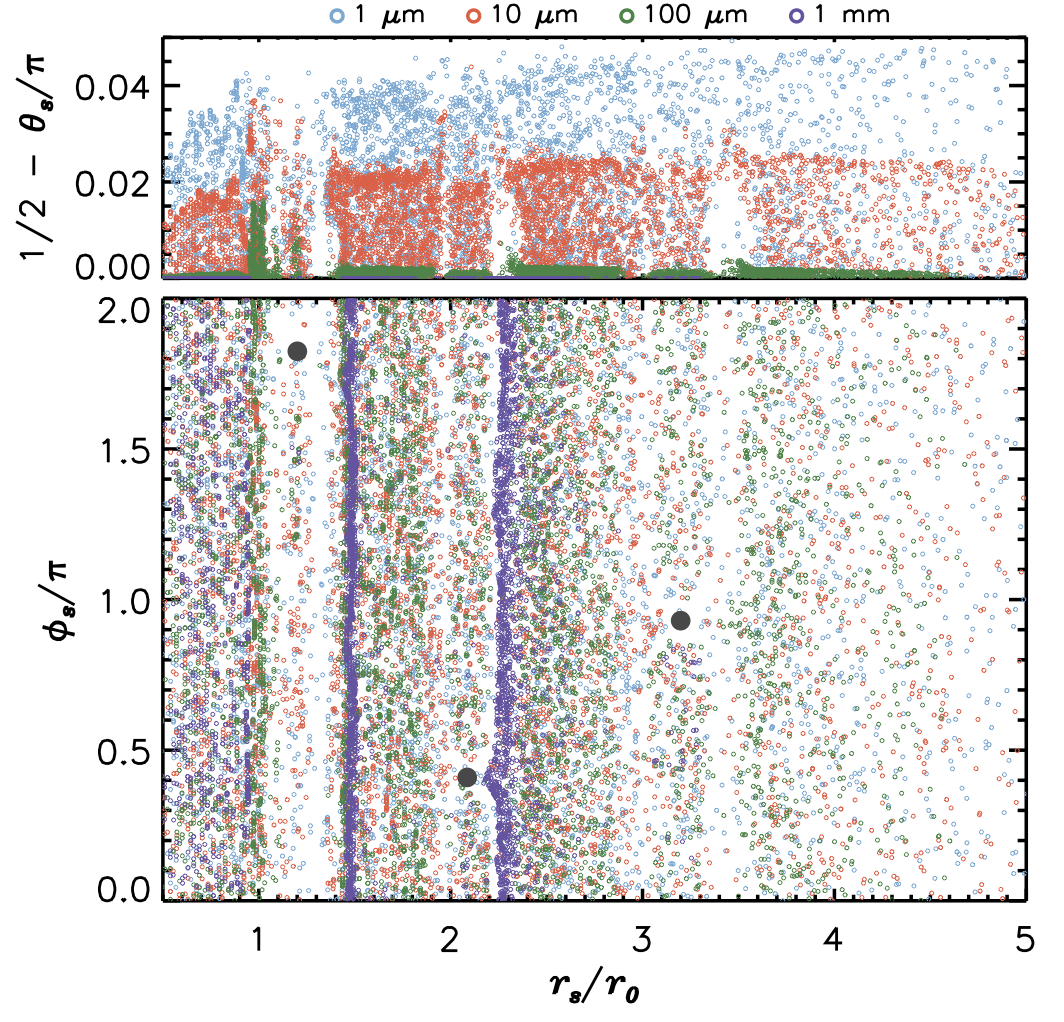}}
\caption{%
        Comparison of the dust distributions, at $t=2750$ orbits, 
        for the simulations of Figure~\ref{fig:sp_hist_comp} (case 
        with smaller number of particles on the left and larger
        number at the centre). 
        Top and bottom panels indicate the particles' positions 
        projected in the $r$-$\theta$ and $r$-$\phi$ planes 
        Particle radii are colour-coded (see legend);
        solid circles represent the planets.
        The right panel refers to a simulation with a setup
        similar to that of model \lmla, but with turbulence
        parameter $\alpha=10^{-3}$.
        (\textit{High-resolution versions of this figure can be found
        \href{https://doi.org/10.6084/m9.figshare.16926040.v1}{here}}.)
        }
\label{fig:sp_dist_comp}
\end{figure*}

The basic assumption of this study is that dusty particles are 
distributed above the disc mid-plane by planetesimals colliding
at heights $\gtrsim H$, as found in previous work \citep{turrini2019}.
Although we did not model collisions among large bodies, we checked
that model configurations used herein are compatible with this 
assumption in terms of inclination and eccentricity excitation
driven by the planets. 
We performed a calculation based on \hmha\ and including bodies 
ranging in radius between $0.1$ and $100\,\mathrm{km}$,
initilized on near-circular ($e_{s}<0.01$) and non-inclined
($i_{s}<0.5^{\circ}$) orbits. Results show that, after about
$600$ orbits (at $r=r_{0}$), $43$\% of the bodies have $e_{s}>0.2$ 
and $46\%$ have $i_{s}>H/a_{s}$, in accord with prior work.

In addition to the initial distribution of grains used in 
the calculations discussed above (see Section~\ref{sec:SE}), 
we also tested an initial distribution comprising $112000$ particles, 
equally divided in four size bins of radius
$R_{s}=1\,\mu\mathrm{m}$, $10\,\mu\mathrm{m}$, $100\,\mu\mathrm{m}$, 
and $1\,\mathrm{mm}$. 
The test is intended to probe the extent to which results may be 
affected by small-number statistics in some disc regions.

The orbital characteristics of the initial (normalized) distributions
are displayed in the top panels of Figure~\ref{fig:sp_hist_comp}.
As discussed above, the initial distribution of eccentricities is 
not important since orbits are circularised relatively quickly.
The details of the initial inclination distribution are not expected 
to affect much the results of the calculations, as long as dust 
grains are initially present at heights $\gtrsim H$ above the disc
mid-plane. 
The middle and bottom rows of the Figure show 
the evolution at $t=2750$ orbits (at $r=r_{0}$), that is, $250$ 
orbits after the deployment of the solids, for the calculations 
with lower (middle) and higher (bottom) number of grains.
The normalized histograms, including all particles, use the same
bin sizes (in the middle and bottom panels) and display similar
features for all orbital quantities. Therefore, small-number statistics
does not significantly affect the results presented and discussed above.

Two-dimensional dust distributions for the same cases are reported
in Figure~\ref{fig:sp_dist_comp}. Results from the calculation with
the smaller number of particles are displayed on the left and those
with the larger number of particles at the centre. The distributions
in the $r$-$\theta$ planes show very similar characteristics.
Similar traits also appear in the $r$-$\phi$ distributions, which
clearly trace the accumulation of grains inside three vortices,
exterior to each of the planets' orbit.
At this epoch, enhanced concentrations mainly involve $1\,\mathrm{mm}$ 
grains, which have the shortest coupling time with the gas.
But smaller grains can also be seen to locally concentrate,
especially in the most external vortex.
As discussed above, over longer timescales, also $\sim 1\,\mu\mathrm{m}$
dust orbiting close to the mid-plane is collected in vortices
(see Figures~\ref{fig:sp_dist_a16}, \ref{fig:sp_dist_hma15}
and \ref{fig:sp_dist_dhma15}).
The emergence of vortices requires low levels of turbulence in
the gas \citep[as are predicted for \HD, e.g.,][]{liu2018}.
The right panel of the Figure shows the dust distributions
in a simulation whose setup follows that of model \lmla, but with
a higher gas viscosity, corresponding to a turbulence parameter
$\alpha=10^{-3}$. In this case, the dust distribution in
the $r$-$\phi$ plane clearly points to the absence of vortices
in the gas, and $1\,\mathrm{mm}$ particles are collected 
at locations in proximity  of the edges of the tidal gaps 
of the two innermost planets, where 
$\langle u_{\phi}/u^{*}_{\phi} \rangle>1$.

\begin{figure*}
\centering%
\resizebox{2.05\columnwidth}{!}{\includegraphics[clip]{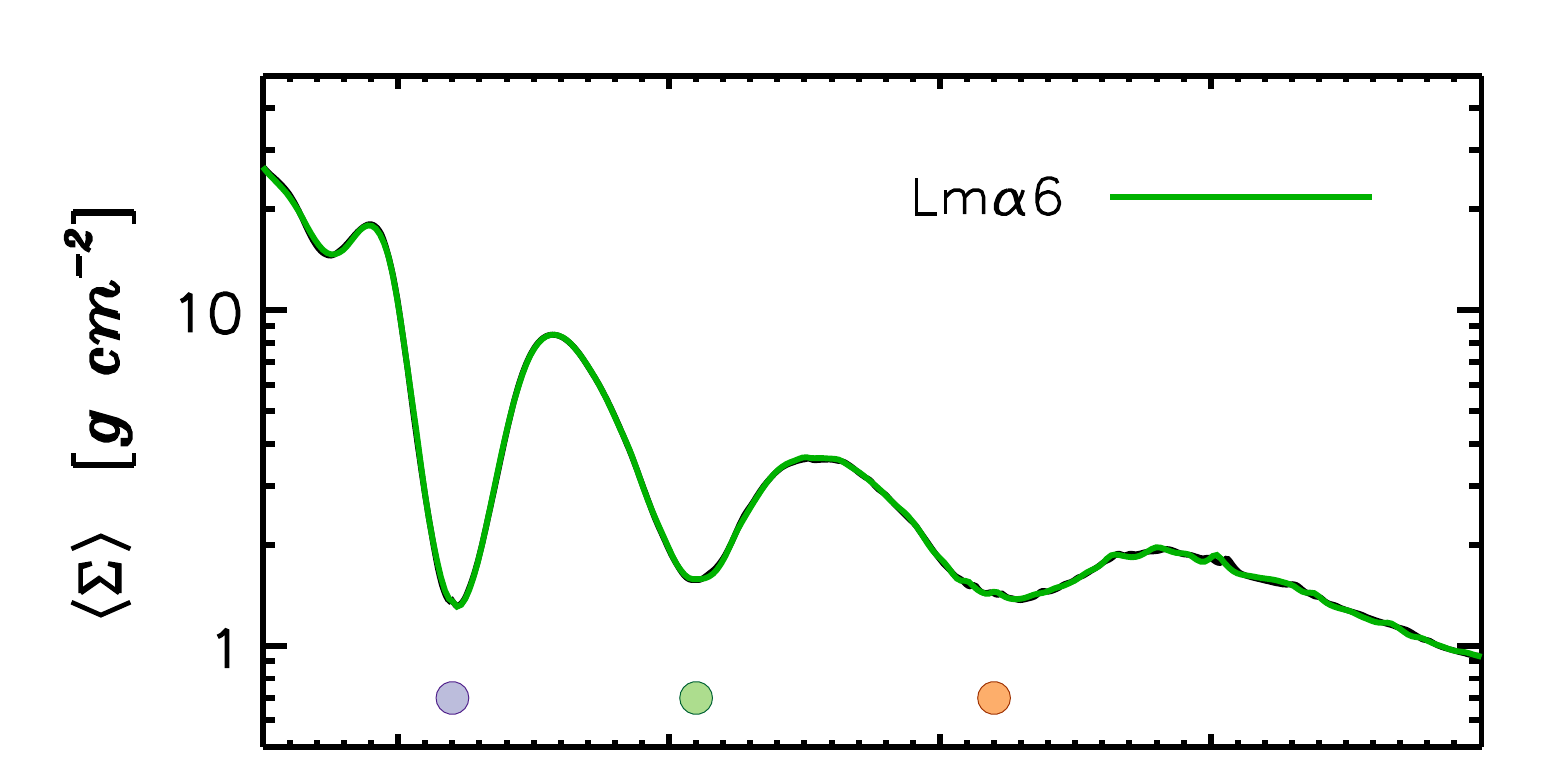}%
                             \includegraphics[clip]{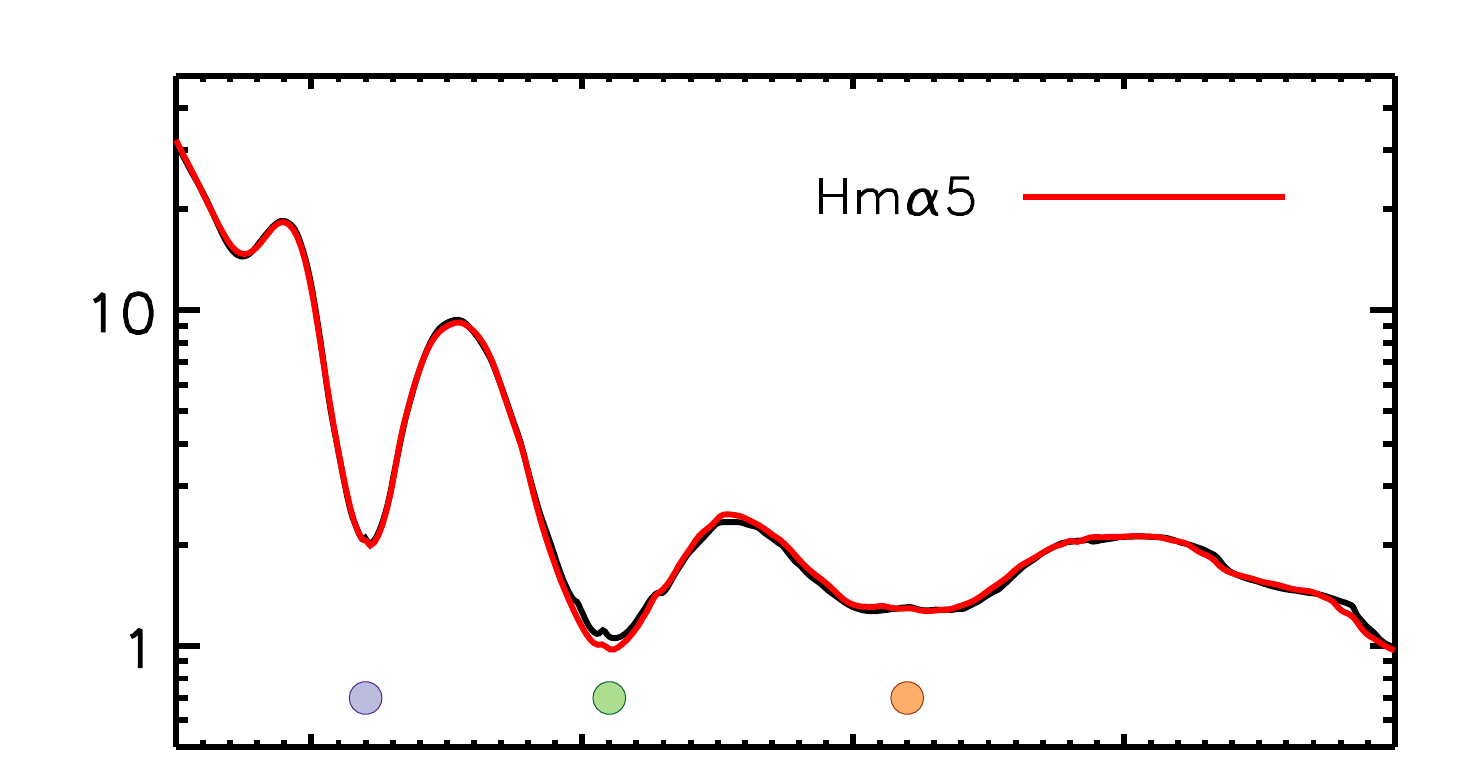}%
                             \includegraphics[clip]{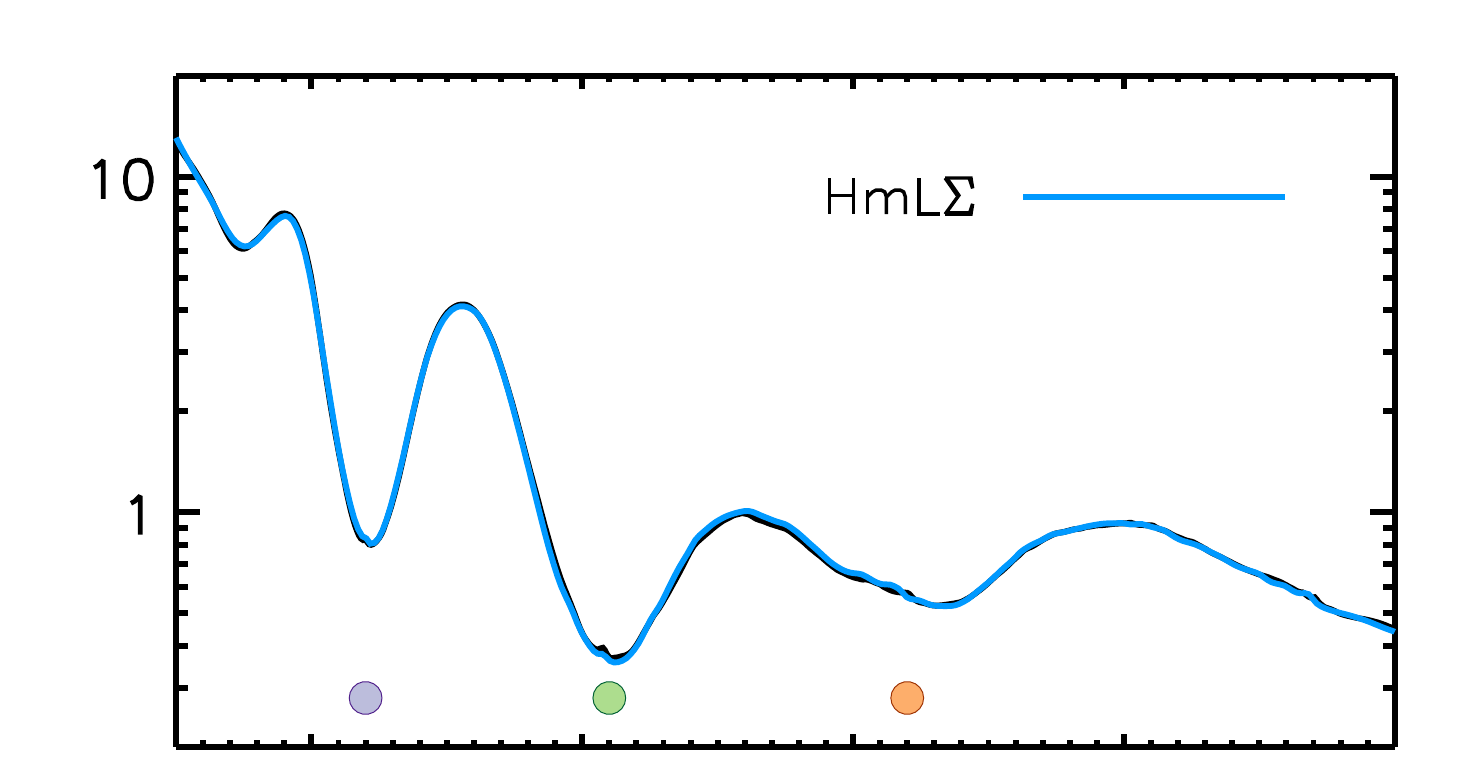}}
\resizebox{2.05\columnwidth}{!}{\includegraphics[clip]{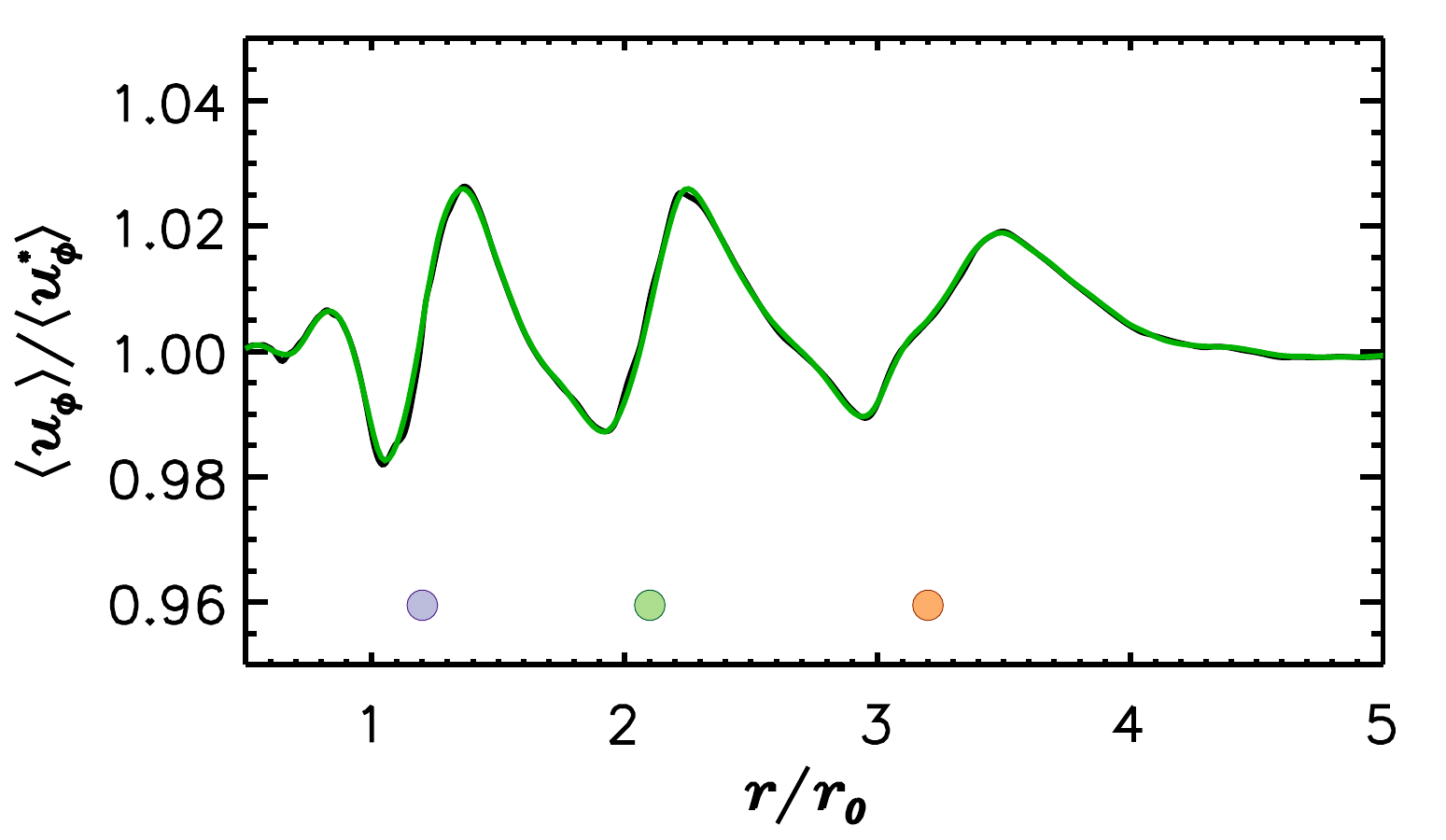}%
                             \includegraphics[clip]{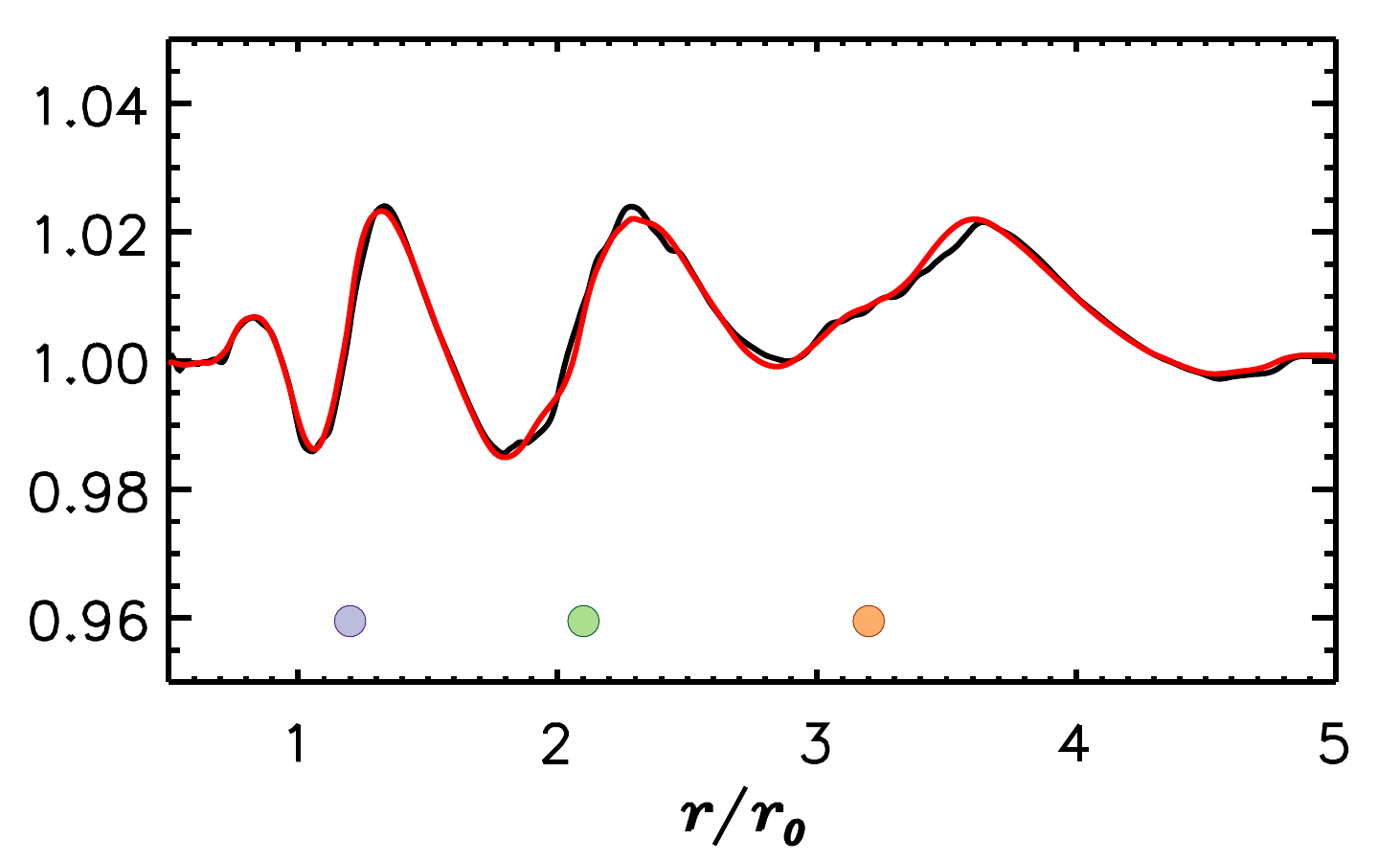}%
                             \includegraphics[clip]{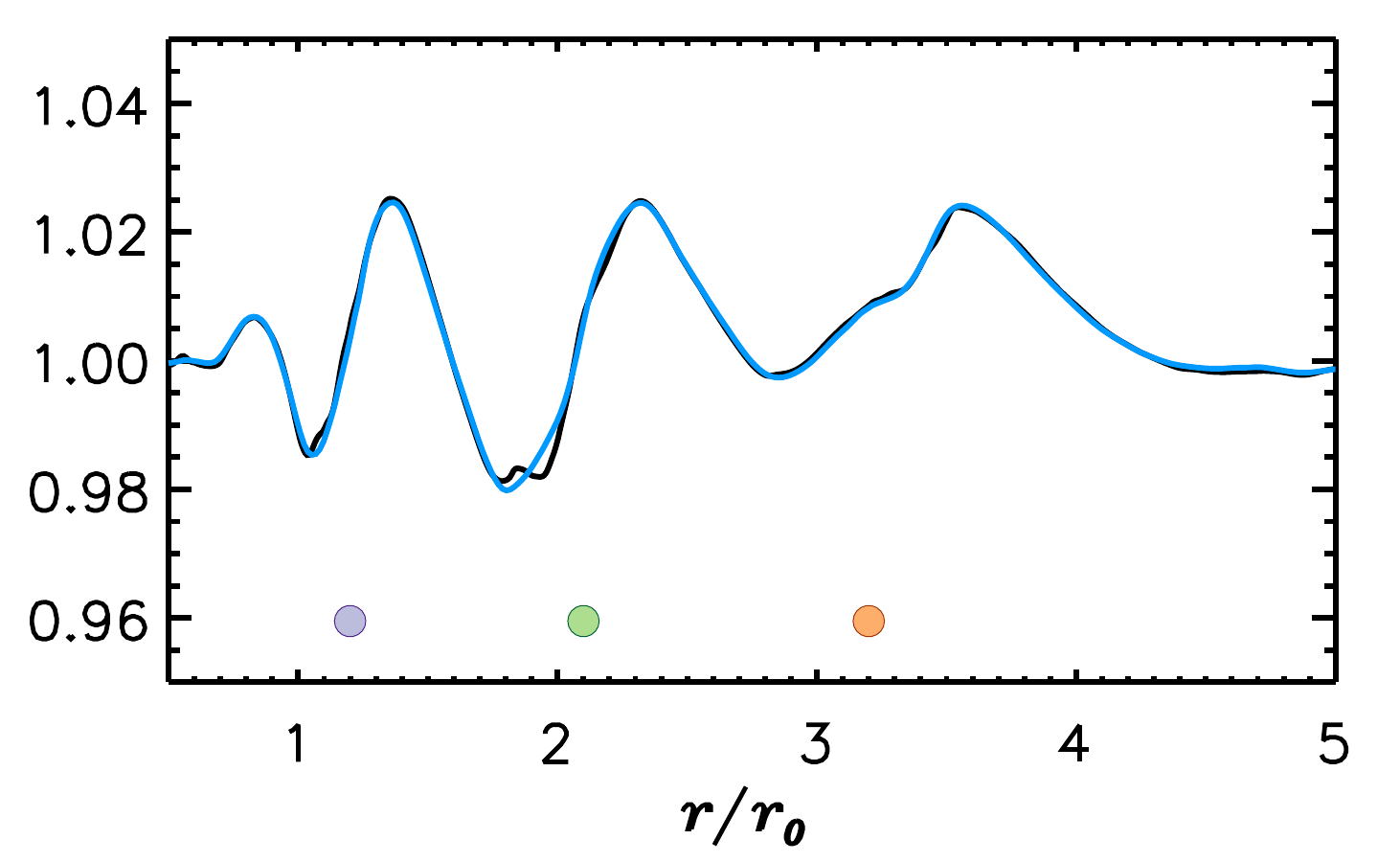}}
\resizebox{2.05\columnwidth}{!}{\includegraphics[clip]{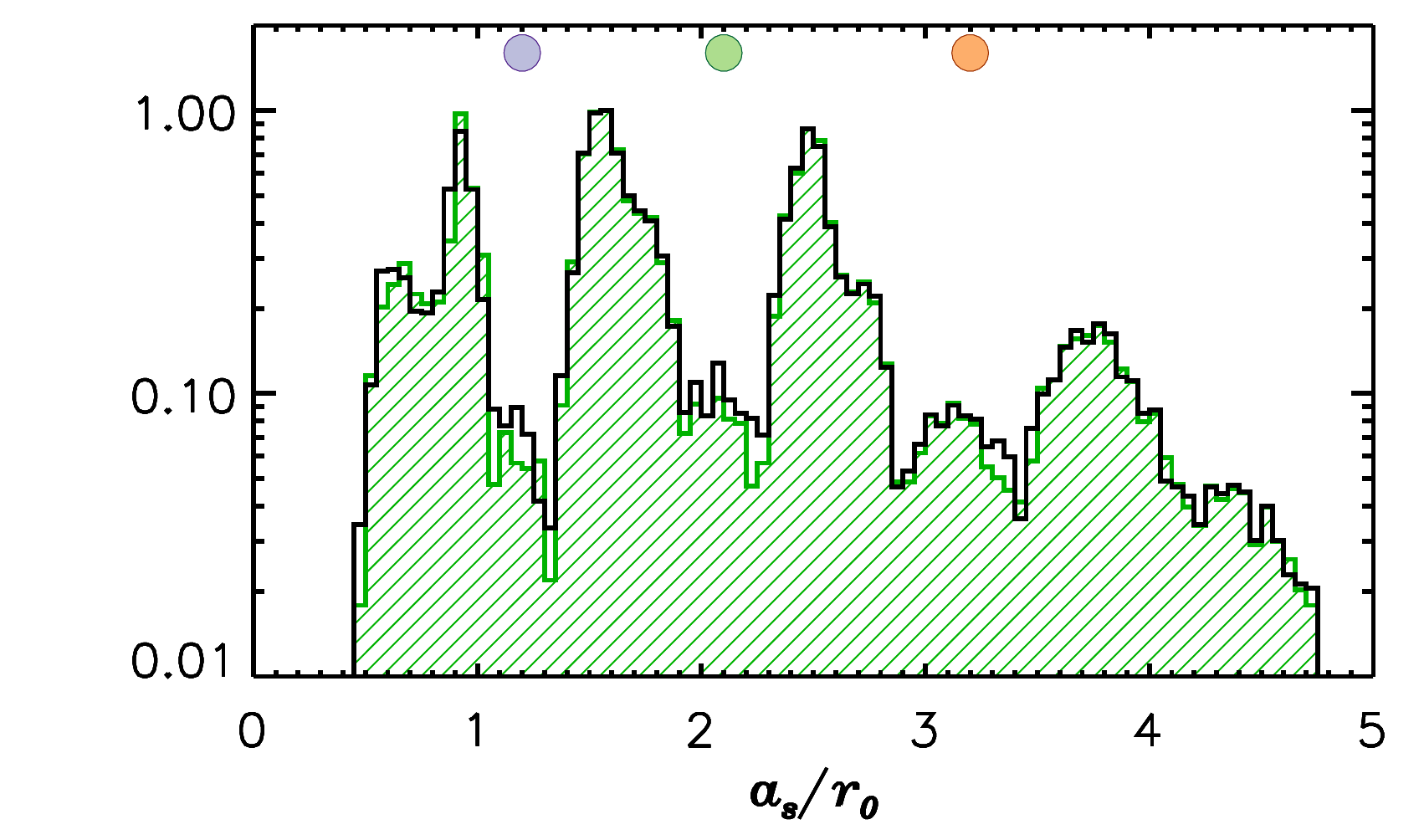}%
                             \includegraphics[clip]{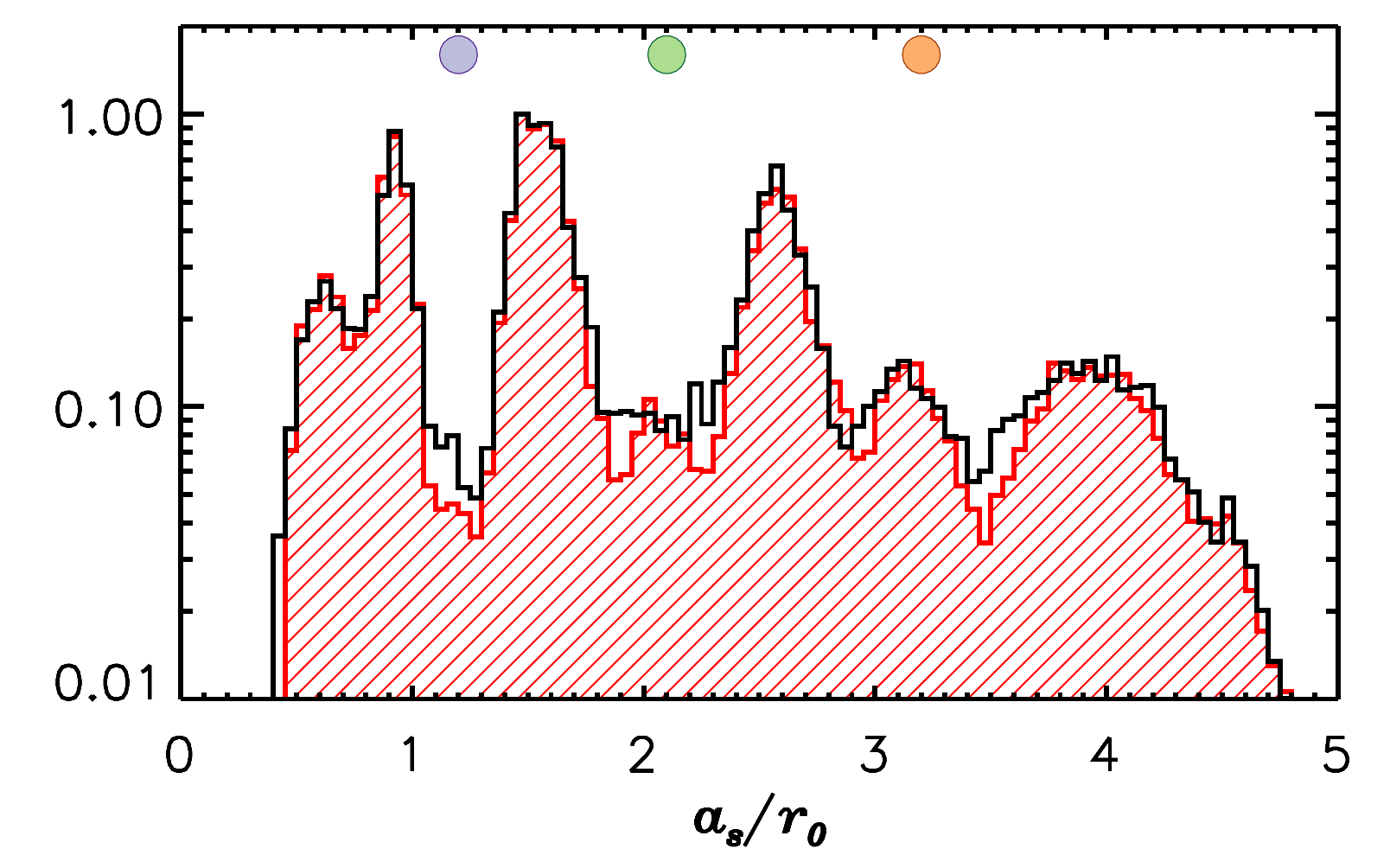}%
                             \includegraphics[clip]{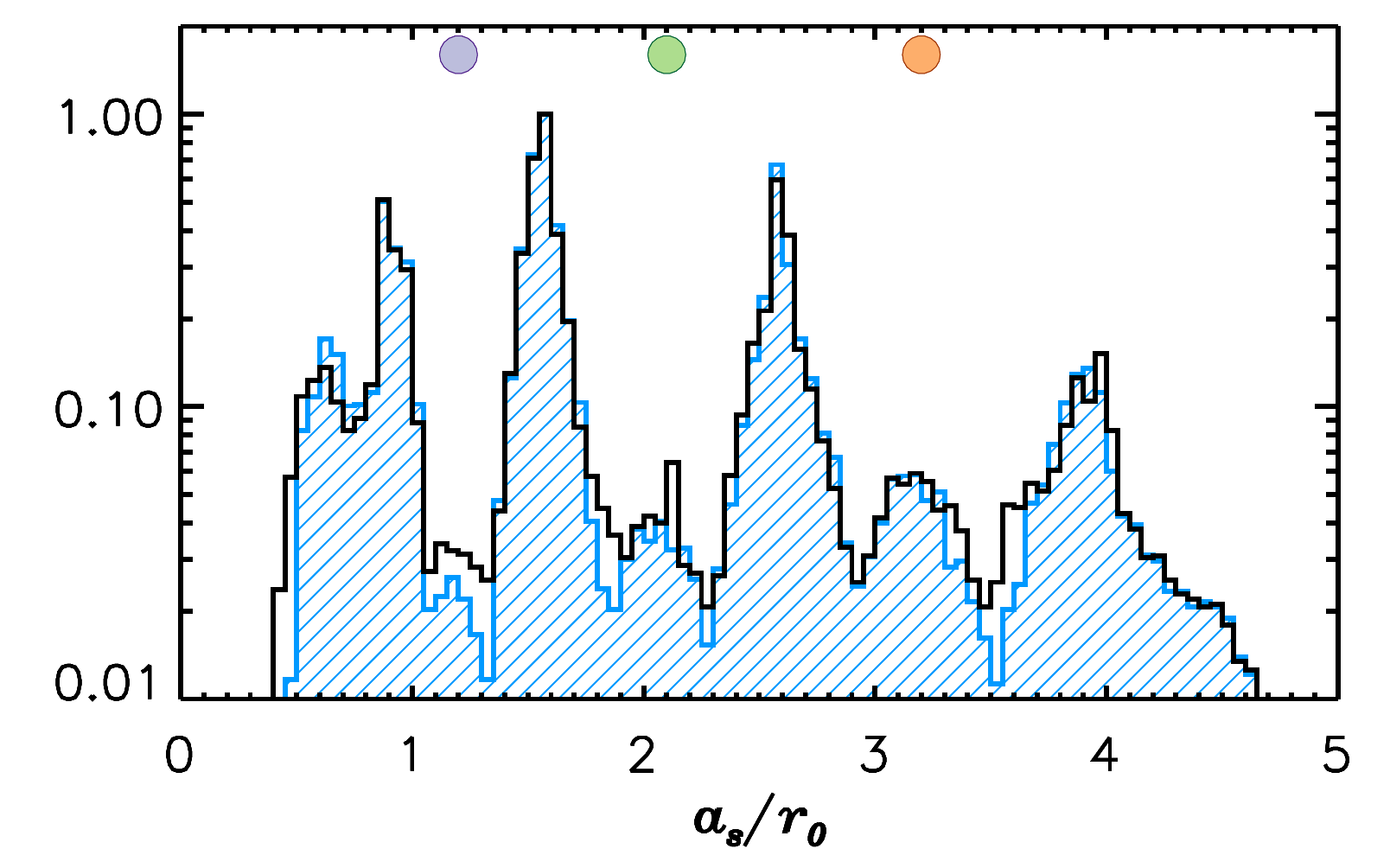}}
\caption{%
        Comparison of models at different resolutions, for the averaged
        surface density (top), the azimuthal velocity perturbation
        (middle), and the total dust distribution (bottom), as indicated.
        The black curves represent the high-resolution calculations 
        (see text for details).
        }
\label{fig:res_comp}
\end{figure*}
The three models in Table~\ref{tab:cases} were also tested at a grid 
resolution twice as high in each direction ($1332\times 42\times 842$ 
grid elements) as the standard resolution, a factor of $8$ enhancement 
in volume resolution.
The hydrodynamic variables from each model (at standard resolution) were 
interpolated on the high-resolution grid at the time when particles were 
deployed and then the high-resolution models were evolved for 
$\approx 250$ orbits at $r=r_{0}$.
Figure~\ref{fig:res_comp} displays a comparison of bulk quantities
at standard (coloured curves) and high (black curves) resolutions.
Both the surface density ($2\pi$-averaged around the star, top) and 
the perturbation on the azimuthal velocity ($u_{\phi}/u^{*}_{\phi}$, 
see Section~\ref{sec:R}, middle) are very similar in each pair of
simulations, throughout the disc's region where planets orbit. 
The bottom panels report the histograms of the total dust distributions, 
and there is general agreement of the particles' evolution at 
the two resolutions. Therefore, resolution effects are marginal
and likely insignificant in the simulations presented herein.


\bsp	
\label{lastpage}
\end{document}